\documentclass[aps,reprint,nofootinbib,nobibnotes,notitlepage,superscriptaddress,onecolumn,prd,amsmath,amssymb
]{revtex4-2}

\usepackage[caption=false]{subfig}
\usepackage{braket}
\usepackage{float}
\usepackage{lipsum}
\usepackage{graphicx}
\usepackage{dcolumn}
\usepackage{bm}
\usepackage{natbib}
\usepackage{hyperref}
\hypersetup{
	colorlinks = true,
    linkcolor = Red,
    urlcolor  = Red,
    citecolor = Red
}

\usepackage{microtype}
\usepackage{gensymb}
\usepackage{verbatim}
\usepackage{siunitx}
\usepackage{tabularx}
\usepackage[dvipsnames]{xcolor}
\usepackage{wasysym}
\allowdisplaybreaks
\usepackage[skip=4pt plus1pt, indent=10pt]{parskip}

\usepackage{tikz}
\usepackage{afterpage}

\def\be{\begin{equation}}

\usepackage{color}
\definecolor{darkgreen}{RGB}{0,120,0}

\definecolor{darkgreen}{RGB}{0,120,0}

\newcommand{\Mpch}{h^{-1}\mathrm{Mpc}}
\newcommand{\hMpc}{h\,\mathrm{Mpc}^{-1}}

\newcommand{\delD}[1]{(2\pi)^3\delta_\mathrm{D}\left({#1}\right)}

\newcommand{\av}[1]{\left\langle{#1}\right\rangle} 

\newcommand{\vk}{\vec k}
\newcommand{\hk}{\hat{\vec k}}

\newcommand{\vp}{\vec p}

\newcommand{\vq}{\vec q}
\newcommand{\vx}{\vec x}

\newcommand{\hq}{\hat{\vec q}}
\newcommand{\vy}{\vec y}

\newcommand{\hn}{\hat{\vec n}}

\definecolor{darkgreen}{RGB}{0,120,0}
\newcommand{\Pim}[1]{\Pi^{[#1]}}
\newcommand{\Tr}{\mathrm{tr}\,}
\usepackage{mathtools}
\usepackage{enumitem}
\usepackage{longtable}
\usepackage{booktabs}
\newcommand{\spacer}{\qquad\qquad\qquad\qquad\qquad\qquad}
\newcommand{\dK}{\delta^{\rm K}}

\DeclareSymbolFont{toneletters}{T1}{\familydefault}{m}{it}
\SetSymbolFont{toneletters}{bold}{T1}{\familydefault}{bx}{it}

\DeclareMathSymbol\edth{\mathord}{toneletters}{"F0}

\usepackage{adjustbox}
\usepackage{dashbox}
\usepackage[normalem]{ulem}

\def\beq{\begin{eqnarray}}
\def\eeq{\end{eqnarray}}

\let\vec\mathbf

\usepackage{empheq}

\usepackage{xspace}

\begin{document}

\setlength{\parskip}{0pt}

\title{The Redshift-Space Galaxy Power Spectrum at Two-Loop Order}

\author{Oliver~H.\,E.~Philcox}
\email{ohep2@cantab.ac.uk}
\affiliation{Kavli Institute for Particle Astrophysics and Cosmology, Stanford}
\affiliation{Leinweber Institute for Theoretical Physics, Stanford}

\begin{abstract} 
    \noindent Contemporary large-scale structure analyses center around the galaxy power spectrum, modeled with one-loop perturbation theory. This description naturally breaks down at large wavenumbers, prohibiting the extraction of cosmological information from small-scale clustering. In this work, we present a complete model for the redshift-space galaxy power spectrum up to two-loop order, derived in the framework of the Effective Field Theory of Large Scale Structure. This accounts for a wide variety of effects, including fifth-order galaxy bias, redshift-space distortions, small-scale non-linearities, spatial non-locality, long-wavelength bulk flows, coordinate distortions, and stochasticity, involving a total of 49 free coefficients. Our model additionally encodes all deterministic and counterterm contributions to the one-loop bispectrum and trispectrum (as well as the tree-level quadspectrum and pentaspectrum), which depend on many of the same parameters. We implement the model numerically, using the \textsc{vegas} algorithm to precompute the loop integrals; this allows the full likelihood to be evaluated in milliseconds. The approach is validated using power spectra measured from high-resolution simulations, considering both dark matter and realistic galaxy samples. The extension to two-loop order greatly increases the range of scales over which the power spectrum can be modeled, with $k_{\rm max}$ rising from $0.20h\,\mathrm{Mpc}^{-1}$ to $0.34h\,\mathrm{Mpc}^{-1}$ for a luminous red galaxy sample at $z=0.5$, or $0.27h\,\mathrm{Mpc}^{-1}$ to $0.42h\,\mathrm{Mpc}^{-1}$ for an emission line galaxy sample at $z=1.1$. Much of the additional information is absorbed by the numerous fifth-order biases and counterterms; for this reason, constraints on $\sigma_8$ are not significantly enhanced by the inclusion of two-loop terms. Future work will assess the full implications of the two-loop model for cosmological parameter inference, both alone and in combination with higher-order statistics.
\end{abstract}

\maketitle
\setlength{\parskip}{4pt}

\section{Introduction}
\noindent The power spectrum is the cornerstone of modern cosmology. Measurements of the two-point functions of galaxy positions, galaxy shapes, microwave background anisotropies, lensing potentials and beyond have been instrumental in the development of the standard cosmological model, $\Lambda$CDM, and remain the leading tool for the analysis of observational datasets. As with any statistic, the extraction of information from the power spectrum is limited by the accuracy of our theoretical models, particularly their ability to model the complex non-linear processes of dark matter collapse and galaxy formation. By improving the models, we analyze a wider range of scales and thus learn more about a wealth of physical processes in the early and late Universe.

For spectroscopic galaxy surveys, the state-of-the-art model is the one-loop power spectrum, derived with the effective field theory (EFT) of large-scale structure \citep{Baumann:2010tm,Carrasco:2012cv}. By expressing the overdensity of galaxies as a perturbative expansion in the initial conditions and accommodating small-scale uncertainties via renormalization, this provides an accurate description of the two-point statistic down to intermediate scales ($k\approx 0.2\hMpc$). Since its first application to data \citep{Ivanov:2019pdj,DAmico:2019fhj,Chen:2021wdi}, this has been used in a wide variety of studies, probing both the standard cosmological model and numerous extensions \citep[e.g.,][]{DESI:2024jxi,DESI:2024hhd,Chudaykin:2025aux,Chudaykin:2025lww,Chudaykin:2025vdh,Ivanov:2026dvl,Chudaykin:2026nls,Chudaykin:2026amr,Smith:2026bmb}, and has become the backbone of DESI and \textit{Euclid} full-shape analyses \citep{DESI:2016fyo,Laureijs:2011gra,Euclid:2024yrr}.

To learn more from the galaxy power spectrum, we must push the analysis to smaller scales. Within perturbation theory, this requires extending the EFT model to two-loop order. This is challenging: we must self-consistently model all contributions to the galaxy density field up to fifth order in the initial conditions, including bias, redshift-space distortions, and stochasticity, and carefully account for the impacts of both infrared and ultraviolet physics. Though the full model has not yet been constructed, a number of the ingredients have already been developed. \citep{Crocce:2005xy,Bernardeau:2008fa,Taruya:2012ut} derived the two-loop clustering of real-space dark matter within standard perturbation theory, while \citep{Carrasco:2013mua,Baldauf:2015aha,Carrasco:2013sva,Foreman:2015lca,Foreman:2015uva} obtained the renormalized EFT result (with corrections for neutrinos and exact time-dependence presented in \citep{Garny:2022fsh,Fasiello:2022lff} and extension to three-loops developed in \citep{Blas:2013aba,Konstandin:2019bay}), whose efficient implementation was considered in \citep{Schmittfull:2016yqx,Simonovic:2017mhp,Bakx:2025jwa,Anastasiou:2025jsy}. In real-space, the complete two-loop model for galaxies was derived in \citep{Bakx:2025cvu,Bakx:2026rmd,Ivanov:2026zos}, though the redshift-space theory has been presented only for dark matter \citep{Taule:2023izt}. Thus far, there has been no applications to data, except for cosmic shear \citep{Chen:2026usz,Saraivanov:2026sxc}.

In this work, we build the first complete two-loop model for the galaxy power spectrum in redshift-space, extending the real-space models of \citep{Bakx:2026rmd,Ivanov:2026zos}. By building an efficient numerical implementation and comparing to high-resolution simulations, we show that the two-loop model can describe the data across a much larger range of scales than the one-loop theory, which validates our approach and motivates future exploration of its dependence on cosmological parameters beyond $\sigma_8$. This work also facilitates joint analyses of the power spectrum and higher-order statistics (building on \citep{Philcox:2022frc,DAmico:2022osl,Bakx:2025pop} for the one-loop bispectrum): our fifth-order model also describes the deterministic contributions to the one-loop bispectrum and trispectrum, as well as the tree-level quadspectrum and pentaspectrum (building on the one-loop bispectrum and trispectrum developments of \citep{Baldauf:2014qfa,Angulo:2014tfa,Eggemeier:2018qae,Eggemeier:2021cam,Steele:2020tak} and \citep{Bertolini:2016bmt,Steele:2021lnz,Bakx:2026rmd}). The combination of statistics is expected to yield tight constraints on cosmological parameters upon future application to data.

The remainder of this work is as follows. In \S\ref{sec: theory}, we derive the two-loop power spectrum model, whose implementation is discussed in \S\ref{sec: implementation}. \S\ref{sec: simulations} presents the application to numerical simulations, before we conclude in \S\ref{sec: conclusions}. Appendix \ref{app: time-nonlocality} describes the effects of time non-locality, while Appendix \ref{app: class-pt} compares our one-loop results to those of \textsc{ClassPT}. In Appendix \ref{app: dm-renorm}, we present a detailed discussion of dark matter renormalization at fifth order, before we list the (lengthy) galaxy kernels and correlators in Appendix \ref{app: kernels}. 

\section{Theoretical Model}\label{sec: theory}
\noindent In this section, we carefully derive each ingredient in the two-loop power spectrum model, including bias (\S\ref{sec: bias}), redshift-space distortions (\S\ref{sec: rsd}), loop integrals (\S\ref{sec: loops}), infrared resummation (\S\ref{sec: ir-resum}), stochasticity (\S\ref{sec: stoch}), and coordinate distortions (\S\ref{sec: ap}). We pay particular attention to renormalization in both real- and redshift-space (\S\ref{sec: renorm-real}\,\&\,\ref{sec: renorm-rsd}), building on the dark matter discussion of Appendix \ref{app: dm-renorm}. Our main results are summarized in \S\ref{sec: theory-summary}, with the real- and redshift-space operators of our model summarized in Tabs.\,\ref{tab: operators}\,\&\,\ref{tab: rsd-table}.

\subsection{Bias Expansion}\label{sec: bias}
\noindent The bias expansion encodes the relation of the galaxy overdensity, $\delta_g$, to the underlying dark matter overdensity and velocity potential, $\delta$ and $\theta$, which are themselves sourced by the linear density field, $\delta_{\rm L}$. In this work, we adopt the Eulerian basis proposed in \citep{Mirbabayi:2014zca} (following \citep{Bakx:2025cvu,Bakx:2026rmd}, see also \citep{Senatore:2014eva,Assassi:2014fva,McDonald:2009dh,Angulo:2015eqa}), which starts from the building blocks
\beq\label{eq: Pi-ij}
	\Pi^{[n]}_{ij}(\vx,z)\equiv\frac{1}{(n-1)!}\big(D_\eta-(n-1)\big)\,\Pi^{[n-1]}_{ij}(\vx,z),\qquad \Pim{1}_{ij}(\vx,z) = \frac{\partial_{ij}}{\nabla^2}\delta(\vx,z)
\eeq
where $D_\eta\equiv\partial_\eta-(\partial_k\theta/\nabla^2)\partial_k$ is a convective derivative, $\eta \equiv \log D$ for growth factor $D(z)$, and the $n=1$ term is proportional to the tidal field, $\partial_{ij}\Phi$, for gravitational potential $\Phi$. Note that all terms involve the non-linear density and velocity fields, such that $\Pim{n}$ contains contributions at $m$-th order in $\delta_{\rm L}$ for all $m\geq n$. 

To form the galaxy density field, we sum all possible scalar operators formed from $\Pi_{ij}^{[n]}$ up to a given order in fields and derivatives, subject to a set of symmetries: homogeneity, isotropy, parity, locality-in-space, and the Einstein equivalence principle. Due to the convective time derivatives (which trace the evolution along the matter fluid flow), the basis naturally incorporates non-locality in time \citep[cf.][]{Donath:2023sav,Edison:2025bmj}; this is demonstrated explicitly in Appendix \ref{app: time-nonlocality}. Up to second order, the basis contains three terms:
\beq\label{eq: delta-g-o2}
	\delta_g(\vx,z) = c_{1,0}\Tr\Pim{1} + c_{2,0}\Tr[(\Pim{1})^2] + c_{3,0}(\Tr[\Pim{1}])^2+\cdots,
\eeq
where the bias coefficients, $c_{a,b}$, are free functions of redshift, and we have excluded the degenerate operator $\Tr\Pim{2}$. This is equivalent to the commonly-adopted `Galilean' basis \citep{Assassi:2014fva} (used in \textsc{class-pt} \cite[e.g.,][]{Chudaykin:2020aoj}):
\beq
	\delta_g(\vx,z) = b_1\delta + b_2\frac{\delta^2}{2} + b_{\mathcal{G}_2}\mathcal{G}_2+\cdots,
\eeq
identifying $b_1 = c_{1,0}$, $b_2 = 2(c_{3,0}+c_{2,0})$ and $b_{\mathcal{G}_2} = c_{2,0}$. To compute the one-loop power spectrum, \eqref{eq: delta-g-o2} must be supplemented by two third-order terms, $\Tr[\Pim{1}\Pim{2}]$ and $\nabla^2\Tr\Pim{1}$, which map to the $\Gamma_3$ and $\nabla^2\delta$ operators from \citep{Assassi:2014fva}.\footnote{The $\Gamma_3$ operator is conventionally defined as $\Gamma_3 = \mathcal{G}_2[\Phi_\delta]-\mathcal{G}_2[\Phi_\theta]$, where $\mathcal{G}_2[\Phi] \equiv (\partial_{ij}\Phi)^2 - (\nabla^2\Phi)^2$ and $\Phi_\delta,\Phi_\theta$ are the potentials associated to $\delta$ and $\theta$. At cubic order, this is equal to $\tfrac{4}{5}\Tr[\Pim{1}\Pim{2}] - \tfrac45\Tr[(\Pim{1})^3]+\tfrac47\delta\,\Tr[(\Pim{1})^2]-\tfrac47\delta^3$ (but differs at higher order). Given that only the first term contributes to the one-loop power spectrum, we can set $c_{4,0}=\tfrac54b_{\Gamma_3}$ when comparing one-loop calculations.}

Here, we require the bias expansion up to fifth-order. Following \citep{Bakx:2025cvu}, this can be written
\beq\label{eq: bias-expansion}
	\delta_{g,\rm det}(\vx,z) =  \sum_{a,b} c_{a,b}(z)\,\mathcal{O}_{a,b}(\vx,z),
\eeq
where $\mathcal{O}_{a,b}$ is the $a$-th operator at $b$-th order in derivatives and $c_{a,b}$ is its corresponding bias parameter. These operators are given explicitly in Tab.\,\ref{tab: operators} (using a different ordering to \citep{Bakx:2026rmd} to streamline the discussion) and constitute every non-trivial scalar combination of the $\Pim{n}$ tensors and $\partial_i$ spatial gradients up to fifth order (obtained as described in \S\ref{subsec: impl-ops}). In contrast to \citep{Bakx:2026rmd}, we assume that the relation is local-in-space, and thus omit any operators involving inverse Laplacians (e.g., formed from $(\partial_{kl}/\nabla^2)\Pim{2}_{kl}$); the importance of such operators for renormalization will be discussed in \S\ref{sec: renorm-rsd}. In total, there are 29 local-in-space operators with $b=0$ (denoted `leading-gradient', or LG), 17 with $b=1$ (next-to-leading-gradient, NLG) and 1 with $b=2$ (next-to-next-to-leading-gradient, NNLG). While all of the above terms are required if one wishes to model the full spectrum of correlators up to the two-loop power spectrum, the one-loop trispectrum and the tree-level pentaspectrum, only 17 (renormalized) LG operators, 7 NLG operators and 1 NNLG operator appear in the two-loop power spectrum (in addition to the redshift-space and non-local contributions discussed below).\footnote{We can straightforwardly extend this basis to include parity-violating operators. This adds three terms to \eqref{eq: bias-expansion}: $\mathcal{O}^{\rm PV}_{1}=\epsilon^{abc}\Pim{1}_{bd}\partial_a\Pim{2}_{cd}$, $\mathcal{O}^{\rm PV}_{2} = \epsilon^{abc}(\partial_a\Pim{2}_{bd})\Pim{2}_{cd}$ and $\mathcal{O}^{\rm PV}_{3} = \epsilon^{abc}(\partial_a\delta)\Pim{1}_{bd}\Pim{2}_{cd}$, all of which contribute at fifth-order and above (with $\mathcal{O}_{3}^{\rm PV}$ matching \citep{Cabass:2022oap}). Since these terms are not sourced by conventional physics, and, moreover, do not contribute to the power spectrum or bispectrum \citep{Coulton:2023oug}, we omit them in this work.\label{foot: parity}} 

{\small
 \afterpage{%
  \clearpage
 \begin{longtable}{@{}l|llcccccc@{}}
    \caption[Deterministic bias operators used in this work]{\textbf{Scalar bias operators}. This table lists the deterministic bias operators used in this work. In each row, we show the bias coefficient, $c_{a,b}$, the operator definition (written in terms of the convective distortion tensors of \eqref{eq: Pi-ij} with $\Tr\Pim{1}=\delta$), the perturbative order at which it first arises, and the list of statistics it contributes to (with $nL$ indicating $n$-loop). The final section shows the non-local-in-space operators sourced by renormalization (defined in \eqref{eq: bakx-nonlocal-ops}), which are discussed in \S\ref{sec: renorm-real}. Considering all non-degenerate combinations of $\Pim{n}$ tensors and spatial gradients, we find 29 operators with zero derivatives (LG), 17 with two derivatives (NLG) and 1 with four derivatives (NNLG), with two additional non-local-in-space NLG operators. These are sufficient to describe the real-space power spectrum up to two-loop order, the bispectrum and trispectrum up to one-loop order, and the quadspectrum and pentaspectrum up to tree-level order. Additional tensor structures arise from redshift-space mappings; these are presented in Tab.\,\ref{tab: rsd-table}. Note that we rotate the $\mathcal{O}_{17,1}$ and $\tilde{O}_{2,1}$ operators to remove degenerate contributions from the two-loop power spectrum.
    }\label{tab: operators}\\

    \toprule
    Operator & Coefficient & Definition & Order
      & $P^{\rm tree}$
      & $P^{\rm 1L}$
      & $P^{\rm 1L}{+}B^{\rm 1L}$
      & $P^{\rm 2L}$
      & $P^{\rm 2L}{+}B^{\rm 1L}{+}T^{\rm 1L}$\\
    \midrule
    \endfirsthead

    \toprule
    Operator & Coefficient & Definition & Order
      & $P^{\rm tree}$
      & $P^{\rm 1L}$
      & $P^{\rm 1L}{+}B^{\rm 1L}$
      & $P^{\rm 2L}$
      & $P^{\rm 2L}{+}B^{\rm 1L}{+}T^{\rm 1L}$\\
    \midrule
    \endhead

    \bottomrule
    \endfoot

    \addlinespace[2pt]
    \multicolumn{9}{@{}l}{\quad\textbf{LG} ($b=0$)}\\[1pt]

    $\mathcal{O}_{1,0}$ & $c_{1,0}$ & $\delta$
      & 1 & $\checkmark$ & $\checkmark$ & $\checkmark$ & $\checkmark$ & $\checkmark$\\
    $\mathcal{O}_{2,0}$ & $c_{2,0}$ & $\Tr[(\Pim{1})^2]$
      & 2 & & $\checkmark$ & $\checkmark$ & $\checkmark$ & $\checkmark$\\
    $\mathcal{O}_{3,0}$ & $c_{3,0}$ & $\delta^2$
      & 2 & & $\checkmark$ & $\checkmark$ & $\checkmark$ & $\checkmark$\\
    $\mathcal{O}_{4,0}$ & $c_{4,0}$ & $\Tr[\Pim{1}\Pim{2}]$
      & 3 & & $\checkmark$ & $\checkmark$ & $\checkmark$ & $\checkmark$\\
    $\mathcal{O}_{5,0}$ & $c_{5,0}$ & $\delta^3$
      & 3 & & & $\checkmark$ & $\checkmark$ & $\checkmark$\\
    $\mathcal{O}_{6,0}$ & $c_{6,0}$ & $\delta\,\Tr[(\Pim{1})^2]$
      & 3 & & & $\checkmark$ & $\checkmark$ & $\checkmark$\\
    $\mathcal{O}_{7,0}$ & $c_{7,0}$ & $\Tr[(\Pim{1})^3]$
      & 3 & & & $\checkmark$ & $\checkmark$ & $\checkmark$\\
    $\mathcal{O}_{8,0}$ & $c_{8,0}$ & $\Tr[\Pim{1}\Pim{1}\Pim{2}]$
      & 4 & & & $\checkmark$ & $\checkmark$ & $\checkmark$\\
    $\mathcal{O}_{9,0}$ & $c_{9,0}$ & $\delta\,\Tr[\Pim{1}\Pim{2}]$
      & 4 & & & $\checkmark$ & $\checkmark$ & $\checkmark$\\
    $\mathcal{O}_{10,0}$ & $c_{10,0}$ & $\Tr[\Pim{1}\Pim{3}]$
      & 4 & & & $\checkmark$ & $\checkmark$ & $\checkmark$\\
    $\mathcal{O}_{11,0}$ & $c_{11,0}$ & $\Tr[\Pim{2}\Pim{2}]$
      & 4 & & & $\checkmark$ & $\checkmark$ & $\checkmark$\\
    $\mathcal{O}_{12,0}$ & $c_{12,0}$ & $\Tr[\Pim{1}\Pim{2}\Pim{2}]$
      & 5 & & & & $\checkmark$ & $\checkmark$\\
    $\mathcal{O}_{13,0}$ & $c_{13,0}$ & $\delta\,\Tr[\Pim{2}\Pim{2}]$
      & 5 & & & & $\checkmark$ & $\checkmark$\\
    $\mathcal{O}_{14,0}$ & $c_{14,0}$ & $\Tr[\Pim{1}\Pim{1}\Pim{3}]$
      & 5 & & & & $\checkmark$ & $\checkmark$\\
    $\mathcal{O}_{15,0}$ & $c_{15,0}$ & $\delta\,\Tr[\Pim{1}\Pim{3}]$
      & 5 & & & & $\checkmark$ & $\checkmark$\\
    $\mathcal{O}_{16,0}$ & $c_{16,0}$ & $\Tr[\Pim{2}\Pim{3}]$
      & 5 & & & & $\checkmark$ & $\checkmark$\\
    $\mathcal{O}_{17,0}$ & $c_{17,0}$ & $\Tr[\Pim{1}\Pim{4}]$
      & 5 & & & & $\checkmark$ & $\checkmark$\\
    $\mathcal{O}_{18,0}$ & $c_{18,0}$ & $\delta^4$
      & 4 & & & & & $\checkmark$\\
    $\mathcal{O}_{19,0}$ & $c_{19,0}$ & $\delta\,\Tr[(\Pim{1})^3]$
      & 4 & & & & & $\checkmark$\\
    $\mathcal{O}_{20,0}$ & $c_{20,0}$ & $\delta^2\,\Tr[(\Pim{1})^2]$
      & 4 & & & & & $\checkmark$\\
    $\mathcal{O}_{21,0}$ & $c_{21,0}$ & $(\Tr[(\Pim{1})^2])^2$
      & 4 & & & & & $\checkmark$\\
    $\mathcal{O}_{22,0}$ & $c_{22,0}$ & $\delta^2\,\Tr[\Pim{1}\Pim{2}]$
      & 5 & & & & & $\checkmark$\\
    $\mathcal{O}_{23,0}$ & $c_{23,0}$ & $\Tr[\Pim{1}\Pim{1}]\Tr[\Pim{1}\Pim{2}]$
      & 5 & & & & & $\checkmark$\\
    $\mathcal{O}_{24,0}$ & $c_{24,0}$ & $\delta\,\Tr[\Pim{1}\Pim{1}\Pim{2}]$
      & 5 & & & & & $\checkmark$\\
    $\mathcal{O}_{25,0}$ & $c_{25,0}$ & $\delta^5$
      & 5 & & & & &\\
    $\mathcal{O}_{26,0}$ & $c_{26,0}$ & $\delta^2\,\Tr[(\Pim{1})^3]$
      & 5 & & & & &\\
    $\mathcal{O}_{27,0}$ & $c_{27,0}$ & $\delta^3\,\Tr[(\Pim{1})^2]$
      & 5 & & & & &\\
    $\mathcal{O}_{28,0}$ & $c_{28,0}$ & $\Tr[(\Pim{1})^3]\Tr[(\Pim{1})^2]$
      & 5 & & & & &\\
    $\mathcal{O}_{29,0}$ & $c_{29,0}$ & $\delta\,(\Tr[(\Pim{1})^2])^2$
      & 5 & & & & &\\

    \addlinespace[2pt]
    \multicolumn{9}{@{}l}{\quad\textbf{NLG: Local} ($b=1$)}\\[1pt]

    $\mathcal{O}_{1,1}$ & $c_{1,1}$ & $\nabla^2\delta$
      & 3 & & $\checkmark$ & $\checkmark$ & $\checkmark$ & $\checkmark$\\
    $\mathcal{O}_{2,1}$ & $c_{2,1}$ & $\nabla^2\Tr[(\Pim{1})^2]$
      & 4 & & & $\checkmark$ & $\checkmark$ & $\checkmark$\\
    $\mathcal{O}_{3,1}$ & $c_{3,1}$ & $\nabla^2\delta^2$
      & 4 & & & $\checkmark$ & $\checkmark$ & $\checkmark$\\
    $\mathcal{O}_{4,1}$ & $c_{4,1}$ & $\partial^{kl}[\delta\,\Pim{1}_{kl}]$
      & 4 & & & $\checkmark$ & $\checkmark$ & $\checkmark$\\
    $\mathcal{O}_{5,1}$ & $c_{5,1}$ & $\delta\,\nabla^2\delta$
      & 4 & & & $\checkmark$ & $\checkmark$ & $\checkmark$\\
    $\mathcal{O}_{6,1}$ & $c_{6,1}$ & $\nabla^2\Tr[\Pim{1}\Pim{2}]$
      & 5 & & & & $\checkmark$ & $\checkmark$\\
    $\mathcal{O}_{7,1}$ & $c_{7,1}$ & $\partial^{kl}[\Pim{1}\Pim{2}]_{kl}$
      & 5 & & & & $\checkmark$ & $\checkmark$\\
    $\mathcal{O}_{8,1}$ & $c_{8,1}$ & $\nabla^2\delta^3$
      & 5 & & & & & $\checkmark$\\
    $\mathcal{O}_{9,1}$ & $c_{9,1}$ & $\partial^{kl}[\delta^2\Pim{1}_{kl}]$
      & 5 & & & & & $\checkmark$\\
    $\mathcal{O}_{10,1}$ & $c_{10,1}$ & $\nabla^2[\delta\,\Tr[(\Pim{1})^2]]$
      & 5 & & & & & $\checkmark$\\
    $\mathcal{O}_{11,1}$ & $c_{11,1}$ & $\partial^{kl}[\Tr[(\Pim{1})^2]\Pim{1}_{kl}]$
      & 5 & & & & & $\checkmark$\\
    $\mathcal{O}_{12,1}$ & $c_{12,1}$ & $\nabla^2\Tr[(\Pim{1})^3]$
      & 5 & & & & & $\checkmark$\\
    $\mathcal{O}_{13,1}$ & $c_{13,1}$ & $\partial^{kl}[\delta\,(\Pim{1}\Pim{1})_{kl}]$
      & 5 & & & & & $\checkmark$\\
    $\mathcal{O}_{14,1}$ & $c_{14,1}$ & $\delta^2\,\nabla^2\delta$
      & 5 & & & & & $\checkmark$\\
    $\mathcal{O}_{15,1}$ & $c_{15,1}$ & $\delta\,\Tr[\Pim{1}\nabla^2\Pim{1}]$
      & 5 & & & & & $\checkmark$\\
    $\mathcal{O}_{16,1}$ & $c_{16,1}$ & $\partial^k\delta\,\Tr[\Pim{1}\partial_k\Pim{1}]$
      & 5 & & & & & $\checkmark$\\
    $\mathcal{O}_{17,1}$ & $c_{17,1}$ & $\Tr[\Pim{2}\nabla^2\Pim{1}]-2\mathcal{O}_{7,1}+\mathcal{O}_{6,1}$
      & 5 & & & & & $\checkmark$\\

    \addlinespace[2pt]
    \multicolumn{9}{@{}l}{\quad\textbf{NLG: Non-Local} ($b=1$)}\\[1pt]

    $\tilde{\mathcal{O}}_{1,1}$ & $\tilde c_{1,1}$ & $\hat\Pi^{[3]}$
      & 5 & & & & $\checkmark$ & $\checkmark$\\
    $\tilde{\mathcal{O}}_{2,1}$ & $\tilde c_{2,1}$ & $\Pi_{B1}-\Pi_{B2}-\tfrac{7}{5}\mathcal{O}_{7,1}$
      & 5 & & & & & $\checkmark$\\

    \addlinespace[2pt]
    \multicolumn{9}{@{}l}{\quad\textbf{NNLG} ($b=2$)}\\[1pt]

    $\mathcal{O}_{1,2}$ & $c_{1,2}$ & $\nabla^4\delta$
      & 5 & & & & $\checkmark$ & $\checkmark$\\

  \end{longtable}
}
}

Introducing bias-dependent convolution kernels, $K_n$, the deterministic contributions to the galaxy density field can be expanded in powers of the linear density field:
\beq\label{eq: delta-g-expansion-real}
    \delta_{g,\rm det}(\vk,z)=\sum_{n\ge1} \int_{\vq_1\cdots\vq_n}\delD{\vk-\vq_1\cdots-\vq_n}K_n(\vq_1,\dots,\vq_n,z)\,\delta_{\rm L}(\vq_1,z)\cdots\delta_{\rm L}(\vq_n,z),
\eeq
for $\int_{\vq}\equiv \int d\vq/(2\pi)^3$, where $\delta_{\rm L}$ is related to the linear power spectrum by $\av{\delta_{\rm L}(\vk,z)\delta_{\rm L}(\vk',z)}=\delD{\vk+\vk'}P_{\rm L}(k,z)$. Up to third-order, the unsymmetrized LG kernels are given by
\beq
	&&K^{\rm LG}_1[c_{1,0}](\vk) = c_{1,0}\\\nonumber
	&&K^{\rm LG}_2[c_{1,0},c_{2,0},c_{3,0}](\vq_1,\vq_2) = c_{1,0}F_2(\vq_1,\vq_2) + c_{2,0}(\hq_1\cdot\hq_2)^2+c_{3,0}\\\nonumber
	&&K^{\rm LG}_3[c_{1,0},\ldots,c_{7,0}](\vq_1,\vq_2,\vq_3) = c_{1,0}F_3(\vq_1,\vq_2,\vq_3) + 2c_{2,0}(\hq_1\cdot\hq_{23})^2F_2(\vq_2,\vq_3)+2c_{3,0}F_2(\vq_2,\vq_3)\\\nonumber
    &&\qquad\spacer\,+\,c_{4,0}\big[(\hq_{12}\cdot\hq_3)^2F_2(\vq_1,\vq_2)-\frac{\vq_1\cdot\vq_2}{q_1^2}(\hq_2\cdot\hq_3)^2\big]\\\nonumber
	&&\qquad\spacer\,+\,c_{5,0}+c_{6,0}(\hq_1\cdot\hq_2)^2+c_{7,0}(\hq_1\cdot\hq_2)(\hq_2\cdot\hq_3)(\hq_3\cdot\hq_1),
\eeq
where $\vk\equiv \vq_1+\cdots +\vq_n$, $F_n$ is the usual density kernel \citep[e.g.,][]{2002PhR...367....1B} and we drop the redshift argument for clarity. Similarly, the first two NLG kernels are
\beq
	&&K_1^{\rm NLG}[c_{1,1}](\vk) =  -c_{1,1}k^2\\\nonumber
	&&K_2^{\rm NLG}[c_{1,1},\ldots, c_{5,1}](\vq_1,\vq_2) = -c_{1,1}k^2F_2(\vq_1,\vq_2)-c_{2,1}k^2(\hq_1\cdot\hq_2)^2-c_{3,1}k^2-c_{4,1}\frac{(\vk\cdot\vq_2)^2}{q_2^2}-c_{5,1}q_2^2
\eeq
while the single NNLG kernel is
\beq
	K_1^{\rm NNLG}[c_{1,2}](\vk) &=& c_{1,2}k^4.
\eeq
These kernels (whose full forms are given in Appendix \ref{app: kernels-real}, including the additional terms generated by renormalization) can be used to compute the tree-level, one-loop, and two-loop power spectra; this is discussed in \S\ref{sec: loops}.

\subsection{Redshift-Space Mapping}\label{sec: rsd}
\noindent Redshift-space distortions impart anisotropy to the observed galaxy distribution, adding dependence on the velocity field, $\vec v_g(\vx)$ \citep[e.g.,][]{Kaiser:1987qv,Scoccimarro:2004tg}.\footnote{Throughout this work, we ignore velocity bias and line-of-sight selection effects, such that $\vec v_g \to \vec v$ on large-scales \citep[cf.,][]{Desjacques:2018pfv}.} For line-of-sight direction $\hn$, the redshift-space field, $\delta_g^{(s)}$, is related to the real-space field, $\delta_g$, by
\beq
    \delD{\vk}+\delta_g^{(s)}(\vk,z) &=& \int d\vx\,e^{-i\vk\cdot\vx}\left(1+\delta_g(\vx,z)\right)e^{-if(z)k_{\parallel}u_{g,\parallel}(\vx,z)}
\eeq
where $u_g^i \equiv v_g^i/(afH)$, $f\equiv d\log D/d\log a$ for Hubble parameter $H$, growth factor $D$, and scale factor $a$, denoting line-of-sight projection by $v_\parallel \equiv \vec v\cdot\hn$. Expanding the exponential, we find
\beq\label{eq: rsd-mapping}
    \delD{\vk}+\delta_g^{(s)}(\vk,z) &=& \sum_{M\geq 0}\frac{(-if(z)k_\parallel)^M}{M!}\int d\vx\,e^{-i\vk\cdot\vx}\left(1+\delta_g(\vx,z)\right)\left(u_\parallel(\vx,z)\right)^M.
\eeq
The deterministic contributions to $\delta_g^{(s)}$ can be written as a perturbative expansion for the redshift-space field (cf.\,\eqref{eq: delta-g-expansion-real}) by expressing the velocity in terms of $\theta \equiv -\partial_i u^i$ and expanding both pieces in powers of $\delta_{\rm L}$:\footnote{Note that we assume $u_g = u$ and drop transverse velocity contributions, which are not sourced within SPT. Corrections to both assumptions are provided by redshift-space counterterms (\S\ref{sec: renorm-rsd}).}
\beq\label{eq: delta-g-expansion-rsd}
    \delta_{g,\rm det}^{(s)}(\vk,z)=\sum_{n\ge1} \int_{\vq_1\cdots\vq_n}\delD{\vk-\vq_1-\cdots-\vq_n}Z_n(\vq_1,\dots,\vq_n,z)\,\delta_{\rm L}(\vq_1)\cdots\delta_{\rm L}(\vq_n),
\eeq
where the redshift-space kernels, $Z_n$, depend on both bias coefficients, $c_{a,b}$, and $f(z)$. Up to second order, the unsymmetrized LG kernels take the form
\beq
    &&Z^{\rm LG}_1(\vk) = K^{\rm LG}_1(\vk)+f\mu^2\\\nonumber
    &&Z^{\rm LG}_2(\vq_1,\vq_2) = K^{\rm LG}_{2}(\vq_1,\vq_2)+fk\mu\,K^{\rm LG}_1(\vq_1)\frac{\nu_2}{q_2}+f\mu^2G_2(\vq_1,\vq_2)+\frac{(fk\mu)^2}{2}\frac{\nu_1}{q_1}\frac{\nu_2}{q_2}
\eeq
for $\nu_{i}\equiv\hq_{i}\cdot\hn$ and $\mu\equiv\hk\cdot\hn$. Similarly, the leading NLG and NNLG kernels are
\beq
    &&Z^{\rm NLG}_1(\vk) = K^{\rm NLG}_1(\vk), \quad Z^{\rm NLG}_2(\vq_1,\vq_2) = K^{\rm NLG}_{2}(\vq_1,\vq_2)+fk\mu\,K^{\rm NLG}_1(\vq_1)\frac{\nu_2}{q_2}, \quad Z^{\rm NNLG}_1(\vk) = K^{\rm NNLG}_1(\vk).
\eeq
The explicit forms of all required kernels (including counterterm corrections, cf.\,\S\ref{sec: renorm-rsd}) are given in Appendix \ref{app: kernels-rsd}; in the limit of $f\to 0$, these reduce to the real-space kernels, $K_n$.

\subsection{Renormalization: Real-Space}\label{sec: renorm-real}
\noindent Non-linear processes induce corrections to the large-scale clustering of dark matter and galaxies \citep{Baumann:2010tm,Carrasco:2012cv}. Within the EFT formalism, this is captured by renormalization, which supplements the above description with counterterms encoding the dependence of $\delta_g$ on ultraviolet (UV) physics. Practically, this ensures that the loop integrals are cut-off-independent in a Wilsonian manner \citep[e.g.,][]{Cabass:2019lqx,Rubira:2023vzw}. Three sources of UV dependence are relevant to this work; these arise from dark matter microphysics, galaxy formation, and redshift-space velocities. Below, we present the deterministic pieces (involving $\delta_{\rm L}$) in real-space; redshift-space contributions and stochastic terms will be described in \S\ref{sec: renorm-rsd}\,\&\,\ref{sec: stoch} respectively.

\subsubsection{Dark Matter}
\noindent As discussed in \citep{Baumann:2010tm,Carrasco:2012cv,Carroll:2013oxa,Pajer:2013jj,Abolhasani:2015mra}, the renormalization of the dark matter density and velocity fields, $\delta$ and $v^i$, can be obtained directly from the fluid equations. To solve the system perturbatively, it is necessary to smooth the continuity, Euler and Poisson equations, which necessarily induces non-ideal effects, such as sound-speeds and viscosities. Such effects are encapsulated by the stress tensor, $\tau^{ij}$, which itself can be expanded in long-wavelength fields. Degrees of freedom contained within $\tau^{ij}$ source new terms in $\delta$; for example, a piece proportional to $\partial^{ij}\Phi$ generates a density counterterm of the form $\nabla^2\delta$, which is degenerate with the NLG operator $\mathcal{O}_{1,1}$ from Tab.\,\ref{tab: operators}. A detailed discussion of dark matter renormalization up to fourth-order in fields and derivatives was presented in \citep{DAmico:2022ukl}; in Appendix \ref{app: dm-renorm}, we extend this to fifth-order for both densities and velocities, additionally allowing for arbitrary time-dependence in $\tau^{ij}$ (extending \citep{DAmico:2022ukl,Anastasiou:2025jsy,Raikman:2026gcx}). 

The net result of this calculation is a set of counterterms for the dark matter density and velocity fields, which are derived self-consistently from a local-in-space expansion of the stress tensor. Including all contributions relevant to the two-loop power spectrum, the one-loop bispectrum, and the one-loop trispectrum, we find the following counterterms for the real-space dark matter field at NLG and NNLG order:
\beq\label{eq: dm-renorm-main}
	\delta^{\rm NLG}_{\rm ct}(\vx,z) &=& D^2(z)\sum_{a=1}^{13}T^{(\delta)}_{a,1}(z)\mathcal{O}_{a,1}(\vx,z)+D^2(z)\sum_{a=1}^{2}\tilde{T}^{(\delta)}_{a,1}(z)\tilde{\mathcal{O}}_{a,1}(\vx,z)+\cdots\\\nonumber
	\delta^{\rm NNLG}_{\rm ct}(\vx,z) &=& D^4(z)T^{(\delta)}_{1,2}(z)\mathcal{O}_{1,2}(\vx,z)+\cdots,
\eeq
where $T^{(\delta)}_{a,b}(z)$ are unknown time-dependent functions with $T^{(\delta)}_{5,1}=0$.\footnote{Note that only operators with external gradients contribute to the dark matter renormalization, thus $\nabla^2\delta^2$ is present but not $\delta\nabla^2\delta$ (or any $\mathcal{O}_{a,1}$ with $a=5$ or $a>13$).}  In addition to the regular $\mathcal{O}_{a,b}$ operators listed in Tab.\,\ref{tab: operators}, the dark matter renormalization sources two \textit{non-local-in-space operators}, $\tilde{\mathcal{O}}_{a,b}$, involving inverse Laplacians. These are presented in the final panel of Tab.\,\ref{tab: operators}, and can be written in terms of the $\hat\Pi^{[n]}$, $\Pi_{B1}$ and $\Pi_{B2}$ operators introduced in \citep{Bakx:2026rmd}:
\beq\label{eq: bakx-nonlocal-ops}
	\hat\Pi^{[n]}&=&\frac{1}{(n-1)!}\,\nabla^2\Big[(D_\eta-(n{-}1))\,\frac{\partial_{kl}}{\nabla^2}\Pi^{[n-1]}_{kl}\Big]
-\partial^{kl}\Pi^{[n]}_{kl} \quad (n\geq 2),\\\nonumber
    \Pi_{B1}&=&\nabla^2\left[\delta\frac{\partial^{kl}}{\nabla^2}\Pi^{[2]}_{kl}\right], \qquad \Pi_{B2}=\partial^{ij}\left[\Pim{1}_{ij}\frac{\partial^{kl}}{\nabla^2}\Pi^{[2]}_{kl}\right].
\eeq
As discussed in Appendix \ref{app: dm-renorm} (and found empirically in \citep{Bakx:2026rmd}), the latter two operators appear only in the combination $\Pi_{B1}-\Pi_{B2}$ (which can be written in terms of the projector, $\delta_{\rm K}^{ij}-\partial^{ij}/\nabla^2$).\footnote{In practice, we use the combination $\Pi_{B1}-\Pi_{B2}-\tfrac75\mathcal{O}_{7,1}$ to eliminate a degeneracy present in the two-loop power spectrum.} Furthermore, if one fixes the time-dependence of $\tau^{ij}$ to that required to capture the cut-off dependence of the loop integrals (\textit{i.e.}\ ignores finite contributions to the stress-tensor), only the combination $13\,\hat{\Pi}^{[3]}-11(\Pi_{B1}-\Pi_{B2})$ appears; moreover, its amplitude is fixed by $T^{(\delta)}_{4,1}$. Physically, these terms are sourced by the backreaction of lower-order density and velocity counterterms through the dark matter equations of motion, and contribute only at fifth order and above. In total, the dark matter density counterterm contains 15 degrees of freedom at fifth order, reducing to $13$ if the time-dependence of $\tau^{ij}$ is fixed.

\subsubsection{From Dark Matter to Galaxies}
\noindent The dark matter counterterms contribute to the galaxy density field through the deterministic bias expansion of \eqref{eq: bias-expansion}. This induces shifts in the NLG and NNLG coefficients; for example, inserting the leading-order dark matter counterterm $D^2T^{(\delta)}_{1,1}\nabla^2\delta$ in the galaxy expansion $\delta_g = c_{1,0}\delta+\cdots$ leads to a redefinition of the gradient bias: 
\beq
	c_{1,1}(z) \to  c_{1,1}(z)+c_{1,0}(z)D^2(z)T^{(\delta)}_{1,1}(z)
\eeq
(\textit{i.e.} $b_{\nabla^2}\to b_{\nabla^2}+b_1c_s^2$ in the \textsc{class-pt} notation).\footnote{Since the bias expansion contains non-linear terms, this also sources NLG operators that were absent in \eqref{eq: dm-renorm-main}. For example, the $\delta\nabla^2\delta$ term is generated by the LG operators $(\Tr{\Pim{1}})^2$ and $\Tr{[\Pim{1}\Pim{2}]}$ combined with the leading counterterm operator $\nabla^2\delta$.} 
Since only the renormalized biases can be probed with experiments, such shifts are practically unobservable. 

In addition, dark matter renormalization induces non-locality-in-space in the galaxy bias expansion. This has two sources: (1) fifth-order non-local terms in $\delta_{\rm ct}$ coupling through the LG bias operator $\mathcal{O}_{1,0}$:
\beq\label{eq: nonlocal-g-1}
    \delta_g(\vx,z)\supset \left.c_{1,0}(z) \delta_{\rm ct}(\vx,z)\right|_{\rm non-local} = c_{1,0}(z)D^2(z) \sum_{a=1}^2\tilde{T}^{(\delta)}_{a,1}(z)\tilde{\mathcal{O}}_{a,1}(\vx,z)+\cdots;
\eeq
(2) the fourth-order local operator $\mathcal{O}_{4,1}$ in $\delta_{\rm ct}$ coupling through the (non-monomial) LG bias operators $\mathcal{O}_{2,0}$ and $\mathcal{O}_{4,0}$:
\beq\label{eq: nonlocal-g-2}
    \delta_g(\vx,z)&\supset& 2c_{2,0}(z)\Tr[\Pim{1}\Pim{1}_{\rm ct}](\vx,z)+c_{4,0}(z)\Tr[\Pim{1}\Pim{2}_{\rm ct}](\vx,z)\\\nonumber
    &=&-2c_{2,0}(z)D^2(z)T^{(\delta)}_{4,1}(z)\tilde{\mathcal{O}}_{2,1}(\vx,z)-c_{4,0}(z)D^2(z)(\partial_\eta+3)T^{(\delta)}_{4,1}(z)\tilde{\mathcal{O}}_{2,1}(\vx,z)+\left(\text{local terms}\right),
\eeq
where the time derivative arises from $\Pim{2}_{ij,\rm ct}\supset (\partial_\eta-1)\Pim{1}_{ij,\rm ct}$ \eqref{eq: Pi-ij}.
If the time-dependence of the stress tensor is known (which is usually true only for the divergent piece), the ratio of the two terms in \eqref{eq: nonlocal-g-1} is fixed; this relation is broken for galaxies due to the additional sourcing via \eqref{eq: nonlocal-g-2}.

In combination, we find the following non-local addition to the real-space galaxy density field:
\beq\label{eq: delta-g-nonlocal}
	\left.\delta_{g}^{\rm NLG}(\vx,z)\right|_{\rm non{-}local} &=& \sum_{a=1}^2\tilde{c}_{a,1}(z)\tilde{\mathcal{O}}_{a,1}(\vx,z),
\eeq
where the non-locality coefficients, $\tilde{c}_{a,1}(z)$, are set by combinations of (usually unobservable) dark matter properties and bias coefficients. Since we have assumed that galaxy formation is a local-in-space (but non-local-in-time) process, all of the spatial non-locality arises as a consequence of the non-linear physics of dark matter evolution (noting that we have assumed the stress tensor, $\tau^{ij}$, to be a local function of space). Allowing for non-locality in the bias expansion itself would lead to a number of additional terms featuring inverse Laplacians (generalizing \eqref{eq: bias-expansion}): we have verified that these are not required to renormalize the one-loop trispectrum or two-loop power spectrum.\footnote{Our basis contains one fewer term than \citep{Bakx:2026rmd}, which allowed $\Pi_{B1}$ and $\Pi_{B2}$ to have independent coefficients. We exclude the additional term since it is not generated by the dark matter equations of motion nor required for renormalization.} The Fourier-space representation of the non-local terms, which form part of the $K_3^{\rm NLG}$ kernel, is presented in Appendix \ref{app: kernels-real}. 

\subsubsection{Bias Renormalization}
\noindent As discussed in \citep[e.g.,][]{McDonald:2006mx,McDonald:2009dh,Assassi:2014fva}, the bias expansion of \eqref{eq: bias-expansion} is UV-sensitive. For example, the $\delta^3$ term has the large-scale limit $3\sigma^2\delta(\vx)$, which rescales the linear bias by an amount dependent on the (possibly-divergent) variance $\sigma^2 = \av{\delta^2}$. To address this, one can replace the $\mathcal{O}_{a,b}$ operators by their renormalized equivalents, $[\mathcal{O}_{a,b}]$, defined via
\beq\label{eq: renorm-operators}
	[\mathcal{O}_{a,b}] = \mathcal{O}_{a,b} + \sum_{a',b'}\mathcal{Z}^{a',b'}_{a,b}\mathcal{O}_{a',b'}
\eeq
(up to stochastic terms), where the $\mathcal{Z}$ matrix is fixed by a suitable renormalization condition. A simple choice is to demand that the loop contributions to the correlation functions of each operator vanish in the squeezed limit, \textit{i.e.}
\beq\label{eq: renorm-scheme-lg}
    \lim_{\alpha\to0}\left(\frac{\av{[\mathcal{O}_{a,b}]\delta_{\rm L}(\alpha\vk_1)\cdots \delta_{\rm L}(\alpha\vk_m)}'_{L\text{-loop}}}{P_{\rm
  L}(\alpha k_1)\cdots P_{\rm L}(\alpha k_m)}\right)=0
\eeq
for all $L\geq 1$ and $m\geq 0$ \citep{Assassi:2014fva}. In this work, we use an alternative scheme, defined by:
\beq\label{eq: renorm-scheme}
      \lim_{\alpha\to0}\big(\frac{d^2}{d\alpha^2}\big)^N\left(\frac{\av{[\mathcal{O}_{a,b}]\delta_{\rm L}(\alpha\vk_1)\cdots \delta_{\rm L}(\alpha\vk_m)}'_{L\text{-loop}}}{P_{\rm
  L}(\alpha k_1)\cdots P_{\rm L}(\alpha k_m)}\right) = 0 \text{  for } N \leq L+b-1,
\eeq
across all $L\geq 1$ and $m\geq 0$. For LG operators, this requires that the squeezed one-loop contributions cancel at $\mathcal{O}(k_i^0)$, while the two-loop pieces vanish through $\mathcal{O}(k_i^2)$. Similarly, the NLG one-loop terms must cancel up to $\mathcal{O}(k_i^2)$.\footnote{This differs from the renormalization scheme proposed in \citep{Bakx:2026rmd}, which required the loop contributions to cancel through $\mathcal{O}(k_i^2)$ for all operators at all loop orders. Though both renormalization schemes are valid, we adopt \eqref{eq: renorm-scheme} in this work since (a) it facilitates direct comparison between the one-loop results and those of other codes (which assume \eqref{eq: renorm-scheme-lg}), and (b) it avoids absorbing large infrared-dominated contributions ($\propto k^2\sigma_v^2P_{\rm L}(k)$) into the leading-order counterterm parameters.}
This ensures that the one- and two-loop galaxy power spectra (defined below) scale as $k^2P_{\rm L}(k)$ and $k^4P_{\rm L}(k)$ in the squeezed limit(up to stochastic contributions). As an example, the LG operator $\delta^3$ is renormalized according to
\beq
	\delta^3 \to [\delta^3] = \delta^3 - 3\sigma^2\delta - \frac{68}{7}\sigma^2\delta^2+\cdots
\eeq
removing the $\mathcal{O}(k^0)$ piece \citep[e.g.,][]{Assassi:2014fva}, while the NLG operator $\delta\nabla^2\delta$ becomes
\beq
	\delta\nabla^2\delta \to [\delta\nabla^2\delta] = \delta\nabla^2\delta + \frac{82}{21}\sigma_\nabla^2\delta - \frac{16}{7}\,\sigma^2\,\nabla^2\delta +
  \cdots,
\eeq
for $\sigma_\nabla^2 \equiv \av{(\nabla\delta)^2}$, eliminating both the $\mathcal{O}(k^0)$ and $\mathcal{O}(k^2)$ contamination. By carefully computing the squeezed limits of each correlation function, we thus obtain a renormalized set of operators, which are accompanied by the renormalized bias coefficients, $[c_{a,b}]$ (with $\delta_{g,\rm det} = \sum_{a,b}c_{a,b}\mathcal{O}_{a,b}=\sum_{a,b}[c_{a,b}][\mathcal{O}_{a,b}]$). 

Following \citep{Bakx:2026rmd} we can equivalently implement the renormalization scheme by modifying the bias kernels directly. Explicitly, kernels involving a pair of momenta, such as $K_n(\vk_1,\ldots,\vk_{n-2},\vq,-\vq)$, are replaced by a UV-subtracted form, $[K_n]$, defined via:
\beq\label{eq: renorm-kernel1}
	[K^{\rm LG}_n](\vk_1,\ldots,\vk_{n-2},\vp,-\vp) &\equiv& K^{\rm LG}_n(\vk_1,\ldots,\vk_{n-2},\vp,-\vp)  - \left[\lim_{p\to \infty}K^{\rm LG}_n(\vk_1,\ldots,\vk_{n-2},\vp,-\vp)\right]_{\mathcal{O}(p^0)}\\\nonumber
	[K^{\rm NLG}_n](\vk_1,\ldots,\vk_{n-2},\vp,-\vp) &\equiv& K^{\rm NLG}_n(\vk_1,\ldots,\vk_{n-2},\vp,-\vp)  - \left[\lim_{p\to \infty}K^{\rm NLG}_n(\vk_1,\ldots,\vk_{n-2},\vp,-\vp)\right]_{\mathcal{O}(p^0,p^2)},
\eeq
for LG and NLG operators respectively. For example, the one-loop power spectrum contains $K^{\rm LG}_3(\vk,\vp,-\vp)$, for which we remove the hard-limit contributions proportional to $(k/p)^{0}$ (thus renormalizing the linear bias), but retain those proportional to $(k/p)^2$. Similarly, for kernels involving two pairs of momenta, such as $K_5(\vk,\vp,-\vp,\vq,-\vq)$, we subtract the single-hard ($p\to\infty$ or $q\to\infty$), nested ($q\to \infty$ then $p\to\infty$ or vice versa), and double-hard ($p,q\to \infty$ at fixed $p/q$) limits \citep[cf.,][]{Bakx:2026rmd,Ivanov:2026zos} (based on \citep{Bogoliubov:1957gp,Hepp:1966eg,Zimmermann:1969jj}):
\beq\label{eq: renorm-kernel2}
	[[K^{\rm LG}_n]](\vk_1,\ldots,\vk_{n-4},\vp,-\vp,\vq,-\vq) &\equiv& K^{\rm LG}_n(\vk_1,\ldots,\vk_{n-4},\vp,-\vp,\vq,-\vq)\\\nonumber
    &&\,-\,\left(\left[\lim_{q\to \infty}K^{\rm LG}_n(\vk_1,\ldots,\vk_{n-4},\vp,-\vp,\vq,-\vq)\right]_{\mathcal{O}(q^0)}+(p\leftrightarrow q)\right)\\\nonumber
    &&\,+\,\left(\left[\lim_{p\to \infty}\left[\lim_{q\to \infty}K^{\rm LG}_n(\vk_1,\ldots,\vk_{n-4},\vp,-\vp,\vq,-\vq)\right]_{\mathcal{O}(q^0)}\right]_{\mathcal{O}(p^0,p^{-2})}+(p\leftrightarrow q)\right)\\\nonumber
    &&\,-\,\left[\lim_{p,q\to \infty, \,p/q=\mathrm{fixed}}K^{\rm LG}_n(\vk_1,\ldots,\vk_{n-4},\vp,-\vp,\vq,-\vq)\right]_{\mathcal{O}(k_i^0,k_i^2)},
\eeq 
where we remove only the leading contributions from the single-hard configurations, in line with our renormalization scheme. The subtraction is performed pointwise (such that we do not average over the orientation of the hard-modes); this leads to improved convergence of the Monte Carlo integrators. Notably, renormalization forces some operators to vanish from the correlators at leading-order: for example, $\mathcal{O}_{5,0}\equiv \delta^3$ does not contribute to the one-loop power spectrum and $\mathcal{O}_{25,0}\equiv \delta^5$ enters neither the two-loop power spectrum nor the one-loop trispectrum. The operators contributing non-trivially to each spectrum are listed in Tab.\,\ref{tab: operators}. 

The NLG and NNLG operators in the galaxy bias expansion play a crucial role in renormalization by absorbing the UV-sensitive contributions excised by the kernel redefinitions. As shown in \citep{Bakx:2026rmd}, the hard limits removed by \eqref{eq: renorm-kernel1}\,\&\,\eqref{eq: renorm-kernel2} can be expressed in terms of the LG, NLG and NNLG operators; if correctly defined, the operator basis is closed under renormalization and the cut-off dependence of the loop integrals is removed. In practice, this is a powerful test: we compute the single- and double-hard limits of all kernels, average over $\hat{\vp}$ and/or $\hat{\vq}$, and verify that the resulting functions can be written in terms of operators at lower perturbative order. For example, the single-hard limits of $K^{\rm NLG}_3$ and the nested and double-hard limits of $K^{\rm LG}_5$ must be expressible in terms of $K_1^{\rm LG}$ and $K_1^{\rm NLG}$ for each operator. We have explicitly verified that the basis shown in Tab.\,\ref{tab: operators} satisfies this property, both for the renormalization scheme used herein and the NLG scheme of \citep{Bakx:2026rmd}.

\subsection{Renormalization: Redshift-Space}\label{sec: renorm-rsd}

{\small
\afterpage{%
  \clearpage
  \begin{longtable}{@{}l|llcccccc@{}}
    \caption[Redshift-space operators used in this work]{\textbf{Higher-rank operators}. List of operators contributing to the redshift-space counterterms, $C_{u^m}^{i_1\cdots i_m}$ (which renormalize the velocity cumulants $(1+\delta_g)u^{i_1}\cdots u^{i_n}$). We include all terms required to compute the redshift-space density field up to fifth order, giving the corresponding bias parameter in the second column. Note that all tensor indices should be symmetrized, e.g.,\ $C_u^{ij}\to (C_u^{ij}+C_u^{ji})/2$. The operators additionally provide a complete local-in-space basis for symmetric rank-$n$ tensors, up to order $(5-n)$ in fields and derivatives (assuming homogeneity, isotropy, Galilean invariance, the equivalence principle and parity invariance). The `non-local' terms indicate non-local-in-space contributions; we include only those inherited from the dark matter field (which is sufficient to renormalize the two-loop power spectrum, one-loop bispectrum, and one-loop trispectrum). As in Tab.\,\ref{tab: operators}, the fourth column indicates the minimum order at which a term contributes to $\delta_g$, and the remaining columns show which statistics the operator contributes to. Operators marked with $*$ are formally independent, but functionally degenerate at fixed $f$. The coefficient of the $\tilde{\mathcal{O}}^i_{8,1}$ operator is fixed to $\tilde{c}_{8,1}^{(1)}=c_{1,0}\tilde{c}_{1,1}^{(1)}$ by the equivalence principle.}\label{tab: rsd-table}\\

    \toprule
    Operator & Coefficient & Definition & Order
      & $P^{\rm tree}$
      & $P^{\rm 1L}$
      & $P^{\rm 1L}{+}B^{\rm 1L}$
      & $P^{\rm 2L}$
      & $P^{\rm 2L}{+}B^{\rm 1L}{+}T^{\rm 1L}$\\
    \midrule
    \endfirsthead

    \toprule
    Operator & Coefficient & Definition & Order
      & $P^{\rm tree}$
      & $P^{\rm 1L}$
      & $P^{\rm 1L}{+}B^{\rm 1L}$
      & $P^{\rm 2L}$
      & $P^{\rm 2L}{+}B^{\rm 1L}{+}T^{\rm 1L}$\\
    \midrule
    \endhead

    \bottomrule
    \endfoot

    \addlinespace[2pt]
    \multicolumn{9}{@{}l}{\quad\textbf{$C_u^i$: Local} ($b=1$)}\\[1pt]

    $\mathcal{O}^i_{1,1}$ & $c^{(1)}_{1,1}$ & $\partial_i\delta$
      & 3 & & $\checkmark$ & $\checkmark$ & $\checkmark$ & $\checkmark$\\
    $\mathcal{O}^i_{2,1}$ & $c^{(1)}_{2,1}$ & $\partial_i\Tr[(\Pim{1})^2]$
      & 4 & & & $\checkmark$ & $\checkmark$ & $\checkmark$\\
    $\mathcal{O}^i_{3,1}$ & $c^{(1)}_{3,1}$ & $\partial_i\delta^2$
      & 4 & & & $\checkmark$ & $\checkmark$ & $\checkmark$\\
    $\mathcal{O}^i_{4,1}$ & $c^{(1)}_{4,1}$ & $\partial_a\Pim{2}_{ia}$
      & 4 & & & $\checkmark$ & $\checkmark$ & $\checkmark$\\
    $\mathcal{O}^i_{5,1}$ & $c^{(1)}_{5,1}$ & $\partial_i\Tr[\Pim{1}\Pim{2}]$
      & 5 & & & & $\checkmark$ & $\checkmark$\\
    $\mathcal{O}^i_{6,1}$ & $c^{(1)}_{6,1}$ & $\partial_a\Pim{3}_{ia}$
      & 5 & & & & $\checkmark$ & $\checkmark$\\
    $\mathcal{O}^i_{7,1}$ & $c^{(1)}_{7,1}$ & $\partial_i\delta^3$
      & 5 & & & & & $\checkmark$\\
    $\mathcal{O}^i_{8,1}$ & $c^{(1)}_{8,1}$ & $\partial_i(\Tr[(\Pim{1})^2]\delta)$
      & 5 & & & & & $\checkmark$\\
    $\mathcal{O}^i_{9,1}$ & $c^{(1)}_{9,1}$ & $\partial_i\Tr[(\Pim{1})^3]$
      & 5 & & & & & $\checkmark$\\
    $\mathcal{O}^i_{10,1}$ & $c^{(1)}_{10,1}$ & $\partial_i\Tr[(\Pim{1})^2]\delta$
      & 5 & & & & & $\checkmark$\\
    $\mathcal{O}^i_{11,1}$ & $c^{(1)}_{11,1}$ & $\partial_a\Tr[(\Pim{1})^2]\Pim{1}_{ia}$
      & 5 & & & & & $\checkmark$\\
    $\mathcal{O}^i_{12,1}$ & $c^{(1)}_{12,1}$ & $\delta\,\partial_b\Pim{2}_{ib}$
      & 5 & & & & & $\checkmark$\\
    $\mathcal{O}^i_{13,1}$ & $c^{(1)}_{13,1}$
      & $\Pim{2}_{ab}\partial_b\Pim{1}_{ia}
         -\tfrac{35}{39}\mathcal{O}^{i}_{5,1}
         -\tfrac{10}{13}\mathcal{O}^{i}_{6,1}$
      & 5 & & & & & $\checkmark$\\
    $\mathcal{O}^i_{14,1}$ & $c^{(1)}_{14,1}$ & $\Pim{1}_{ia}\partial_b\Pim{2}_{ab}$
      & 5 & & & & & $\checkmark$\\

    \addlinespace[2pt]
    \multicolumn{9}{@{}l}{\quad\textbf{$C_u^i$: Local} ($b=2$)}\\[1pt]

    $\mathcal{O}^i_{1,2}$ & $c^{(1)}_{1,2}$ & $\nabla^2\partial_i\delta$
      & 5 & & & & $\checkmark$ & $\checkmark$\\

    \addlinespace[2pt]
    \multicolumn{9}{@{}l}{\quad\textbf{$C_u^i$: Non-Local} ($b=1$)}\\[1pt]

    $\tilde{\mathcal{O}}^i_{1,1}$ & $\tilde{c}^{(1)}_{1,1}$
      & $\frac{\partial^{iab}}{\nabla^2}\Pim{2}_{ab}$
      & 4 & & & $\checkmark$ & $\checkmark$ & $\checkmark$\\
    $\tilde{\mathcal{O}}^i_{2,1}$ & $\tilde{c}^{(1)}_{2,1}$
      & $\frac{\partial^{ia}}{\nabla^2}
         \big[\Pim{1}_{ab}\partial_c
         (\delta_{\rm K}^{bd}-\frac{\partial^{bd}}{\nabla^2})
         \Pim{2}_{cd}\big]$
      & 5 & & & & $\checkmark$ & $\checkmark$\\
    $\tilde{\mathcal{O}}^i_{3,1}$ & $\tilde{c}^{(1)}_{3,1}$
      & $\frac{\partial^{iab}}{\nabla^2}\Pim{3}_{ab}-\mathcal{O}^{i}_{6,1}$
      & 5 & & & & & $\checkmark$\\
    $\tilde{\mathcal{O}}^i_{4,1}$ & $\tilde{c}^{(1)}_{4,1}$
      & $\frac{\partial^{iab}}{\nabla^2}[\delta\,\Pim{2}_{ab}]
         -\tfrac{31}{39}\mathcal{O}^{i}_{5,1}
         -\tfrac{20}{13}\mathcal{O}^{i}_{6,1}
         -2\mathcal{O}^{i}_{13,1}$
      & 5 & & & & & $\checkmark$\\
    $\tilde{\mathcal{O}}^i_{5,1}$ & $\tilde{c}^{(1)}_{5,1}$
      & $\frac{\partial^{iab}}{\nabla^2}[\Pim{1}_{ac}\Pim{2}_{cb}]
         -\tfrac{35}{39}\mathcal{O}^{i}_{5,1}
         -\tfrac{10}{13}\mathcal{O}^{i}_{6,1}
         -\mathcal{O}^{i}_{13,1}$
      & 5 & & & & & $\checkmark$\\
    $\tilde{\mathcal{O}}^i_{6,1}$ & $\tilde{c}^{(1)}_{6,1}$
      & $\frac{\partial^{iab}}{\nabla^2}
         [\Pi^{[1]}_{ab}\Tr[(\Pim{1})^2]]$
      & 5 & & & & & $\checkmark$\\
    $\tilde{\mathcal{O}}^i_{7,1}$ & $\tilde{c}^{(1)}_{7,1}$
      & $\frac{\partial^i}{\nabla^2}(\Pi_{B1}-\Pi_{B2})
         -\tfrac{49}{39}\mathcal{O}^{i}_{5,1}
         -\tfrac{14}{13}\mathcal{O}^{i}_{6,1}
         -\tfrac{7}{5}\mathcal{O}^{i}_{13,1}$
      & 5 & & & & & $\checkmark$\\
    $\tilde{\mathcal{O}}^i_{8,1}$ & $\tilde{c}^{(1)}_{8,1}$
      & $\delta\frac{\partial^{iab}}{\nabla^2}\Pim{2}_{ab}$
      & 5 & & & & $\dagger$ & $\dagger$\\

    \midrule
    \addlinespace[2pt]
    \multicolumn{9}{@{}l}{\quad\textbf{$C_{u^2}^{ij}$: Local} ($b=1$)}\\[1pt]

    $\mathcal{O}^{ij}_{1,1}$ & $c^{(2)}_{1,1}$ & $\dK_{ij}$
      & 3 & & $\checkmark$ & $\checkmark$ & $\checkmark$ & $\checkmark$\\
    $\mathcal{O}^{ij}_{2,1}$ & $c^{(2)}_{2,1}$ & $\delta\,\dK_{ij}$
      & 3 & & $*$ & $\checkmark$ & $\checkmark$ & $\checkmark$\\
    $\mathcal{O}^{ij}_{3,1}$ & $c^{(2)}_{3,1}$ & $\Pim{1}_{ij}$
      & 3 & & $*$ & $\checkmark$ & $\checkmark$ & $\checkmark$\\
    $\mathcal{O}^{ij}_{4,1}$ & $c^{(2)}_{4,1}$ & $\Pim{2}_{ij}$
      & 4 & & & $\checkmark$ & $\checkmark$ & $\checkmark$\\
    $\mathcal{O}^{ij}_{5,1}$ & $c^{(2)}_{5,1}$ & $\delta\,\Pim{1}_{ij}$
      & 4 & & & $*$ & $\checkmark$ & $\checkmark$\\
    $\mathcal{O}^{ij}_{6,1}$ & $c^{(2)}_{6,1}$ & $\Pim{1}_{ia}\Pim{1}_{aj}$
      & 4 & & & $*$ & $\checkmark$ & $\checkmark$\\
    $\mathcal{O}^{ij}_{7,1}$ & $c^{(2)}_{7,1}$ & $\Tr[(\Pim{1})^2]\,\dK_{ij}$
      & 4 & & & $*$ & $*$ & $\checkmark$\\
    $\mathcal{O}^{ij}_{8,1}$ & $c^{(2)}_{8,1}$ & $\delta^2\,\dK_{ij}$
      & 4 & & & $*$ & $*$ & $\checkmark$\\
    $\mathcal{O}^{ij}_{9,1}$ & $c^{(2)}_{9,1}$ & $\Pim{1}_{ia}\Pim{2}_{aj}$
      & 5 & & & & $*$ & $\checkmark$\\
    $\mathcal{O}^{ij}_{10,1}$ & $c^{(2)}_{10,1}$ & $\delta\,\Pim{2}_{ij}$
      & 5 & & & & $*$ & $\checkmark$\\
    $\mathcal{O}^{ij}_{11,1}$ & $c^{(2)}_{11,1}$ & $\Tr[\Pim{1}\Pim{2}]\,\dK_{ij}$
      & 5 & & & & $*$ & $*$\\
    $\mathcal{O}^{ij}_{12,1}$ & $c^{(2)}_{12,1}$
      & $\Pim{3}_{ij}
         +\tfrac{101}{30}\mathcal{O}^{ij}_{9,1}
         -\tfrac{7}{3}\mathcal{O}^{ij}_{10,1}
         -\tfrac{7}{6}\mathcal{O}^{ij}_{11,1}$
      & 5 & & & & & $*$\\
    $\mathcal{O}^{ij}_{13,1}$ & $c^{(2)}_{13,1}$ & $\delta\,\Pim{1}_{ia}\Pim{1}_{aj}$
      & 5 & & & & & $*$\\
    $\mathcal{O}^{ij}_{14,1}$ & $c^{(2)}_{14,1}$ & $\delta^3\,\dK_{ij}$
      & 5 & & & & & $*$\\
    $\mathcal{O}^{ij}_{15,1}$ & $c^{(2)}_{15,1}$ & $\delta\,\Tr[(\Pim{1})^2]\,\dK_{ij}$
      & 5 & & & & & $*$\\
    $\mathcal{O}^{ij}_{16,1}$ & $c^{(2)}_{16,1}$ & $\Tr[(\Pim{1})^3]\,\dK_{ij}$
      & 5 & & & & & $*$\\
    $\mathcal{O}^{ij}_{17,1}$ & $c^{(2)}_{17,1}$ & $\Tr[(\Pim{1})^2]\,\Pim{1}_{ij}$
      & 5 & & & & & $*$\\
    $\mathcal{O}^{ij}_{18,1}$ & $c^{(2)}_{18,1}$ & $\delta^2\,\Pim{1}_{ij}$
      & 5 & & & & & $*$\\

    \addlinespace[2pt]
    \multicolumn{9}{@{}l}{\quad\textbf{$C_{u^2}^{ij}$: Local} ($b=2$)}\\[1pt]

    $\mathcal{O}^{ij}_{1,2}$ & $c^{(2)}_{1,2}$ & $\nabla^2\Pim{1}_{ij}$
      & 5 & & & & $\checkmark$ & $\checkmark$\\
    $\mathcal{O}^{ij}_{2,2}$ & $c^{(2)}_{2,2}$ & $\nabla^2\delta\,\dK_{ij}$
      & 5 & & & & $*$ & $*$\\

    \midrule
    \addlinespace[2pt]
    \multicolumn{9}{@{}l}{\quad\textbf{$C_{u^3}^{ijk}$: Local} ($b=2$)}\\[1pt]

    $\mathcal{O}^{ijk}_{1,2}$ & $c^{(3)}_{1,2}$ & $\partial_k\Pim{1}_{ij}$
      & 5 & & & & $\checkmark$ & $\checkmark$\\
    $\mathcal{O}^{ijk}_{2,2}$ & $c^{(3)}_{2,2}$ & $\partial_k\delta\,\dK_{ij}$
      & 5 & & & & $*$ & $*$\\

    \midrule
    \addlinespace[2pt]
    \multicolumn{9}{@{}l}{\quad\textbf{$C_{u^4}^{ijkl}$: Local} ($b=2$)}\\[1pt]

    $\mathcal{O}^{ijkl}_{1,2}$ & $c^{(4)}_{1,2}$ & $\dK_{ij}\dK_{kl}$
      & 5 & & & & $*$ & $*$\\
    $\mathcal{O}^{ijkl}_{2,2}$ & $c^{(4)}_{2,2}$ & $\delta\,\dK_{ij}\dK_{kl}$
      & 5 & & & & $*$ & $*$\\
    $\mathcal{O}^{ijkl}_{3,2}$ & $c^{(4)}_{3,2}$ & $\Pim{1}_{kl}\dK_{ij}$
      & 5 & & & & $*$ & $*$\\

  \end{longtable}
}
}
\noindent In the previous section, we discussed the renormalize of the real-space density field up to fifth order, following \citep{Bakx:2025cvu,Bakx:2026rmd}. In redshift-space, additional sources of UV-dependence arise due to contact terms generated by the redshift-space mapping \eqref{eq: rsd-mapping}, such as $(1+\delta_g)u_{g,\parallel}^2$ \citep{Pajer:2013jj,Mercolli:2013bsa,Senatore:2014vja,Perko:2016puo,Lewandowski:2015ziq,Chen:2020zjt}. Each combination of density and velocity must be separately renormalized; this can be achieved by introducing a set of higher-rank counterterms, $C_{u^m}^{i_1\cdots i_m}$, defined by
\beq\label{eq: ctmaster}
    [(1+\delta_g)u^{i_1}_g\cdots u_g^{i_M}]=(1+\delta_g)u^{i_1}\cdots u^{i_M}
+\sum_{m=0}^{M}\binom{M}{m}C_{u^m}^{(i_1\cdots i_m}u^{i_{m+1}}\cdots u^{i_M)},
\eeq
generalizing \citep{DAmico:2022ukl,Philcox:2022frc,Bakx:2025pop}. Here $C_{u^0}\equiv[\delta_{g}]-\delta_g$ is the real-space counterterm, $u^i$ is the scaled dark matter velocity, and the combination of operators on the right-hand-side ensures that each $C^{i_1\cdots i_m}_{u^m}$ is invariant under Galilean transformations, and the combination is symmetric under index permutations (denoting $M^{(ij)}=(M^{ij}+M^{ji})/2$ \textit{et cetera}). 

\subsubsection{Higher-Rank Operators}
\noindent To compute the galaxy density field up to fifth-order in fields and derivatives, we require $C^{i_1\cdots i_m}_{u^m}$ at order $(5-m)$ in fields and derivatives. As in \S\ref{sec: renorm-real}, these counterterms can have both local- and non-local-in-space contributions. To obtain the former, we write down every non-degenerate combination of $\Pim{n}$ and spatial derivatives up to the required order, which (a) are a symmetric function of $m$ spatial indices, (b) are scalars under Galilean transformations, and (c) are compatible with the symmetries of the system (homogeneity, isotropy, parity, and the equivalence principle). For example, the simplest rank-1 and rank-2 operators are $\partial^i\delta$ and $\delta_{\rm K}^{ij}$ (since $u^i$ breaks Galilean invariance); these renormalize $(1+\delta_g)u^i_g$ and $(1+\delta_g)u^i_gu^j_g$ respectively. Up to order $(5-m)$, we find the local-in-space expansion:
\beq\label{eq: Ou-expansion-local}
	\left.C_{u^m}^{i_1\cdots i_m}(\vx,z)\right|_{\rm local} = \sum_{a,b}c^{(m)}_{a,b}(z)\mathcal{O}^{i_1\cdots i_m}_{a,b}(\vx, z),
\eeq
where $c^{(m)}_{a,b}(z)$ are unknown coefficients and $\mathcal{O}^{i_1\cdots i_m}_{a,b}(\vx,z)$ are the rank-$m$ operators listed explicitly in Tab.\,\ref{tab: rsd-table}.\footnote{$b$ specifies the gradient order in the galaxy density field, such that all counterterms have $b\geq 1$.} Combining NLG and NNLG contributions, we obtain $15$ $C_u^i$ operators, $20$ $C_{u^2}^{ij}$ operators, $2$ $C_{u^3}^{ijk}$ operators and $3$ $C_{u^4}^{ijkl}$ operators.\footnote{As for the scalar case (see footnote \ref{foot: parity}), we can extend the basis to include parity-violating operators. This adds five rank-one tensors (starting at third- and fourth-order):  $\mathcal{O}^{i,\rm PV}_{1}=\epsilon^{iab}\Pim{1}_{ac}\Pim{2}_{bc}$, $\mathcal{O}^{i,\rm PV}_{2}=\epsilon^{iab}\Pim{1}_{ac}\Pim{3}_{bc}$, $\mathcal{O}^{i,\rm PV}_{3}=\epsilon^{iab}(\Pim{1}\Pim{1})_{ac}\Pim{2}_{bc}$, $\mathcal{O}^{i,\rm PV}_{4}=\delta\epsilon^{iab}\Pim{1}_{ac}\Pim{2}_{bc}$, and $\mathcal{O}^{i,\rm PV}_{5} =\epsilon^{ijk}\partial_j\partial^l\Pim{2}_{kl}$ (with the first introduced in \citep{Cabass:2022oap}), as well as two rank-two operators (starting at third-order): $\mathcal{O}^{ij,\rm PV}_{1}=\epsilon^{iab}\partial_a\Pim{2}_{bj}+(i\leftrightarrow j)$ and $\mathcal{O}^{ij,\rm PV}_{2}=\epsilon^{iab}(\partial_a\delta)\Pim{1}_{bj}+(i\leftrightarrow j)$. These contribute only to the trispectrum and beyond.}
Notably, the same set of operators can be used to describe the non-ideal stress tensor, $\tau^{ij}$, as well as the dark matter velocity counterterm, $u^i_{\rm ct}$ (see Appendix \ref{app: dm-renorm}).

\subsubsection{Spatial Non-Locality}
\noindent The non-local-in-space terms can be derived following the argument of \citep{DAmico:2022ukl}, which we extend to arbitrary order. Due to the equivalence principle, the difference between the renormalized galaxy and dark matter velocities is a higher-derivative Galilean-invariant vector, $C^i_{\Delta u}$, which must be expressable in terms of $\Pim{n}$ and derivatives \citep{Perko:2016puo}. Noting also that $[u_g^i]$ can be expressed in terms of the renormalized galaxy momentum, $[(1+\delta_g)u_g^i]$ and density, $[\delta_g]$, up to a higher-derivative contact term $C_{\rm contact}^i$ \citep{Carrasco:2013mua}, we find
\beq
    \frac{[(1+\delta_g)u_g^i]}{[1+\delta_g]} = [u_g^i]+C^i_{\rm contact} = [u^i] + C^i_{\Delta u}  + C^i_{\rm contact}.
\eeq
Using \eqref{eq: ctmaster} and extracting the counterterms via $\delta_{g,\rm ct}\equiv [\delta_g]-\delta_g$, $u^i_{\rm ct}\equiv [u^i]-u^i$, we find
\beq
    C^i_u = (1+\delta_g+\delta_{g,\rm ct})(u^i_{\rm ct}+C^i_{\Delta u}+C^i_{\rm contact}).
\eeq
This has an important interpretation: since $C_{\Delta u}^i$ and $C_{\rm contact}^i$ are local-in-space, all non-locality in $C_u^i$ must be inherited either from $\delta_{g,\rm ct}$ (starting at fifth-order in fields and derivatives) or $u_{\rm ct}^i$ (starting at third-order). In this work, we require $C_u^i$ up to fourth-order; in this case, the non-local piece collapses to $\left.C^i_u\right|_{\rm non-local}= (1+\delta_g)\left.u^i_{\rm ct}\right|_{\rm non-local}$,\footnote{This approximation is valid up to fifth-order in fields and derivatives. At sixth-order, this remains true for the NLG component, while the NNLG piece acquires contributions from $\delta_{g, \rm ct}\left.u^i_{\rm ct}\right|_{\rm non-local}$. Starting from seventh-order, non-locality arises from non-local terms in $\delta_{g,\rm ct}$ multiplying all three local-in-space vectors (which start at second-order). Finally, at eighth-order, new NNLG contributions arise from the product $\left.\delta_{g,\rm ct}\right|_{\rm non-local}\left.u_i^{\rm ct}\right|_{\rm non-local}$. Of course, such contributions will probably never be measured.} giving
\beq\label{eq: vel-nonlocal-o4}
    \left.C^{i,(3)}_u\right|_{\rm non-local} = \left.u^{i,(3)}_{\rm ct}\right|_{\rm non-local}, \qquad \left.C^{i,(4)}_u\right|_{\rm non-local} = \left.u^{i,(4)}_{\rm ct}\right|_{\rm non-local} + \delta_g^{(1)}\left.u^{i,(3)}_{\rm ct}\right|_{\rm non-local},
\eeq
with the first term matching \citep{DAmico:2022ukl}. 

In Appendix \ref{app: dm-renorm}, we derive the dark matter velocity counterterm, finding
\beq
    u^i_{\rm ct}(\vx,z) &=&-D^2(z)\left[\sum_{a=1}^{14}T^{(v)}_{a,1}(z)\mathcal{O}^i_{a,1}(\vx,z)+\sum_{a=1}^{7}\tilde{T}^{(v)}_{a,1}(z)\tilde{\mathcal{O}}^i_{a,1}(\vx,z)\right]-D^4(z)T^{(v)}_{1,2}(z)\mathcal{O}^i_{1,2}(\vx,z)+\cdots,
\eeq
where $\mathcal{O}_{a,b}^i$ and $\tilde{\mathcal{O}}_{a,b}^i$ are (some of) the local- and non-local operators given in Tab.\,\ref{tab: rsd-table}, and $T^{(v)}_{a,b}$ and $\tilde{T}^{(v)}_{a,b}$ are unknown amplitudes. Combining the non-local part of $u^i_{\rm ct}$ with \eqref{eq: vel-nonlocal-o4}, we obtain the final expression for $C_u^i$, including both local- and non-local-in-space deterministic contributions:
\beq\label{eq: Ou-expansion-rank1}
	C_{u}^{i}(\vx,z) = \left(\sum_{a=1}^{14}c^{(1)}_{a,1}(z)\mathcal{O}^{i}_{a,1}(\vx, z)+c^{(1)}_{1,2}(z)\mathcal{O}^{i}_{1,2}(\vx, z)\right)+\left(\sum_{a=1}^{7}\tilde{c}^{(1)}_{a,1}(z)\tilde{\mathcal{O}}^{i}_{a,1}(\vx, z)+c_{1,0}(z)\tilde{c}^{(1)}_{1,1}\tilde{\mathcal{O}}^i_{8,1}(\vx,z)\right)+\cdots,
\eeq
defining $\tilde{c}_{a,1}^{(1)}(z)=-D^2(z)\tilde{T}^{(v)}_{a,1}(z)$ and introducing the composite operator $\tilde{\mathcal{O}}^i_{8,1}(\vx,z)=\delta(\vx,z)\tilde{\mathcal{O}}^i_{1,1}(\vx,z)$, whose coefficient is fixed. Interestingly, the non-local coefficients $\tilde{c}_{a,1}^{(1)}$ are inherited from the dark matter field; this leads to a reduction in dimensionality if multiple tracers are analyzed at the same redshift.\footnote{As discussed in Appendix \ref{app: dm-renorm}, fixing the time-dependence of the stress tensor to that required to cancel the UV-dependence of the loop integrals reduces the number of degrees of freedom in $\tilde{T}^{(v)}_{a,1}(z)$ from seven to five,  and the remaining coefficients are fixed by the local sector. To remain agnostic to the small-scale properties of the dark matter fluid, we do not impose such restrictions in this work.} 

Similar relations can be obtained for the higher-rank counterterms. At rank-$m$, we find
\beq
    C_{u^m}^{i_1\cdots i_m} = \frac{C_u^{i_1}\cdots C_u^{i_m}}{(1+\delta_g+\delta_{g,\rm ct})^{m-1}}
   + \sum_{p=2}^{m}\binom{m}{p} C^{(i_1\cdots i_p}_{\rm contact}\,C_u^{i_{p+1}}\cdots C_u^{i_{m})}(1+\delta_g+\delta_{g,\rm ct})^{1-m+p}
\eeq
where the contact term $C^{i_1\cdots i_p}_{\rm contact}$ arises from expanding the product $[(1+\delta_g)u_g^{i_1}\cdots u_g^{i_p}]$ in terms of density and velocities.\footnote{Formally, the contact terms are defined by
\beq
    \frac{[(1+\delta_g)u_g^{i_1}\cdots u_g^{i_m}]}{[1+\delta_g]}
   = U^{i_1}\cdots U^{i_m}
   + \sum_{p=2}^{m}\binom{m}{p}\,C^{(i_1\cdots i_p}_{\rm contact}U^{i_{p+1}}\cdots U^{i_m)},
\eeq
for $m\geq 2$, where $U^i\equiv [(1+\delta_g)u_g^i]/[1+\delta_g]$. This recursive definition is necessary to preserve Galilean invariance.}
This is both local-in-space and invariant under Galilean transformations, and starts at order zero for even $p$ (due to $\delta_{\rm K}^{ij}$ contributions) and order two else. Since the non-locality in $C_{u}^i$ starts at third-order and the local-in-space terms start at order zero (for even rank) or two (for odd rank), we conclude that non-local contributions to the rank-$m$ tensor start at third order for odd $m$ and fifth order for even $m\geq 2$. To describe the one-loop trispectrum and the two-loop power spectrum, we can thus omit non-local-in-space contributions to $C^{ij}_{u^2}$, $C^{ijk}_{u^3}$ and $C^{ijkl}_{u^4}$, as in \eqref{eq: Ou-expansion-local}.

\subsubsection{Renormalized Kernels}
\noindent Combining \eqref{eq: ctmaster} with \eqref{eq: rsd-mapping}, we obtain the redshift-space galaxy counterterm 
\beq\label{eq: rsd-ct-map}
    \delta_{g,\rm ct}^{(s)}(\vk,z)=\sum_{m\ge 0}\frac{(-if(z)k_\parallel)^m}{m!}\,\int d\vx\,e^{-i\vk\cdot\vx}C_{u^m}^{i_1\cdots i_m}(\vx,z)\hat{n}_{i_1}\cdots \hat{n}_{i_m}\,e^{-if(z)k_\parallel u_\parallel(\vx,z)},
\eeq
resumming the deterministic contributions. By expanding the exponential, we can obtain the counterterm kernel up to arbitrary order: at third-order, we find the NLG contribution 
\beq\label{eq: O3-rsd-ct}
      \delta_{g,\rm ct}^{(s)}(\vk,z)\supset -k^2\left[c_{1,1}(z)+\left(-c^{(1)}_{1,1}(z)
      +\frac{1}{2}f(z)c^{(2)}_{2,1}(z)\right)f(z) \mu^2+\frac{1}{2}\left(c^{(2)}_{3,1}(z)
      +f(z)c^{(2)}_{1,1}(z)\right)f^2(z)\mu^4\right]\delta_{\rm L}(\vk,z)
\eeq
for $\mu\equiv\hk\cdot\hn$.\footnote{This is equivalent to the form used in \citep{Chudaykin:2020aoj}, defining $\tilde c^{\textsc{class-pt}}_0=c_{1,0}c_{1,1}$, $\tilde c^{\textsc{class-pt}}_2=c_{1,1}+c_{1,0}(-c^{(1)}_{1,1}+fc^{(2)}_{2,1}/2)$, $\tilde c^{\textsc{class-pt}}_4=-c^{(1)}_{1,1}+fc^{(2)}_{2,1}/2+c_{1,0}(c^{(2)}_{3,1}+fc^{(2)}_{1,1})/2$ and $\tilde c^{\textsc{class-pt}}=-c^{(4)}_{1,2}/12$ (fixing $c^{(4)}_{2,2}=c_{1,0}c^{(4)}_{1,2}$). The additional factors of linear bias ($c_{1,0}$) arise since we define counterterms at the level of the density field rather than the power spectrum.} 
Practically, these terms are captured by modifying the NLG and NNLG redshift-space kernels; for example, the first-order NLG term becomes:
\beq
	Z_1^{\rm NLG}(\vk,z) &=& K^{\rm NLG}_1(\vk,z) + fk^2\mu^2\left(c_{1,1}^{(1)}-\frac{1}{2}fc_{2,1}^{(2)}\right) - \frac{1}{2}f^2k^2\mu^4\left(c_{3,1}^{(2)}+fc_{1,1}^{(2)}\right).
\eeq
The full set of kernels up to fifth-order in fields and derivatives is given in Appendix \ref{app: kernels-rsd}.

Following \S\ref{sec: renorm-real}, the $C_{u^m}^{i_1\cdots i_m}$ counterterms should be rewritten in terms of renormalized operators, $[\mathcal{O}_{a,b}^{i_1\cdots i_m}]$, to account for short-scale effects. As in real-space, this is equivalent to redefining the $Z_n$ kernels as in \eqref{eq: renorm-kernel1}\,\&\,\eqref{eq: renorm-kernel2}, explicitly subtracting the single-hard, nested and double-hard limits (which now depend explicitly on the line-of-sight direction). This removes certain higher-rank operators from the two-loop power spectrum (e.g., $\partial_i\delta^3$), though all are present in the one-loop trispectrum.\footnote{While some operators cancel identically from the kernels, others drop out only after angular integration. Formally, the latter fails in the presence of anisotropic infrared resummation (\S\ref{sec: ir-resum}), implying that six operators ($\mathcal{O}^i_{13,1}$, $\tilde{\mathcal{O}}^{i}_{3,1},\tilde{\mathcal{O}}^{i}_{4,1},\tilde{\mathcal{O}}^{i}_{5,1},\tilde{\mathcal{O}}^{i}_{7,1}$ and $\mathcal{O}^{ij}_{12,1}$) give small contributions to the power spectrum and the degeneracy of a seventh ($\mathcal{O}^{ij}_{9,1}$) is lifted. Since these effects scale as three-loop contributions, we ignore them in this work.} Renormalization again provides a powerful test of the formalism: the pieces subtracted from $Z_3$ through $Z_5$ must be exactly expressible as combinations of lower-order kernels (after angle averaging).\footnote{Only a subset of terms are required to absorb the subtracted limits. This is clearly seen for dark matter, which needs only a single combination of operators to capture each hard-limit structure (since there are no free bias coefficients). For galaxies in redshift-space, 12 combinations are required for simultaneous renormalization of the two-loop power spectrum and one-loop trispectrum, or 9 for the two-loop power spectrum alone (across all orders in $f$).} We have verified explicitly that this holds exactly at every power of $f$ (with the same coefficients across all orders).

Up to fifth-order in fields and derivatives, the NLG counterterm combines $17+14+18+0+0=49$ local-in-space operators (from rank-0 through rank-4) and $2+8+0+0+0=10$ non-local-in-space operators, while the NNLG term contains $1+1+2+2+3=9$ local-in-space operators. In total, this yields $68$ counterterm coefficients (of which $67$ are free, following \eqref{eq: Ou-expansion-rank1}), compared to $20$ present in real-space. Several combinations are degenerate at fixed $f$; for example, the $Z_1^{\rm NLG}$ kernel, which defines the one-loop power spectrum, depends on $c_{2,1}^{(2)}$ only through the combination $c_{1,1}^{(1)}-\tfrac{f}{2}c_{2,1}^{(2)}$. After renormalization, the one-loop power spectrum is sensitive to the following two combinations of redshift-space counterterms:
\beq\label{eq: ctr-comb-1loop}
    \bar{c}^{(1)}_{1,1} \equiv c^{(1)}_{1,1} - \frac{f}{2}c^{(2)}_{2,1}, \qquad \bar{c}^{(2)}_{1,1} \equiv c^{(2)}_{1,1} + \frac{1}{f}c^{(2)}_{3,1},
\eeq
while the two-loop power spectrum is sensitive to fifteen:
\beq\label{eq: ctr-comb-2loop}
    &&\hat c^{(1)}_{1,1} \equiv c^{(1)}_{1,1} - \frac{25}{46}\,c^{(2)}_{9,1} - \frac{5}{4}\,c^{(2)}_{10,1}, \qquad  \hat c^{(1)}_{2,1} \equiv c^{(1)}_{2,1} - \frac{f}{2}\,c^{(2)}_{7,1}, \qquad \hat c^{(1)}_{3,1} \equiv c^{(1)}_{3,1} - \frac{f}{2}\,c^{(2)}_{8,1}, \\\nonumber
    &&\hat c^{(1)}_{4,1} \equiv c^{(1)}_{4,1}, \qquad \hat  c^{(1)}_{5,1} \equiv c^{(1)}_{5,1} - \frac{5f}{46}\,c^{(2)}_{9,1} - \frac{f}{2}\,c^{(2)}_{11,1}, \qquad \hat c^{(1)}_{6,1} \equiv c^{(1)}_{6,1}, \\\nonumber
    &&\hat c^{(2)}_{1,1} \equiv c^{(2)}_{1,1} - \frac{75}{92f}\,c^{(2)}_{9,1}, \qquad \hat c^{(2)}_{2,1} \equiv c^{(2)}_{2,1} - \frac{25}{23f}\,c^{(2)}_{9,1} - \frac{5}{2f}\,c^{(2)}_{10,1}, \qquad \hat c^{(2)}_{3,1} \equiv c^{(2)}_{3,1} + \frac{75}{92}\,c^{(2)}_{9,1}\\\nonumber
    &&\hat c^{(2)}_{4,1} \equiv c^{(2)}_{4,1} - \frac{15}{46}\,c^{(2)}_{9,1}, \qquad \hat c^{(2)}_{5,1} \equiv c^{(2)}_{5,1} + \frac{25}{23}\,c^{(2)}_{9,1} + \frac{5}{2}\,c^{(2)}_{10,1}, \qquad \hat c^{(2)}_{6,1} \equiv c^{(2)}_{6,1} + \frac{15}{46}\,c^{(2)}_{9,1}, \\\nonumber
    &&\hat c^{(1)}_{1,2} \equiv c^{(1)}_{1,2} - \frac{f}{2}\,c^{(2)}_{2,2}, \qquad \hat c^{(2)}_{1,2} \equiv c^{(2)}_{1,2} - \frac{f}{3}\,c^{(3)}_{2,2} + \frac{f^2}{12}\,c^{(4)}_{2,2}, \qquad \hat c^{(3)}_{1,2} \equiv c^{(3)}_{1,2} - \frac{f^2}{4}\,c^{(4)}_{1,2} - \frac{f}{4}\,c^{(4)}_{3,2},
\eeq
in addition to the non-local terms. Adding the real-space terms, this gives a total of $26$ independent NLG and NNLG parameters in the two-loop power spectrum, or $54$ when combining with the one-loop bispectrum and trispectrum. For reference, the operators present at each order are marked in Tabs.\,\ref{tab: operators}\,\&\,\ref{tab: rsd-table}.

\subsection{Loop Integrals}\label{sec: loops}
\noindent Given the redshift-space kernels, $Z_n$, we can construct the galaxy power spectrum. Up to two-loop order, the deterministic contributions to the power spectrum are:
\beq\label{eq: Pdef}
    &&P^{\rm det}_{gg}(\vk)=P^{\rm det}_{\rm tree}(\vk)+P^{\rm det}_{\rm 1-loop}(\vk)+P^{\rm det}_{\rm 2-loop}(\vk)+\dots\\\nonumber
    &&P^{\rm det}_{\rm tree}(\vk) = |Z_{1}^{\rm LG}(\vk)|^2P_{\rm L}(\vk)\\\nonumber
    &&P^{\rm det}_{\rm 1-loop}(\vk) = \int_{\vp}\left[6Z_{1}^{\rm LG}(\vk)[Z_3^{\rm LG}](\vk,\vp,-\vp)P_{\rm L}(\vk)P_{\rm L}(\vp) + 2|Z_{2}^{\rm LG}(\vp,\vk-\vp)|^2P_{\rm L}(\vp)P_{\rm L}(\vk-\vp)\right]\\\nonumber
    &&\qquad\qquad\qquad+\,2Z_{1}^{\rm NLG}(\vk)Z_{1}^{\rm LG}(\vk)P_{\rm L}(\vk)\\\nonumber
    &&P^{\rm det}_{\rm 2-loop}(\vk) = \int_{\vp\,\vq}\Big[30Z_{1}^{\rm LG}(\vk)[[Z_5^{\rm LG}]](\vk,\vp,-\vp,\vq,-\vq)P_{\rm L}(\vk)P_{\rm L}(\vp)P_{\rm L}(\vq)\\\nonumber
    &&\qquad\qquad\qquad\qquad\,\,+\,24Z_2^{\rm LG}(\vp,\vk-\vp)Z_4^{\rm LG}(-\vp,\vp-\vk,\vq,-\vq)P_{\rm L}(\vk-\vp)P_{\rm L}(\vp)P_{\rm L}(\vq)\\\nonumber
    &&\qquad\qquad\qquad\qquad\,\,+\,9[Z_3^{\rm LG}](\vk,\vp,-\vp)[Z_3^{\rm LG}](\vk,\vq,-\vq)P_{\rm L}(\vk)P_{\rm L}(\vp)P_{\rm L}(\vq)\\\nonumber
    &&\qquad\qquad\qquad\qquad\,\,+\,6\left|Z_3^{\rm LG}(\vp,\vq,\vk-\vp-\vq)\right|^2P_{\rm L}(\vk-\vp-\vq)P_{\rm L}(\vp)P_{\rm L}(\vq)\Big]\\\nonumber
    &&\qquad\qquad\qquad+\,\int_{\vp}\Big[6\left(Z_{1}^{\rm NLG}(\vk)[Z_3^{\rm LG}](\vk,\vp,-\vp)+Z_{1}^{\rm LG}(\vk)[Z_3^{\rm NLG}](\vk,\vp,-\vp)\right)P_{\rm L}(\vk)P_{\rm L}(\vp)\\\nonumber
    &&\qquad\qquad\qquad\qquad\,\,+\,4Z_{2}^{\rm LG}(\vp,\vk-\vp)Z_{2}^{\rm NLG}(\vp,\vk-\vp)P_{\rm L}(\vp)P_{\rm L}(\vk-\vp)\Big]\\\nonumber
    &&\qquad\qquad\qquad+\,\left(2Z_{1}^{\rm NNLG}(\vk)Z_{1}^{\rm LG}(\vk)+Z_1^{\rm NLG}(\vk)Z_1^{\rm NLG}(\vk)\right)P_{\rm L}(\vk)
\eeq
\citep[e.g.,][]{Carrasco:2013mua}, where $[Z_n]$ and $[[Z_n]]$ are the renormalized kernels of \eqref{eq: renorm-kernel1}\,\&\,\eqref{eq: renorm-kernel2}. Note that $P^{\rm 1-loop}$ contains a tree-level NLG contribution (encoding the counterterms) and $P^{\rm 2-loop}$ includes a one-loop integral sourced by NLG contributions, and a tree-level component from both the NLG and NNLG kernels. The above expression ignores infrared resummation, stochastic contributions, and coordinate distortions, which will be discussed below.

Importantly, the $Z_n$ kernels defined above additionally specify the deterministic contributions to the redshift-space bispectrum and trispectrum up to one-loop order \citep[cf.][]{Bakx:2026rmd}, as well as the quadspectrum and pentaspectrum at tree-level. The explicit forms of each statistic are given in Appendix \ref{app: kernels-correlators} (including infrared resummation effects); the combination involves all of the operators listed in Tab.\,\ref{tab: operators}. 

\subsection{Infrared Resummation}\label{sec: ir-resum}
\noindent In the above sections, we have derived an Eulerian perturbative model for the galaxy overdensity field. The convergence of this expansion is controlled by a number of dimensionless parameters, including those characterizing the contribution of modes above ($>$) and below ($<$) a given scale $k$ to the linear displacement field \citep{Senatore:2014via}:
\beq
    \epsilon_{s >}(k,z) = k^2\int^\infty_k\frac{q^2dq}{2\pi^2}\frac{P_{\rm L}(q,z)}{q^2}, \qquad \epsilon_{s<}(k,z) = k^2\int_0^k \frac{q^2dq}{2\pi^2}\frac{P_{\rm L}(q,z)}{q^2}.
\eeq
The first coefficient encodes the impact of small-scale physics, which renormalization absorbs into the EFT coefficients. The second describes infrared (IR) displacements: at low redshift, $\epsilon_{s<}$ can reach $\mathcal{O}(1)$, indicating a breakdown of the Eulerian expansion \citep{Senatore:2014via,Baldauf:2015xfa,Blas:2016sfa}. Due to the equivalence principle, these bulk flows principally impact the oscillatory parts of the density field \citep{Baldauf:2015xfa}, which implies that truncating the expansion at too low an order results in underdamping of the baryon acoustic oscillations (and any other oscillatory features \citep[e.g.,][]{Chen:2024pyp}).

This effect can be ameliorated through \textit{infrared resummation}, which resums the long-wavelength displacements in a self-consistent framework \citep[e.g.,][]{Senatore:2014via,Porto:2013qua,Vlah:2015zda,Lewandowski:2018ywf}. Here, we adopt the `wiggle-no-wiggle' approach presented in \citep{Baldauf:2015xfa,Chudaykin:2020aoj,Ivanov:2018gjr}, which can be formalized using time-sliced perturbation theory \citep{Blas:2015qsi,Blas:2016sfa}. This splits the linear power spectrum into a smooth component, $P_{\rm nw}$, and an oscillatory residual, $P_{\rm w} \equiv P_{\rm L}-P_{\rm nw}$, and applies damping only to the latter, with the leading-order result
\beq
    P_{\rm L}(q,z)\to \mathcal{P}_{\rm L}^{(0)}(\vq,z) &\equiv& P_{\rm nw}(q,z) + e^{-q^2\Sigma^2_{\rm IR}(\mu_q,z)}P_{\rm w}(q,z)
\eeq
for $\mu_q\equiv \hq\cdot\hn$. The direction-dependent damping scale is given by
\beq
    \Sigma^2_{\rm IR}(\mu,z) &=& \Sigma^2(z)[1+f(z)\mu^2(2+f(z))] + \delta\Sigma^2(z)f^2(z)\mu^2(\mu^2-1)\\\nonumber
    \Sigma^2(z) &=& \int^{\Lambda_{\rm IR}}_{0} \frac{k^2dk}{6\pi^2}\frac{P_{\rm nw}(k,z)}{k^2}[1-j_0(kr_{\rm BAO})+2j_2(kr_{\rm BAO})], \qquad \delta\Sigma^2(z) = \int_0^{\Lambda_{\rm IR}} \frac{k^2dk}{2\pi^2}\frac{P_{\rm nw}(k,z)}{k^2}j_2(kr_{\rm BAO}).
\eeq
Here, $r_{\rm BAO}\approx 100\Mpch$ is the (cosmology-dependent) scale of the baryon acoustic oscillations and $\Lambda_{\rm IR}$ is a cut-off scale, which we fix to $0.2\hMpc$ \citep{Chudaykin:2020aoj}.

Inserted into the (deterministic) galaxy power spectrum, we find the tree-level IR-resummed spectrum:
\beq\label{eq: Pir-tree}
	P_{gg}^{{\rm IR},(0)}(\vk,z) &=& P_{\rm tree}^{\rm det}\left[\mathcal{P}_{\rm L}^{(0)}\right],
\eeq
treating $P^{\rm det}_{\rm tree}$ (defined in \eqref{eq: Pdef}) as a functional of the input power spectrum. Similarly, the power spectra up to one- and two-loop orders become (see Appendix \ref{app: kernels-correlators} for an explicit form):
\beq\label{eq: Pir-loop}
	P_{gg}^{{\rm IR},(1)}(\vk,z) &=& P^{\rm det}_{\rm 1{-}loop}\left[\mathcal{P}_{\rm L}^{(0)}\right]+P^{\rm det}_{\rm tree}\left[\mathcal{P}_{\rm L}^{(1)}\right]\\\nonumber
    P_{gg}^{{\rm IR},(2)}(\vk,z) &=& P^{\rm det}_{\rm 2{-}loop}\left[\mathcal{P}_{\rm L}^{(0)}\right]+P^{\rm det}_{\rm 1{-}loop}\left[\mathcal{P}_{\rm L}^{(1)}\right]+P^{\rm det}_{\rm tree}\left[\mathcal{P}_{\rm L}^{(2)}\right],
\eeq
defining
\beq
    \mathcal{P}_{\rm L}^{(n)}(\vq,z) &\equiv& P_{\rm nw}(q,z) + \left(\sum_{m=0}^n\frac{q^{2m}\Sigma^{2m}_{\rm IR}(\mu_q,z)}{m!}\right)e^{-q^2\Sigma^2_{\rm IR}(\mu_q,z)}P_{\rm w}(q,z).
\eeq
The factors in parentheses (which are just a partial expansion of $e^{q^2\Sigma^2_{\rm IR}}$), ensure that IR modes are not double counted. As discussed in \citep{Ivanov:2018gjr,Chudaykin:2020aoj}, errors from the wiggle-no-wiggle decomposition arise only at higher-loop order.

In practice, we implement IR resummation by replacing the linear power spectrum with its damped equivalent wherever it appears in the loop integrals. This induces exponential dependence on the angles between the loop momenta and the line-of-sight (due to the anisotropies in $\Sigma^2_{\rm IR}$) in contrast to the polynomial dependence found in the $Z_n$ kernels. This complicates the evaluation of the loop integrals, prohibiting exact computation of the power spectrum multipoles using tensor reduction \citep{Chudaykin:2020aoj} (requiring explicit angular integration). For this reason, many works have adopted an approximate form for the resummed loop integrals, taking the exponential damping outside the integrals and dropping higher order terms in $P_{\rm w}$. For completeness, we will not make such assumptions in this work (except in Appendix \ref{app: class-pt} to compare our pipeline to \textsc{class-pt}), though we have confirmed that, at least at one loop, they induce negligible bias in cosmological parameter inference.

\subsection{Stochasticity}\label{sec: stoch}
\noindent Short-wavelength fluctuations uncorrelated with the large-scale modes generate stochastic contributions to the galaxy density field. These can be modeled by supplementing the bias expansion with additional fields, for example:
\beq
    \delta_g(\vx,z) &\supset& \epsilon(\vx,z) + \epsilon_\delta(\vx,z)\delta(\vx,z) + \nabla^2\epsilon_{\nabla^2}(\vx,z)+\cdots
\eeq
\citep[e.g.,][]{Desjacques:2016bnm}. These satisfy $\av{\epsilon_\mathcal{O}}=\av{\epsilon_\mathcal{O}\mathcal{O}'}=0$ (for deterministic operator $\mathcal{O}'$) and $\av{\epsilon_\mathcal{O}(\vx)\epsilon_{\mathcal{O}''}(\vy)} \propto \delta_{\rm D}(\vx-\vy)$ due to locality. The redshift-space counterterms of \S\ref{sec: renorm-rsd} are similarly modified:
\beq
    \mathcal{O}_u^i(\vx,z) \supset \partial^i\epsilon_u(\vx,z)+\cdots, \quad \mathcal{O}_{u^2}^{ij}(\vx,z) \supset \delta_{\rm K}^{ij}\epsilon_{u^2}(\vx,z)+\partial^{ij}\epsilon_{u^2}(\vx,z)+\cdots
\eeq
\citep[e.g.,][]{Perko:2016puo}, which induces dependence on the line-of-sight direction. 

For the bispectrum and beyond, the stochastic sector is highly non-trivial, with mixed deterministic-stochastic loop integrals arising at fourth order and above \citep[e.g.,][]{Philcox:2022frc,DAmico:2022ukl,Bakx:2025pop}. For the power spectrum, however, the contributions reduce to a Taylor expansion in $k^2$ and $k_\parallel^2\equiv k^2\mu^2$, yielding the redshift-space result
\beq\label{eq: stoch-Pk}
    &&\qquad\qquad\qquad P_{gg}^{\rm stoch}(\vk,z) = P^{\rm stoch}_{\rm tree}(\vk,z) + P^{\rm stoch}_{\rm 1-loop}(\vk,z) + P^{\rm stoch}_{\rm 2-loop}(\vk,z)+\,\cdots\\\nonumber
    &&P^{\rm stoch}_{\rm tree}(\vk,z) = a_{1,0}, \quad P^{\rm stoch}_{\rm 1-loop}(\vk,z) = k^2\left(a_{1,1} + a_{2,1}\mu^2\right), \quad P^{\rm stoch}_{\rm 2-loop}(\vk,z) = k^4(a_{1,2} + a_{2,2}\mu^2 + a_{3,2}\mu^4),
\eeq
where $a_{i,j}(z)$ are time-dependent coefficients.\footnote{To see this, note that composite operators such as $\epsilon_\delta\delta$ contribute only by rescaling the amplitudes of non-composite terms, e.g., $\av{(\epsilon_\delta\delta)(\vx)(\epsilon_\delta\delta)(\vy)} = \av{\epsilon_\delta(\vx)\epsilon_\delta(\vy)}\av{\delta(\vx)\delta(\vy)}\propto\delta_{\rm D}(\vx-\vy)$, which is degenerate with $\av{\epsilon(\vx)\epsilon(\vy)}$ (and removed by renormalization). The line-of-sight dependence arises from both the $-ifk_\parallel$ and $\hn_i$ factors in \eqref{eq: rsd-ct-map}.} The real-space form is obtained by setting $a_{2,1}=a_{2,2}=a_{3,2}=0$, with $a_{1,0}$ associated with the (non-Poissonian) shot-noise term. The stochastic sector also captures contributions to $P_{gg}$ from velocity vorticity, which scale as $k^4\mu^2(1-\mu^2)$ at leading order \citep{Mercolli:2013bsa}.

\subsection{Projection \& Alcock-Paczy\'{n}ski Effects}\label{sec: ap}
\noindent Experiments typically measure the Legendre moments of the power spectrum, rather than the continuous quantity, $P^{\rm theory}_{gg}(k,\mu)$, derived above. These can be obtained as
\beq\label{eq: legendre-projection}
	P^{\rm theory}_{gg,\ell}(k) &=& (2\ell+1)\int_{-1}^{1}\frac{d\mu}{2}P^{\rm theory}_{gg}(k,\mu)\mathcal{L}_{\ell}(\mu),
\eeq
where $\mathcal{L}_\ell$ is the (regular) Legendre polynomial of order $\ell$ and $\mu\equiv \hk\cdot\hn$. At $n$-loop order, the deterministic contributions of \S\ref{sec: rsd} source all even $\ell\leq 4(n+1)$; the exponential dependence on angle induced by IR resummation propagates this to all even $\ell$. Odd multipoles are generated only by relativistic and wide-angle effects and are ignored in this work.

An additional source of anisotropy arises from Alcock--Paczy\'{n}ski effects, which account for the fiducial cosmology used to convert the observed galaxy redshifts and angles into physical positions. Following previous works \citep[e.g.,][]{Alcock:1979mp,Ballinger:1996cd,Philcox:2021kcw}, these can be treated exactly by rescaling the parallel and perpendicular coordinates:
\beq
	(k_\parallel, k_\perp) \to (k'_\parallel, k'_\perp) \equiv (k_\parallel/\alpha_\parallel, k_\perp/\alpha_\perp)
\eeq
where $k_\parallel \equiv k\mu$, $k_\perp \equiv k\sqrt{1-\mu^2}$ and
\beq
	\alpha_\parallel = \frac{H_{\rm fid}(z)H_{0,\rm true}}{H_{\rm true}(z)H_{0,\rm true}}, \qquad \alpha_\perp = \frac{D_{A,\rm true}(z)H_{0,\rm true}}{D_{A,\rm fid}(z)H_{0,\rm fid}}.
\eeq
Here, $H$ and $D_A$ are the Hubble rate and angular diameter distance in $\Mpch$ units, $z$ is the sample redshift, and `fid' indicates quantities computed in the fiducial cosmology. This gives the combined power spectrum multipoles
\beq\label{eq: multipole-AP}
    P_{gg,\ell}^{\rm AP}(k) &=& \frac{2\ell+1}{\alpha_\perp^2\alpha_\parallel}\int_{-1}^1\frac{d\mu}{2}P_{gg}^{\rm theory}(k'(k,\mu),\mu'(\mu))\mathcal{L}_\ell(\mu),
\eeq
for $k' = k\sqrt{(1-\mu^2)/\alpha_\perp^2+\mu^2/\alpha_\parallel^2}$ and $\mu' = \mu/\alpha_\parallel/\sqrt{(1-\mu^2)/\alpha_\perp^2+\mu^2/\alpha_\parallel^2}$. This form can additionally be integrated over $k^2dk$ to compare to a datavector binned in wavenumber.

\subsection{Summary}\label{sec: theory-summary}
\noindent Our model for the two-loop redshift-space galaxy power spectrum contains the following contributions:
\begin{enumerate}
	\item \textbf{Bias Operators}: The real-space density field is expanded in terms of all possible local-in-space (but non-local-in-time) bias operators, following \eqref{eq: bias-expansion}. For the two-loop power spectrum, we require 17 LG, 7 NLG, and 1 NNLG operator(s), each with a free bias coefficient, $c_{a,b}$, \textit{i.e.}
	\beq
		\delta^{(\rm 2-loop)}_{g,\rm det}(\vx,z) = \sum_{a=1}^{17}c_{a,0}(z)\mathcal{O}_{a,0}(\vx,z)+\sum_{a=1}^{7}c_{a,1}(z)\mathcal{O}_{a,1}(\vx,z)+c_{1,2}(z)\mathcal{O}_{1,2}(\vx,z).
	\eeq
	\item \textbf{Non-Local Operators}: Renormalization generates new non-local-in-space counterterms via the dark matter equations of motion (assuming that the stress-tensor is local-in-space). For the two-loop power spectrum, this adds a single operator:
	\beq
	 	\delta^{(\rm 2-loop)}_{g,\rm non{-}local}(\vx,z) = \tilde{c}_{1,1}(z)\tilde{\mathcal{O}}_{1,1}(\vx,z).
	\eeq
	\item \textbf{Redshift-space}: The galaxy field is transformed to redshift-space via \eqref{eq: rsd-mapping}, with the general relation
    \beq
        \delD{\vk}+\delta_g^{(s)}(\vk,z) &=& \int d\vx\,e^{-i\vk\cdot\vx}\Big[1+\delta_g(\vx,z)+\sum_{m\ge 0}\frac{(-ifk_\parallel)^m}{m!}C_{u^m}^{\parallel\cdots\parallel}(\vx,z)\Big]e^{-if(z)k_\parallel u_\parallel(\vx,z)}.
    \eeq
    This adds dependence on the velocity field, $\vec u$, in addition to higher-rank counterterms, $C_u^\parallel\equiv C_u^i\hat{n}_i$, $C_{u^2}^{\parallel\parallel}\equiv C_{u^2}^{ij}\hat{n}_i\hat{n}_j$, \textit{et cetera}. For the two-loop power spectrum, we require 13 NLG and 2 NNLG local-in-space operators with coefficients $\hat{c}^{(n)}_{a,b}$ (defined via \eqref{eq: ctr-comb-2loop}), as well as 3 non-local-in-space NLG terms, $\tilde{c}^{(1)}_{a,1}$ (one of which carries a fixed coefficient).
    \item \textbf{Correlators}: Combining (1)-(3), we obtain the deterministic contributions to the redshift-space field, which can be expressed in terms of the kernels, $Z_n$, listed in Appendix \ref{app: kernels-rsd}. These are used to compute the deterministic loop integrals following \eqref{eq: Pdef}.
	\item \textbf{Infrared Resummation}: We damp the linear power spectra to account for the effects of long-wavelength displacements, using the wiggle-no-wiggle prescription. This is summarized in \eqref{eq: Pir-tree}\,\&\,\eqref{eq: Pir-loop}, and does not add free parameters.
	\item \textbf{Stochasticity}: We add stochastic counterterms to the power spectra via \eqref{eq: stoch-Pk}. This depends on six amplitudes, $a_{a,b}$ (or three in real-space).
	\item \textbf{Coordinate Distortions}: We project the combined power spectrum, $P_{gg}^{(s)}(k,\mu)$, onto a Legendre basis, redefining the $k$ and $\mu$ coordinates to account for the difference in true and fiducial cosmological parameters following \eqref{eq: multipole-AP}. 
\end{enumerate}

In total, the two-loop redshift-space power spectrum depends on 49 independent coefficients, compared to 10 for the one-loop power spectrum. In real-space, this reduces to 29 at two-loop and 7 at one-loop. Notably, all of the deterministic terms (except the NNLG pieces) appear also in the one-loop trispectrum model, and 18 are constrained by the redshift-space one-loop bispectrum. We stress that many of these parameters are highly degenerate (as found for the one-loop bispectrum in \citep{Philcox:2022frc,DAmico:2022ukl}).

\section{Numerical Implementation}\label{sec: implementation}
\noindent Above, we have outlined our model for the galaxy power spectrum in terms of loop integrals over Fourier-space kernels and (damped) linear power spectra. We now discuss the numerical implementation of this theory, as well as the pipeline used to analyze simulated data in \S\ref{sec: simulations}.

\subsection{Operators \& Kernels}\label{subsec: impl-ops}
\noindent First, we discuss the enumeration of the local-in-space bias operators. To generate the list shown in Tab.\,\ref{tab: operators}, we first enumerate all possible scalar combinations of $\Pim{n}_{ij}$ operators and derivatives up to a given order, then remove redundant operators by evaluating the underlying kernels on sets of momenta using exact integer arithmetic (modulo a large prime), growing the number until the rank saturates. Starting with $1982$ candidates, this yields a set of $29$ LG, $17$ NLG and $1$ NNLG local-in-space operators, related to those of \citep{Bakx:2026rmd} by a rotation. Following computation of the renormalized kernels (discussed below), the operators are redefined to remove degenerate contributions in the two-loop power spectrum (e.g., subtracting the overlap of $\mathcal{O}_{17,1}$ with $\mathcal{O}_{6,1}$ and $\mathcal{O}_{7,1}$). A similar procedure is used to obtain the higher-rank local-in-space operators listed in Tab.\,\ref{tab: rsd-table}. To obtain the non-local-in-space operators, we solve the dark matter equations of motion perturbatively (see Appendix \ref{app: dm-renorm}), then translate to the galaxy sector following \S\ref{sec: renorm-rsd}. 

We then assemble the real- and redshift-space kernels, $K_n$ and $Z_n$, using the operator definitions and the redshift-space mapping of \eqref{eq: rsd-mapping}. These are implemented independently in both \textsc{mathematica} and \textsc{python}, and verified mechanically against the exact \LaTeX{} definitions from Tabs.\,\ref{tab: operators}\,\&\,\ref{tab: rsd-table}. Each kernel is decomposed into a sum of cosmology-independent templates multiplied by bias coefficients and $f(z)$:
\beq\label{eq: Zn-decomposition}
    Z_n[c_A](\vk_1,\cdots,\vk_n;z,\hn) = \sum_A\sum_{m=0}^nc_A(z)f^m(z)Z_n^{(A,f^m)}(\vk_1,\cdots,\vk_n;\hn),
\eeq
where $\{c_A\}\equiv\{1,c_{1,0},c_{2,0},\ldots\}$ includes all relevant deterministic Wilson coefficients, with the first term multiplying the bias-independent contributions. We further extract the dependence on the line-of-sight $\hn$ by expanding each template in a Cartesian basis:
\beq\label{eq: kernel-los-sum}
    Z_n^{(A,f^m)}(\vk_1,\cdots,\vk_n;\hn) = \sum_{\alpha\geq 0}\sum_{\beta\geq 0}\sum_{\gamma\geq 0}\hat{n}_x^{\alpha}\hat{n}_y^\beta\hat{n}_z^\gamma\,Z_n^{(A,f^m,\alpha\beta\gamma)}(\vk_1,\cdots,\vk_n),
\eeq
with $\alpha+\beta+\gamma\leq 2m$. This facilitates efficient computation of the one- and two-loop integrals (\S\ref{subsec: impl-vegas}). 

\subsection{Renormalization \& Hard Limits}\label{subsec: impl-renorm}
\noindent To compute the one-loop power spectra, we require the limiting behavior of kernels of the form $Z_n(\vk_1,\cdots,\vk_{n-2},\vp,-\vp)$ for $p\equiv |\vp|\gg k_i$ (from \eqref{eq: Pdef}\,\&\,\eqref{eq: renorm-kernel1}). To compute these, we first resolve the $(\vp,-\vp)$ pair via
\beq\label{eq: pair-resolution}
	Z_n(\vk_1,\cdots,\vk_{n-2},\vp,-\vp) \equiv \lim_{\eta\to 0}Z_n(\vk_1,\cdots,\vk_{n-2},\vp,-\vp+\eta^2\vec d),
\eeq
for generic vector $\vec d$; this is necessary to cancel the (formal) divergences arising in the recursive solutions for $F_n$ and $G_n$.\footnote{In practice, we perform a Laurent expansion around $\eta=0$ and check explicitly that the $\eta^{-m}$ terms with $m\geq 1$ cancel. This avoids ordering ambiguities when taking sequential limits.} The single-hard limits are then computed by introducing a second expansion parameter $\epsilon$, giving the renormalized kernel
\beq
	&&[Z_n] = Z_n - Z_n^{p\gg k_i}\\\nonumber
    &&Z^{p\gg k_i}_n(\vk_1,\cdots,\vk_{n-2},\vp,-\vp) \equiv \left[\lim_{\epsilon\to 0}\left(\lim_{\eta\to 0}Z_n(\epsilon\vk_1,\cdots,\epsilon\vk_{n-2},\vp,-\vp+\eta^2\vec d)\right)\right]_{\mathcal{O}(\epsilon^0)},
\eeq
which is decomposed into products of biases and $f(z)$, as in \eqref{eq: Zn-decomposition}. For NLG operators, we additionally remove $\epsilon^2$ contributions, following \eqref{eq: renorm-kernel1}. Practically, we compute the hard limits of the $Z_3$, $Z_4$ and $Z_5$ kernels analytically using \textsc{mathematica} and subtract them pointwise.\footnote{At small $k$, this is numerically unstable since the kernels are dominated by their hard limits. For $Z_3$, this can be avoided by replacing the entire renormalized kernel by its analytic simplification; this is infeasible for higher-order kernels due to their larger complexity. In this case, we replace the kernels by their tenth-order Taylor expansions if one of the loop momenta becomes large.} The hard limits of the kernels have simple functional forms; for example, the contributions of the quadratic and cubic biases to $Z_3^{p\gg k}$ are
\beq
	Z^{p\gg k}_3(\vk,\vp,-\vp) &\supset& \frac{c_{3,0}}{21}\left(8x^2+20+7f\mu_k^2\right)
+c_{5,0}
\eeq
for $x = \hk\cdot\hat{\vec p}$, $\mu_k\equiv \hk\cdot\hn$ and $\mu_p\equiv\hat{\vec p}\cdot\hn$. As discussed in \S\ref{sec: renorm-real}, closure of the basis under renormalization provides a powerful test of the bias expansion: this requires the angle-averaged kernels, which are computed by writing $\mu_p=x\mu_k+\sqrt{1-x^2}\sqrt{1-\mu_k^2}\cos\phi$ and averaging analytically according to $(4\pi)^{-1}\int_{-1}^{1}dx\int_{0}^{2\pi}d\phi$. For the above example, we find
\beq
    \left\langle Z^{p\gg k}_3(\vk,\vp,-\vp)\right\rangle_{\hat{\vec p}} \supset c_{3,0}\left(\frac{68}{63}+\frac{f\mu_k^2}{3}\right)+c_{5,0};
\eeq
this is absorbed by the linear bias and mean density.

The two-loop power spectrum also involves the nested and double-hard limits of $Z_5$. These are computed by first resolving the $(\vp,-\vp)$ and $(\vq,-\vq)$ pairs (using a separate regulator for each), then computing the relevant limits by expanding in small parameters as before. From \eqref{eq: renorm-kernel2}, the full renormalized kernel is given by
\beq\label{eq: forest}
    &&[[Z_5]] = Z_5 - Z^{p\gg q,k}_5 - Z^{q\gg p,k}_5 - Z^{p,q\gg k}_5
    + Z^{p\gg q\gg k}_5 + Z^{q\gg p\gg k}_5\\\nonumber
    &&Z^{p\gg q,k}_5(\vk,\vp,-\vp,\vq,-\vq) \equiv \left[\lim_{\epsilon\to 0}\left(\lim_{\eta\to0}\lim_{\eta'\to0}
    Z_5\left(\epsilon\vk,\vp,-\vp+\eta^2\vec d,\epsilon\vq,\epsilon(-\vq+\eta'^2\vec d)\right)\right)\right]_{\mathcal{O}(\epsilon^0)}\\\nonumber
    &&Z^{p,q\gg k}_5(\vk,\vp,-\vp,\vq,-\vq) \equiv \left[\lim_{\epsilon\to 0}\left(\lim_{\eta\to0}\lim_{\eta'\to0}
    Z_5\left(\epsilon\vk,\vp,-\vp+\eta^2\vec d,\vq,-\vq+\eta'^2\vec d\right)\right)\right]_{\mathcal{O}(\epsilon^0,\epsilon^2)}\\\nonumber
    &&Z^{p\gg q\gg k}_5(\vk,\vp,-\vp,\vq,-\vq) \equiv \left[\lim_{\epsilon'\to 0}\left[\lim_{\epsilon\to 0}
    \left(\lim_{\eta\to0}\lim_{\eta'\to0}
    Z_5\left(\epsilon\epsilon'\vk,\vp,-\vp+\eta^2\vec d,
    \epsilon\vq,\epsilon(-\vq+\eta'^2\vec d)\right)\right)\right]_{\mathcal{O}(\epsilon^0)}\right]_{\mathcal{O}(\epsilon'^0,\epsilon'^2)}.
\eeq
Integration over $\hat{\vec p}$ and $\hat{\vec q}$ (which is required only to prove closure under renormalization) is performed by rewriting $\hq$ as $y\hat{\vec p}+\sqrt{1-y^2}\hat{\vec e}$ for relative angle $y\equiv\hq\cdot\hat{\vp}$ and global orientation $\hat{\vec e}\perp\hat{\vec p}$ (such that $|\vp\pm\vq|$ depends only on $p,q,y$), averaging over $\hat{\vec e}$ and $\hat{\vec p}$, and finally integrating over $y$ (which induces logarithms from regions with $|y|\to 1$ \citep[cf.,][]{Bakx:2026rmd}). We verify explicitly that the subtracted single-hard limit can be expressed in terms of $Z_3^{\rm LG}$ kernels, and the nested and double-hard limits can be written as combinations of $Z_1^{\rm LG}$ and $Z_1^{\rm NLG}$ kernels. As a further test of the method, we have confirmed that the hard limits of the dark matter LG kernels ($Z^{\rm LG}_n$ with $c_{a,b}=0$, except $c_{1,0}=1$) match the analytic expressions presented in \citep{Blas:2013aba,Baldauf:2015aha,Baldauf:2021zlt,Taule:2023izt}.\footnote{We correct a factor of two in the double-hard limit of \citep{Taule:2023izt}.}

\subsection{IR Safety and Tadpoles}\label{subsec: impl-ir-safety}

\noindent As discussed in \citep{Carrasco:2013sva}, na\"ive integration of the above expressions poses difficulties due to IR-sensitive $1/p^2$ poles (encoding bulk flows) contained within the integrands of individual diagrams. By the equivalence principle, such contributions cancel in the combined expressions; for this reason, we assemble the full integrand directly rather than splitting it into component pieces as in \eqref{eq: Pdef} (e.g., we compute only the sum of the `22' and `13' one-loop terms). The single-loop contributions to the two-loop power spectrum are thus given by
\beq
    P^{\rm det, LG}_{\rm 1-loop}(\vk) &=& \int_{\vp}\Big\{
 6 Z^{\rm LG}_1(\vk) [Z^{\rm LG}_3](\vk,\vp,-\vp)\mathcal{P}^{(1)}_{\rm L}(\vk)\mathcal{P}^{(1)}_{\rm L}(\vp)+2\left|Z^{\rm LG}_2(\vp,\vk-\vp)\right|^2\mathcal{P}^{(1)}_{\rm L}(\vp)\mathcal{P}^{(1)}_{\rm L}(\vk-\vp)\Big\}\\\nonumber
    P^{\rm det, NLG}_{\rm 1-loop}(\vk) &=& \int_{\vp}\Big\{
 6\left(Z^{\rm NLG}_1(\vk) [Z^{\rm LG}_3](\vk,\vp,-\vp)+Z^{\rm LG}_1(\vk) [Z^{\rm NLG}_3](\vk,\vp,-\vp)\right)\mathcal{P}^{(0)}_{\rm L}(\vk)\mathcal{P}^{(0)}_{\rm L}(\vp)\\\nonumber
 &&\qquad\,+\,4Z^{\rm LG}_2(\vp,\vk-\vp)Z^{\rm NLG}_2(-\vp,-\vk+\vp)\mathcal{P}^{(0)}_{\rm L}(\vp)\mathcal{P}^{(0)}_{\rm L}(\vk-\vp)\Big\},
\eeq
accounting for IR-resummation effects as in \S\ref{sec: ir-resum}. For the double-loop contributions, we additionally exchange variables to localize the bulk flow contributions such that the divergences cancel pointwise (following \citep{Carrasco:2013sva})::
\beq\label{eq: 2-loop-ir-safe}
    P^{\rm det}_{\rm 2-loop}(\vk) &=& \int_{\vp\,\vq}\Theta(q-p)\Big\{
 60 Z^{\rm LG}_1(\vk) [[Z^{\rm LG}_5]](\vk,\vp,-\vp,\vq,-\vq)\mathcal{P}^{(0)}_{\rm L}(\vk)\mathcal{P}^{(0)}_{\rm L}(\vp)\mathcal{P}^{(0)}_{\rm L}(\vq)\\\nonumber
    &&\qquad\qquad\qquad\,+\,18 [Z^{\rm LG}_3](\vk,\vp,-\vp) [Z_3^{\rm LG}](\vk,\vq,-\vq)\mathcal{P}^{(0)}_{\rm L}(\vk)\mathcal{P}^{(0)}_{\rm L}(\vp)\mathcal{P}^{(0)}_{\rm L}(\vq) \\\nonumber
    &&\qquad\qquad\qquad\,+\,48\Big(Z^{\rm LG}_2(\vp,\vk-\vp) [Z^{\rm LG}_4](-\vp,\vp-\vk,\vq,-\vq) \mathcal{P}^{(0)}_{\rm L}(\vk-\vp)\mathcal{P}^{(0)}_{\rm L}(\vp)\mathcal{P}^{(0)}_{\rm L}(\vq)
 \Theta(|\vk-\vp|-p) \\\nonumber
    &&\quad\qquad\qquad\qquad\qquad+\,(\vp\leftrightarrow\vq)\Big) \\\nonumber
    &&\qquad\qquad\qquad\,+\,36 \left|Z^{\rm LG}_3(\vp,\vq,\vk-\vp-\vq)\right|^2 \mathcal{P}^{(0)}_{\rm L}(\vk-\vp-\vq)\mathcal{P}^{(0)}_{\rm L}(\vp)\mathcal{P}^{(0)}_{\rm L}(\vq) \Theta(|\vk-\vp-\vq|-q)\Big\},
\eeq
where $\Theta$ is a Heaviside function. The integrand above is smooth, allowing for efficient numerical computation (\S\ref{subsec: impl-vegas}).

Following \textsc{class-pt} \citep{Chudaykin:2020aoj}, we subtract the $k\to 0$ tadpole contributions, which are degenerate with the shot-noise amplitude, $a_{1,0}$. These arise from the `22', `24' and `33-II' contributions to \eqref{eq: Pdef} with the explicit forms:\footnote{Due to the Heaviside function, the prefactor of the `24' term differs between here and \eqref{eq: 2-loop-ir-safe}.}
\beq\label{eq: 2-loop-tadpole}
      \lim_{k\to 0}P^{\rm det, LG}_{\rm 1-loop}(\vk) &=& \int_{\vp}2\lim_{k\to 0}\left(\left|Z^{\rm LG}_2(\vp,\vk-\vp)\right|^2\right)[\mathcal{P}^{(1)}_{\rm L}(\vp)]^2\\\nonumber
      \lim_{k\to 0}P^{\rm det, NLG}_{\rm 1-loop}(\vk) &=& \int_{\vp}4\lim_{k\to 0}\left(Z^{\rm LG}_2(\vp,\vk-\vp)Z^{\rm NLG}_2(\vp,\vk-\vp)\right)[\mathcal{P}^{(0)}_{\rm L}(\vp)]^2\\\nonumber
      \lim_{k\to 0}P^{\rm det}_{\rm 2-loop}(\vk) &=& \int_{\vp\,\vq}\Theta(q-p)\Big\{
   24\Big[\lim_{k\to 0}\left(Z^{\rm LG}_2(\vp,\vk-\vp) [Z^{\rm LG}_4](-\vp,\vp-\vk,\vq,-\vq)\right)[\mathcal{P}^{(0)}_{\rm L}(\vp)]^2\mathcal{P}^{(0)}_{\rm L}(\vq)+\,(\vp\leftrightarrow\vq)\Big] \\\nonumber
      &&\qquad\qquad\quad\qquad\,+\,36 \left|Z^{\rm LG}_3(\vp,\vq,-\vp-\vq)\right|^2 \mathcal{P}^{(0)}_{\rm L}(\vp+\vq)\mathcal{P}^{(0)}_{\rm L}(\vp)\mathcal{P}^{(0)}_{\rm L}(\vq) \Theta(|\vp+\vq|-q)\Big\}.
\eeq
Here, the NLG contribution arises from the $\mathcal{O}_{5,1}=\delta\nabla^2\delta$ operator. To ensure numerical accuracy, we subtract this contribution at the integrand level, computing the $k\to 0$ limits of kernel products analytically, following \S\ref{subsec: impl-renorm} (again factorizing into products of bias operators and $f(z)$). We validate this numerically by checking that the low-$k$ limits of $P_{1-\rm loop}^{\rm det}(\vk)$ and $P_{2-\rm loop}^{\rm det}(\vk)$ ares $\mathcal{O}(k^2,k^2P_{\rm L}(k))$ and $\mathcal{O}(k^2,k^4P_{\rm L}(k))$ respectively (noting that we do not remove $k^2$ or $k^2\mu^2$ pieces from the two-loop spectrum in order to avoid expanding $P_{\rm L}$).

\subsection{Infrared Resummation}\label{subsec: impl-ir-res}
\noindent To assemble the damped linear power spectra used in \S\ref{sec: ir-resum}, we require an explicit form for the wiggle and no-wiggle power spectrum, $P_{\rm w}$ and $P_{\rm nw}$. These are obtained from the linear power spectrum, $P_{\rm L}$, using the procedure of \citep{Vlah:2015zda} (used also in \textsc{ClassOneLoop} \citep{Linde:2024uzr}). Given the Einstein-Hu fitting formula for the no-wiggle power spectrum, $P_{\rm EH}$ \citep{Eisenstein:1997ik} (as implemented in \textsc{nbodykit} \citep{Hand:2017pqn}), we decompose the spectra as
\beq
    P_{\rm nw}(k,z) = \mathcal{G}_{\lambda}\left[\frac{P_{\rm L}}{P_{\rm EH}}\right](k,z), \qquad P_{\rm w}(k,z) = P_{\rm L}(k,z) - P_{\rm nw}(k,z),
\eeq
where $\mathcal{G}_\lambda$ indicates Gaussian smoothing in $\log k$ with width $\lambda$. Explicitly, we use the smoothing kernel
\beq
    \mathcal{G}_\lambda[F](k,z) = \int_{-5}^{1} d\log_{10}q\frac{1}{\sqrt{2\pi}\lambda(\log_{10}k)}\exp\left[-\frac{(\log_{10}q-\log_{10}k)^2}{2\lambda^2(\log_{10}k)}\right]F(q,z),
\eeq
where the (numerically optimized) smoothing scale is given by $\lambda(x) = 0.005 + 0.45e^{-0.9(x+0.15)^2}/(1+e^{8(x+0.5)})$. We additionally set $P_{\rm nw} = P_{\rm L}$ for $k<5\times 10^{-5}\hMpc$ and $k>1\hMpc$ to avoid edge effects. We find that this approach yields around $10\times$ smaller oscillatory residuals in $P_{\rm nw}$ than the discrete-sine-transform approach used in \textsc{class-pt} \citep{Chudaykin:2020aoj,Hamann:2010pw}.

\subsection{Numerical Integration}\label{subsec: impl-vegas}
\noindent In this work, we use a Monte Carlo approach to compute the loop integrals. Due to its simplicity and lack of approximation, this is preferred over analytic methods (e.g., \textsc{fftlog} and the massive propagator basis \citep{Simonovic:2017mhp,Anastasiou:2025jsy}). In practice, our numerical scheme can be combined with a template decomposition such as \textsc{cobra} \citep{Bakx:2024zgu} to efficiently assemble the cosmology-dependent spectra in a likelihood analysis. In Appendix \ref{app: class-pt}, we compare the one-loop power spectra described below to those of \textsc{class-pt} \citep{Chudaykin:2020aoj}; once differences in the treatment of IR-resummation are accounted for, we find excellent agreement between the two codes.

Using the bias-factorized kernels discussed in \S\ref{subsec: impl-ops}, we can express the $n$-loop contributions to the power spectrum as a sum over templates multiplied by pairs of Wilson coefficients:\footnote{While we could additionally expand the templates in powers of $f(z)$, as in \eqref{eq: Zn-decomposition}, the series is finite only in the absence of IR resummation (since the damped power spectra depend exponentially on $f(z)$ through $\Sigma^2_{\rm IR}$).}
\beq\label{eq: total-power-spec}
    P^{n-{\rm loop}}_\ell(k,z;c)=\sum_{A\le B}c_A(z)c_B(z)P^{(A,B),n-{\rm loop}}_{\ell}(k,z),
\eeq
where the $n$-loop integrals scale as $\sigma_8^{2n+2}$, up to IR resummation effects. For the redshift-space power spectrum at two-loop order, there are $320$ non-vanishing two-loop integrals with $\ell\leq 8$ (\textit{i.e.}\ combinations of $A$, $B$ and $\ell$, including IR damping) and $475$ one-loop integrals (most of which are sourced by NLG contributions).

The $n=1$ templates require integrating over $\vp$ and averaging over the orientation of $\vk$ with respect to the line-of-sight, $\hn$. Rotating coordinates such that $\vk$ lies along the $z$-axis, the integral over $\hk$ can be exchanged for an integral over $\hn$; parametrizing the remaining degrees of freedom by $\log p$ and $x\equiv \hat{\vp}\cdot\hk$ (setting $\phi_p=0$ by azimuthal symmetry), the full integral becomes
\beq\label{eq: 1-loop-in-practice}
    P^{(A,B),\rm 1-loop}_{\ell}(k,z) = \int_{\log p_{\rm min}}^{\log p_{\rm max}}
      \frac{p^3d\log p}{2\pi^2}\int_{-1}^{1}\frac{dx}{2}(2\ell+1)\int\frac{d\hn}{4\pi}\mathcal{L}_\ell(\hk\cdot\hn)\,
      \mathcal{I}^{(A,B)}_1(\vp,\vk;z,\hn),
\eeq
for integrand $\mathcal{I}_1^{(A,B)}$, setting $p_{\rm min}=10^{-5}\hMpc$ and $p_{\rm max}=50\hMpc$. Following \eqref{eq: kernel-los-sum}, we can express the integrand as a Cartesian sum over $(\hat{n}_x,\hat{n}_y, \hat{n}_z)$, combining all diagrams of the same order, renormalizing, and removing the $k\to 0$ tadpoles as described above. In the absence of anisotropic IR damping, the $\hn$ integral can be performed analytically;\footnote{Explicitly, we use the result $\int d\hn\,\hat{n}_x^\alpha\hat{n}_y^\beta\hat{n}_z^\gamma = 4\pi(\alpha-1)!!(\beta-1)!!(\gamma-1)!!/(\alpha+\beta+\gamma+1)!!$ if $\alpha,\beta,\gamma$ are even and zero else, after expressing $\mathcal{L}_\ell(\hk\cdot\hn)$ as a polynomial in $\hat{n}_z$ (since we restrict $\vk$ to lie along the $z$-axis).} in its presence, we integrate numerically, using 16-node Gauss-Legendre integration in $\hn\cdot\hk$ and uniform sampling in azimuth with $24$ points. Since the $\hn$-integration does not require recomputing the kernels (just the damped power spectra, which are fast), the computational cost is subdominant; moreover, we can assemble all multipoles simultaneously. The remaining integral over $\log p$ and $x$ is performed using the \textsc{vegas} algorithm \citep{Lepage:1977sw,Lepage:2020tgj}, as discussed below.

A similar method can be used to compute the $n=2$ templates. These depend on five parameters in addition to the line-of-sight direction (holding $\hk$ fixed): $\log q$, $\log p$, $x_q=\hat{\vq}\cdot\hk$, $x_p=\hat{\vp}\cdot\hk$ and the relative azimuth $\Delta\phi$. This results in the integral
\beq
    P^{(A,B),\rm 2-loop}_{\ell}(k,z) &=& \int_{\log p_{\rm min}}^{\log p_{\rm max}}\frac{q^3d\log q}{2\pi^2}
    \int_{\log p_{\rm min}}^{\log p_{\rm max}}\frac{p^3d\log p}{2\pi^2}\int_{-1}^{1}\frac{dx_q}{2}\int_{-1}^{1}\frac{dx_p}{2}
    \int_{0}^{2\pi}\frac{d\Delta\phi}{2\pi}\\\nonumber
    &&\,\times\,(2\ell+1)\int\frac{d\hn}{4\pi}\mathcal{L}_\ell(\hk\cdot\hn)\,
    \mathcal{I}_2^{(A,B)}(\vq,\vp,\vk;z,\hn),
\eeq
where $\mathcal{I}_2^{(A,B)}$ is the two-loop integrand. To enforce the IR-safe ordering $p\leq q$ (from \eqref{eq: 2-loop-ir-safe}), we replace $\log p$ by $r = \log(p/p_{\rm min})/\log(q/p_{\rm min})$ for $r\in[0,1]$, and explicitly average the contributions from $\vq$ and $-\vq$. We use $32$ Gauss--Legendre nodes and $24$ uniform points to compute the $\hn$ integral (with an analytic solution obtained for $\Sigma^2_{\rm IR}=0$), and implement the \textsc{vegas} algorithm for the remaining five degrees of freedom.

When computing the integrals, the rate-limiting step is evaluation of the redshift-space kernels. To minimize computational costs, we compute all power spectrum templates simultaneously (stacking bias components, multipoles, redshifts, and cosmologies), ensuring maximal reuse of each $Z_n$ computation. Following \citep{Goroff:1986ep,Jain:1993jh}, the $F_n$ and $G_n$ kernels are obtained using recursion relations (using symmetries to reduce the number of argument permutations that must be evaluated) and are cached to avoid recomputation. The \textsc{vegas} integrals are computed using a semi-adaptive scheme, first learning the structure of the integrand through a warm-up period, then freezing the grid and restarting the integration, progressively increasing the number of iterations until a target precision is reached. Convergence is assessed by the criterion:
\beq\label{eq: conv}
     \sum_{\ell}\frac{1}{\sigma^2_\ell(k,z)}\left\langle\mathrm{Var}[P^{n-{\rm loop}}_\ell(k,z;c)]\right\rangle_{c}\leq \epsilon,
\eeq
which compares the Monte Carlo variance of the summed template, $P_\ell^{n-{\rm loop}}$ \eqref{eq: total-power-spec}, to the statistical power spectrum errors, $\sigma^2_\ell$, averaged over bias parameters according to the priors in Tab.\,\ref{tab: priors}.\footnote{We use a similar form in the \textsc{vegas} adaptation step, replacing $\mathrm{Var}[P_\ell^{n-{\rm loop}}]\to P_\ell^{n-{\rm loop}}$. When assembling templates across multiple cosmologies, we rescale to a fiducial $\sigma_8$, then take the maximal error across all cosmologies.} By default, we set $\epsilon=0.5$, and additionally store the Monte Carlo covariance between all templates (which is highly correlated). The entire approach is embarrassingly parallelized both within and across $k$ points. 

\subsection{Coordinate Rescalings}
\noindent Our code accounts for Alcock--Paczy\'{n}ski effects by remapping the output power spectrum, $P_\ell^{\rm theory}(k)$, according to the $\alpha_\parallel,\alpha_\perp$ parameters, as discussed in \S\ref{sec: ap}. From \eqref{eq: multipole-AP}, the distorted template can be written as a sum over the input multipoles:
\beq\label{eq: AP-in-practice}
    P^{\rm AP}_\ell(k;\alpha_\parallel,\alpha_\perp) &=& \frac{2\ell+1}{\alpha_\perp^2\alpha_\parallel}\int_{-1}^1\frac{d\mu}{2}\left[\sum_{L\geq 0}P_L^{\rm theory}(k'(k,\mu),\mu'(\mu))\mathcal{L}_L(\mu'(\mu))\right]\mathcal{L}_\ell(\mu)\\\nonumber
    &=&\frac{2\ell+1}{\alpha_\perp^2\alpha_\parallel}\sum_{L\geq 0}\int_{0}^1 d\mu\,P_L^{\rm theory}\left(e^{\log k+\sigma(\mu)}\right)\mathcal{L}_L(\mu'(\mu))\mathcal{L}_\ell(\mu),
\eeq
parametrizing the length distortion via $\sigma(\mu) \equiv \tfrac12\log[(1-\mu^2)/\alpha_\perp^2+\mu^2/\alpha_\parallel^2]$. Notably, the length distortion takes the form of a convolution in $\log k$ across a window of width $|\log(\alpha_\perp/\alpha_\parallel)|$. This can be efficiently evaluated by first representing the power spectrum as a spline on a uniform $\log k$ grid, then performing the template-independent $\mu$ integral using Gauss--Legendre quadrature. Using a quintic spline with a spacing of $\Delta \log k = 0.015$ and 16 integration nodes, this can be computed in less than a millisecond, and recovers the directly integrated result to within $1$ part in $10^5$.

\section{Application to Simulations}\label{sec: simulations}
\noindent Finally, we validate our model using numerical simulations. For this purpose, we fit galaxy power spectra using tree-level, one-loop, and two-loop models, testing whether the EFT-based theory models provide an accurate fit to the data, and whether they can recover the input clustering amplitude (leaving a full cosmological parameter inference to future work). As a warm-up exercise, we will also analyze the dark matter power spectrum in both real- and redshift-space.

\subsection{Datasets}\label{subsec: sim-data}
\noindent We use the \textsc{AbacusSummit} simulation suite \citep{Maksimova:2021ynf}, created with the \textsc{Abacus} code \citep{Garrison:2021lfa}. Specifically, we employ the 25 \texttt{c000} realizations at $z=0.5$ and $z=1.1$, generated at the \textit{Planck} 2018 best-fit cosmology: $\omega_b=0.02237$, $\omega_{\rm cdm}=0.12$,
$h=0.6736$, $n_s = 0.9649$, $\sum m_\nu=0.06\,\mathrm{eV}$ and $A_s = 2.083\times 10^{-9}$ (giving $\sigma_8 = 0.8076$ and $r_{\rm BAO}=99.1\Mpch$), with a single massive neutrino species. Each simulation contains $6912^3$ particles in a volume of $8h^{-3}\mathrm{Gpc}^3$, corresponding to a mass resolution of $2.1\times 10^9h^{-1}M_\odot$, with dark matter halos identified using the \textsc{CompaSO} code \citep{Hadzhiyska:2021zbd} (cleaned following \citep{Bose:2021orc}). As usual, redshift-space distortions are incorporated by shifting the relevant tracers along the line-of-sight direction by their projected velocities. 

We analyze two types of clustering sample: the dark matter field (comprising cold dark matter and baryons, obtained by subsampling $0.3\%$ of the simulation particles), and galaxies constructed from the dark matter halos. Using the \textsc{AbacusHOD} code \citep{Yuan:2021izi}, we build two types of catalog, approximating those of the public DESI mocks \citep{DESI:2024uvr}: a luminous red galaxy (LRG) sample at $z=0.5$, constructed using a five-parameter halo occupation distribution model of \citep{Zheng:2007zg} supplemented with velocity bias and incompleteness; (2) an emission line galaxy (ELG) sample at $z=1.1$, obtained from the high-mass quenched prescription of \citep{Rocher:2023zyh}. In both cases, we use hyperparameters fitted to early DESI data \citep{Yuan:2023ezi,Rocher:2023zyh}. The LRG (ELG) sample has a number density of $\bar n=5.2\times10^{-4}h^3\mathrm{Mpc}^{-3}$ ($7.6\times10^{-4}h^3\mathrm{Mpc}^{-3}$) and a satellite fraction of $10\%$ ($2\%$).

\subsection{Statistics}
\noindent Power spectra are measured with the \textsc{abacusutils} package \citep{Maksimova:2021ynf,Garrison:2021lfa,Yuan:2021izi}. We use $100$ linearly-spaced bins across a wavenumber range $k_F\leq k\leq 0.5\hMpc$ (for fundamental mode $k_F=(2\pi/2000)\hMpc\approx 0.003\hMpc$) for each of the monopole ($\ell=0$), quadrupole ($\ell=2$) and hexadecapole ($\ell=4$) moments. These are computed using an interlaced triangle-shaped-cell mass assignment scheme with Nyquist frequency $k_{\rm Nyq}=2.4\hMpc$, and we subtract shot-noise contributions, as measured on the Fourier-space grid. Below, we analyze both the individual realizations and their mean in order to (a) quantify the model error in a typical realization (noting that the entire DESI DR1 dataset has $V_{\rm eff}\approx 5h^{-3}\mathrm{Gpc}^3$ \citep{DESI:2024aax}, smaller than a single \textsc{AbacusSummit} box) and (b) robustly constrain underlying theoretical errors in the low-noise regime.

To accurately compare theory and data, we must account for binning and discreteness effects. Given a set of theoretical power spectrum multipoles, $P_L^{\rm theory}$, the binned power spectra, $P_\ell^{\rm bin}$, can be obtained using a window matrix, $\mathcal{W}$: 
\beq\label{eq: window-function}
    P^{a, \rm bin}_\ell &=& \sum_{L\geq 0}\sum_{\vq\in\mathrm{bin}_a}\mathcal{W}_{\ell L}^a(\vq)P_L^{\rm theory}(|\vq|)\\\nonumber
    \mathcal{W}_{\ell L}^a(\vq) &\equiv& \frac{2\ell+1}{N_a}\mathcal{L}_\ell(\hq\cdot\hn)\mathcal{L}_L(\hq\cdot\hn), \qquad N_a \equiv \sum_{\vq\in\mathrm{bin}_a}1,
\eeq
where $\{\vq\}$ is the discrete lattice of Fourier modes and $\mathrm{bin}_a$ represents the set of modes with magnitudes in wavenumber bin $a$. We compute $\mathcal{W}^a_{\ell L}(\vq)$ by direct summation (truncating to $L\leq 8$) and implement \eqref{eq: window-function} by interpolating the theoretical power spectra onto the Fourier-space grid. 

In the Gaussian limit, the covariance of the binned power spectra is given by
\beq\label{eq: gaussian-cov}
    \mathrm{Cov}^{aa', \rm Gaussian}_{\ell\ell'} &=& (2\ell+1)(2\ell'+1)\frac{2\delta_{\rm K}^{aa'}}{N_a^2}\sum_{\vec q\in \mathrm{bin}_a}P^2_{\rm tot}(\vq)\mathcal{L}_\ell(\hq\cdot\hn)\mathcal{L}_{\ell'}(\hq\cdot\hn),\\\nonumber
    P_{\rm tot}(\vq) &=& \sum_{\ell=0,2,4}P_{\ell}(|\vq|)\mathcal{L}_\ell(\hq\cdot\hn)
\eeq
where $P_{\rm tot}(\vq)$ is the measured power spectrum (averaged over realizations, including shot-noise) evaluated on the lattice. To capture the non-linear mode coupling contributions, we combine this with the sample covariance extracted from around $1800$ small-box simulations (\texttt{small\_c000}, each of volume $0.125h^{-3}\mathrm{Gpc}^3$), via
\beq\label{eq: cov-full}
    \mathrm{Cov}^{\rm full} = \mathrm{Cov}^{\rm Gaussian} + \frac{V_{\rm small}}{V_{\rm full}}\left[\mathrm{Cov}^{\rm sample}_{\rm small} - \mathrm{Cov}^{\rm Gaussian}_{\rm small}\right],
\eeq
where $V_{\rm small}/V_{\rm full}=1/64$ is the ratio of volumes between the two sets of simulations and $\mathrm{Cov}_{\rm small}^{\rm Gaussian}$ is the discrete result \eqref{eq: gaussian-cov} applied to the small boxes. To minimize discreteness and noise artifacts, we include the term in square brackets (which encodes the connected trispectrum) only for $k\geq 0.03\hMpc$, and smooth using a bivariate cubic spline. 
The covariance of the combined power spectra (averaged over all simulation volumes) is obtained by dividing \eqref{eq: cov-full} by $25$.

\subsection{Theory \& Likelihood}\label{sec: sim-priors}

\begin{table}
        \centering
        \begin{tabular}{ll|ccc}
        \textbf{Parameter} & \textbf{Prior} & Tree? & 1-loop? & 2-loop?\\\hline
        \multicolumn{5}{l}{\textit{Bias coefficients}}\\\hline
        $c_{1,0}A$ & $\mathcal{U}(0.3,3.5)$ & \checkmark & \checkmark & \checkmark \\
        $c_{2,0}A^2,\,c_{3,0}A^2$ & $\mathcal{N}(0,1)$ &  & \checkmark & \checkmark \\
        $c_{4,0}A^3$ & $\mathcal{N}(0,1)$ &  & \checkmark & \checkmark \\
        $c_{5,0}A^3,\ldots,c_{7,0}A^3$ & $\mathcal{N}(0,1)$ &  &  & \checkmark \\
        $c_{8,0}A^4,\ldots,c_{11,0}A^4$ & $\mathcal{N}(0,1)$ &  &  & \checkmark \\
        $c_{12,0}A^5,\ldots,c_{17,0}A^5$ & $\mathcal{N}(0,1)$ &  &  & \checkmark \\
        $c_{1,1}A$ & $\mathcal{N}(0,k_{\rm NL}^{-4})$ &  & \checkmark & \checkmark \\
        $c_{2,1}A^2,\ldots,c_{5,1}A^2$ & $\mathcal{N}(0,k_{\rm NL}^{-4})$ &  &  & \checkmark \\
        $c_{6,1}A^3,\,c_{7,1}A^3$ & $\mathcal{N}(0,k_{\rm NL}^{-4})$ &  &  & \checkmark \\
        $c_{1,2}A$ & $\mathcal{N}(0,k_{\rm NL}^{-8})$ &  &  & \checkmark \\
        \hline
        \multicolumn{5}{l}{\textit{Redshift-space counterterms}}\\\hline
        $\bar{c}^{(1)}_{1,1}A,\,\bar{c}^{(2)}_{1,1}A$ & $\mathcal{N}(0,k_{v}^{-4})$ &  & \checkmark &  \\
        $\hat{c}^{(1)}_{1,1}A,\,\hat{c}^{(2)}_{1,1}A,\,\hat{c}^{(2)}_{2,1}A,\,\hat{c}^{(2)}_{3,1}A$ & $\mathcal{N}(0,k_{v}^{-4})$ &  &  & \checkmark \\
        $\hat{c}^{(1)}_{2,1}A^2,\ldots,\hat{c}^{(1)}_{4,1}A^2,\;\hat{c}^{(2)}_{4,1}A^2,\ldots,\hat{c}^{(2)}_{6,1}A^2$ & $\mathcal{N}(0,k_{v}^{-4})$ &  &  & \checkmark \\
        $\hat{c}^{(1)}_{5,1}A^3,\,\hat{c}^{(1)}_{6,1}A^3$ & $\mathcal{N}(0,k_{v}^{-4})$ &  &  & \checkmark \\
        $\hat{c}^{(1)}_{1,2}A,\,\hat{c}^{(2)}_{1,2}A,\,\hat{c}^{(3)}_{1,2}A$ & $\mathcal{N}(0,k_{v}^{-8})$ &  &  & \checkmark \\
        \hline
        \multicolumn{5}{l}{\textit{Non-local counterterms}}\\\hline
        $\tilde{c}_{1,1}A^3$ & $\mathcal{N}(0,k_{\rm NL}^{-4})$ &  &  & \checkmark \\
        $\tilde{c}^{(1)}_{1,1}A^2,\,\tilde{c}^{(1)}_{2,1}A^3$ & $\mathcal{N}(0,k_{v}^{-4})$ &  &  & \checkmark \\
        \hline
        \multicolumn{5}{l}{\textit{Stochastic amplitudes}}\\\hline
        $a_{1,0}$ & $\mathcal{N}(0,\bar{n}^{-2})$ & \checkmark & \checkmark & \checkmark \\
        $a_{1,1}$ & $\mathcal{N}(0,\bar{n}^{-2}k_{\rm NL}^{-4})$ &  & \checkmark & \checkmark \\
        $a_{2,1}$ & $\mathcal{N}(0,\bar{n}^{-2}k_{v}^{-4})$ &  & \checkmark & \checkmark \\
        $a_{1,2}$ & $\mathcal{N}(0,\bar{n}^{-2}k_{\rm NL}^{-8})$ &  &  & \checkmark \\
        $a_{2,2}$ & $\mathcal{N}(0,\bar{n}^{-2}k_{\rm NL}^{-4}k_{v}^{-4})$ &  &  & \checkmark \\
        $a_{3,2}$ & $\mathcal{N}(0,\bar{n}^{-2}k_{v}^{-8})$ &  &  & \checkmark
        \end{tabular}
        \caption{\textbf{Parameters \& Priors}. Coefficients entering the redshift-space galaxy power spectrum model and their fiducial priors. $\mathcal{U}(a,b)$ indicates a uniform distribution across $[a,b]$, while $\mathcal{N}(\mu,\sigma^2)$ represents a Gaussian centered at $\mu$ with variance $\sigma^2$. We rescale parameters by factors of $A\equiv\sigma_8/\sigma_{8,\rm fid}$ to reduce posterior projection effects, and combine degenerate counterterms following \eqref{eq: ctr-comb-1loop}\,\&\,\eqref{eq: ctr-comb-2loop}. We employ order-one priors on all parameters (except $c_{1,0}$), scaling the counterterms by appropriate powers of the non-linear density and velocity scales, $k_{\rm NL}$ and $k_v$. At one loop, all parameters except $c_{1,0}$, $c_{2,0}$ and $c_{3,0}$ enter the model linearly and are marginalized analytically in the likelihood; at two loops, the number of explicitly sampled parameters increases to twelve. We additionally sample $\sigma_8$ with the prior $\mathcal{U}(0.6,1.0)$.}
        \label{tab: priors}
\end{table}

\noindent The most important ingredient in the power spectrum model is the linear power spectrum, $P_{\rm L}(k)$. Here, this is computed using \textsc{class} \citep{Lesgourgues:2011re}, choosing high precision settings across $7000$ logarithmically-spaced points in $k\in[10^{-5},100]\hMpc$, and restricting to the cold dark matter and baryon sector \citep[e.g.,][]{Villaescusa-Navarro:2017mfx}. We evaluate the spectrum across five values of $\sigma_8$ linearly spaced in $[0.6,1.0]$, decomposing each into wiggle- and no-wiggle components using the method discussed in \S\ref{subsec: impl-ir-res}. To match the above catalogs, we use two redshifts: $z=0.5$ and $z=1.1$.

We compute the power spectrum templates, $P_\ell^{(A,B)}(k;\sigma_8)$ for each $\sigma_8$ value across all multipoles up to $\ell_{\rm max}=8$. This uses a fine grid of $189$ points in $k\in[0.0005,0.41]\hMpc$, with logarithmic binning up to $k=0.03\hMpc$ (to efficiently capture large scales) and linear binning with $\Delta k=0.0025\hMpc$ thereafter (to cleanly resolve the BAO peak). After whitening by an appropriate power of $\sigma_8$ (e.g., $P_{\rm tree}\to P_{\rm tree}/\sigma_8^2$),\footnote{The remaining dependence on $\sigma_8$ comes only from the IR-damping exponent, $\Sigma_{\rm IR}$, allowing for efficient interpolation with few $\sigma_8$ nodes.} we build cubic interpolators for power spectrum component, allowing the full spectrum to be efficiently reconstructed at any $\sigma_8$. 
Computing all templates required for the one-loop galaxy power spectrum analysis (across five cosmologies and two redshifts) took 170 CPU-hours, while those for the two-loop analysis required 25\,000 CPU-hours (with somewhat overzealous precision). Restricting to the matter sector, this reduced to 100 and 16\,000 CPU-hours at one- and two-loops respectively. 

Given the templates, we can reconstruct the combined power spectrum by projecting onto bias parameters (using \eqref{eq: total-power-spec}), adding the stochastic sector (defined in \eqref{eq: stoch-Pk}), interpolating in $k$-space, applying Alcock--Paczy\'{n}ski distortions (through \eqref{eq: AP-in-practice}), and convolving with the window function (via \eqref{eq: window-function}). In this work, we fix all cosmological parameters except for $\sigma_8$; this allows us to omit the distortion step and absorb the window function in the templates themselves.

Assuming Gaussian noise, the likelihood for $\sigma_8$ and the Wilson coefficients $\Theta$ is given by
\beq\label{eq: chi2}
    -2\log L(\sigma_8,\Theta) \equiv \chi^2(\sigma_8,\Theta)=\sum_{\ell\ell'aa'}\left(P_\ell^a(\sigma_8,\Theta)-\widehat{P}_\ell^a\right)\left[\mathrm{Cov}^{\rm full}+\mathrm{Cov}^{\rm MC}\right]_{\ell\ell'}^{-1,aa'}\left(P_{\ell'}^{a'}(\sigma_8,\Theta)-\widehat{P}_{\ell'}^{a'}\right),
\eeq
summing over multipoles and bins for data $\widehat{P}$ and statistical covariance $\mathrm{Cov}^{\rm full}$. To guard against residual noise in the templates, we add a Monte Carlo covariance, $\mathrm{Cov}^{\rm MC}$; this is obtained from the \textsc{vegas} samples, and evaluated at the best-fit Wilson coefficients from a preliminary analyses. To expedite sampling, we analytically marginalize over any parameter appearing in the model linearly, using the model derivative $\partial P_\ell^a(\sigma_8,\Theta)/\partial \theta_i$ \citep[e.g.,][]{Philcox:2020zyp}. The likelihood is sampled with \textsc{emcee} \citep{Foreman-Mackey:2012any}, using 64 walkers each with $20\,000$ steps. Since the dependence on parameters appears only through polynomials and interpolation, each likelihood call takes less than a millisecond.

Tab.\,\ref{tab: priors} lists the relevant Wilson coefficients and their priors. To minimize projection effects, we rescale parameters by factors of $A\equiv \sigma_8/\sigma_{8,\rm fid}$, following \citep{Maus:2024sbb,Tsedrik:2025hmj,Chudaykin:2025aux}. Explicitly, for an operator $\mathcal{O}_{a,b}$ starting at $n$-th order in $\delta_{\rm L}$, we redefine $c_{a,b}\to c_{a,b}A^n$ to remove the leading degeneracy directions. Motivated by naturalness arguments, we adopt $\mathcal{O}(1)$ priors for each parameter, scaling the stochastic contributions by $\bar{n}$, the real-space gradient terms by the non-linear scale, $k_{\rm NL}$, and the redshift-space gradients by the velocity scale $k_v$, defined by $k_{\rm NL}^3P_{\rm lin}(k_{\rm NL})=2\pi^2$ and $k_v^{-2}=\int dk\,P_{\rm L}(k)/6\pi^2$ (at the fiducial cosmology). 
The full tree-level, one-loop, and two-loop models are specified by $2$, $10$ and $49$ parameters respectively, with all but $1$, $3$ and $12$ marginalized analytically.

We adopt three metrics to assess how well the theoretical models describe the data:
\begin{itemize}
    \item \textbf{Goodness-of-fit}: Fixing to the true cosmology, we compute $\chi^2$ at the best-fit point for each of the 25 \textsc{AbacusSummit} simulations, analyzed independently. If the model is accurate, each value should follow a $\chi^2$ distribution: $\chi_i^2\sim\chi^2_{\nu_{\rm eff}}$, where $\nu_{\rm eff}$ is the effective number of degrees of freedom (accounting for unconstrained parameters) \citep{Raveri:2018wln}. Here, we judge that a model has broken down if the mean value, $\bar\chi^2=(1/25)\sum_{i=1}^{25}\chi^2_i$, is larger than the $95\%$ upper limit, $\chi^2_{\nu_{\rm eff}}(0.95)$; this tests whether a typical realization shows evidence of model misspecification.
    \item \textbf{Parameter bias}: We infer $\sigma_8$ by analyzing the mean of the 25 power spectrum realizations with the combined covariance (omitting the contribution from Monte Carlo noise). This provides a high-fidelity test of the model with an effective volume of $200h^{-3}\mathrm{Gpc}^3$. We assume the model to have broken down if the best-fit $\sigma_8$ value lies further than $0.5\sigma_{\rm single}$ from the truth, where $\sigma_{\rm single}$ is the errorbar in a single realization (approximating the precision of contemporary datasets).\footnote{To reduce sensitivity to noise fluctuations, we require that this condition is satisfied for two consecutive choices of $k_{\rm max}$.}
    \item \textbf{Parameter stability}: We test for running in $\sigma_8$ by comparing the best-fit values of $\sigma_8$ across two values of $k_{\rm max}$, separated by $0.05\hMpc$. For this test, we use the best-fit values 
    from individual simulations (averaged over the 25 realizations) to properly account for nuisance parameter priors. If the two best-fits differ by more than the nested error, $\sqrt{\sigma_{k_{\rm max}'}^2-\sigma_{k_{\rm max}}^2}$, \citep{Gratton:2019fru}, we judge that the model has strayed beyond its regime of validity.
\end{itemize}
For each dataset, we set $k_{\rm max}$ to the largest value (in steps of $0.01\hMpc$) such that all conditions are satisfied.

\subsection{Results}
\subsubsection{Dark Matter}
\noindent To begin, we validate our model for the dark matter power spectrum in both real- and redshift-space. Due to conservation of mass and momentum, dark matter has a much simpler theoretical description than biased tracers: the linear bias is unity, all higher-order LG biases vanish, all counterterms except total derivatives do not contribute (\textit{i.e.}\ $c_{5,1}=c_{a\geq 14,1}=0$), and the LG and NLG stochasticity parameters ($a_{1,0},a_{1,1},a_{2,1}$) are zero. In total, the one-loop model depends on 1 (3) parameters in real-space (redshift-space), while the two-loop model extends this to 9 (28).\footnote{In real-space, fixing the time-dependence of the dark matter stress tensor reduces the number of two-loop parameters to 8 (cf.\,Appendix \ref{app: dm-renorm}), matching \citep{Ivanov:2026zos}.}

\begin{table}[]
    \centering
    \begin{tabular}{l|ccc|ccc}
    & \multicolumn{3}{c|}{\textit{Real-space}} & \multicolumn{3}{c}{\textit{Redshift-space}}\\
    & \,\,Tree-level\,\, & \,\,One-loop\,\, & \,\,Two-loop\,\, & \,\,Tree-level\,\, & \,\,One-loop\,\, & \,\,Two-loop\,\, \\\hline
    \textbf{Dark matter ($z=0.5$)}  & 0.04& 0.15& $\geq$0.50& 0.03& 0.09& 0.33\\
    \textbf{Dark matter ($z=1.1$)}  & 0.06& 0.17& $\geq$0.50& 0.03& 0.12& $\geq$0.50\\\hline
    \textbf{LRGs ($z=0.5$)}  & -- & -- & -- & 0.04& 0.20& 0.34\\
    \textbf{ELGs ($z=1.1$)}  & -- & -- & -- & 0.04& 0.27& 0.42\\
    \end{tabular}
    \caption{\textbf{Scale-cuts}. Range of scales over which the tree-level, one-loop and two-loop theoretical models for the dark matter and galaxy power spectra. For the dark matter sector, we give the maximal scale, $k_{\rm max}$, for two choices of redshift, working in both real- and redshift-space. For galaxies, we adopt two types of catalog, approximating the DESI LRG and ELG samples, and give results only in redshift-space, due to the large degeneracies present in real-space. Results are obtained from 25 \textsc{AbacusSummit} simulations using the metrics described in \S\ref{sec: sim-priors} (testing for goodness-of-fit, as well as bias and running in $\sigma_8$), and are calibrated for an $8h^{-3}\mathrm{Gpc}^3$ volume. All analyses are truncated at $k_{\rm max}=0.50\hMpc$.}
    \label{tab: kmax}
\end{table}

By comparing the \textsc{AbacusSummit} dark matter power spectra to the predictions from perturbative models and applying the metrics described above, we can infer the range of validity of each theory. These results are presented in the top panel of Tab.\,\ref{tab: kmax}. We find a clear hierarchy between the models: while the tree-level description fails on all but the largest scales, adding loop corrections extends the reach to progressively smaller wavenumbers. At $z=0.5$, the two-loop real-space model is accurate to at least $0.5\hMpc$, whereas the one-loop model breaks down at $0.15\hMpc$ and the tree-level model fails beyond $0.04\hMpc$. This matches expectations: the two-loop theory captures higher-order terms in $k/k_{\rm NL}$, which dominate on small-scales. Due to non-linear velocities, the redshift-space models break down somewhat earlier: we find $k_{\rm max}=0.33\hMpc$ for the two-loop theory at $z=0.5$, but $0.09\hMpc$ at one-loop. As expected, the domains of validity increase somewhat at higher redshift, where non-linearities are suppressed. Our results are broadly consistent with previous works \citep{Carrasco:2013sva,Baldauf:2015aha,Foreman:2015lca,Taule:2023izt}, which found $k_{\rm max}\approx 0.3\hMpc$ ($0.2\hMpc$) for two-loop models in real-space (redshift-space) at redshift zero.

\begin{figure}[t]
    \centering
    \includegraphics[width=\linewidth]{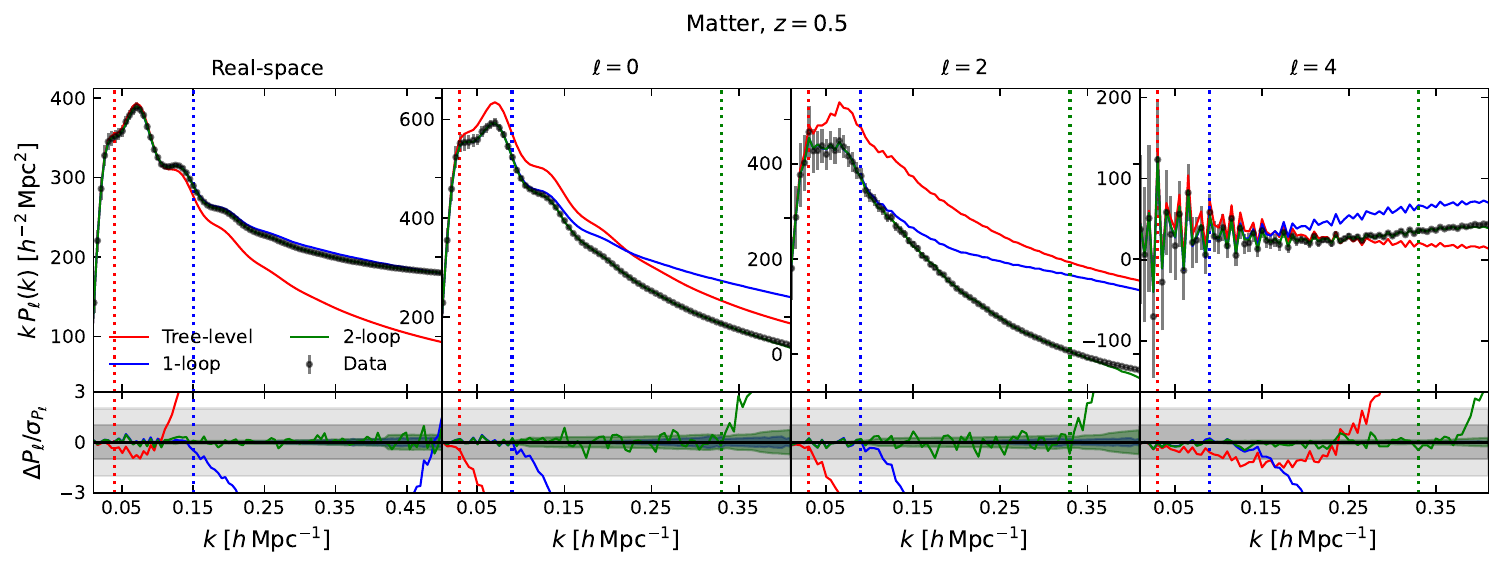}
    \caption{\textbf{Dark Matter Power Spectra}. Comparison of theoretical models for the dark matter power spectrum at $z=0.5$ in real-space (first panel) and redshift-space (subsequent panels). We show the tree-level, one-loop and two-loop results, fitted to the \textsc{AbacusSummit} dark matter power spectrum at fixed cosmology, with the relevant $k_{\rm max}$ values (listed in Tab.\,\ref{tab: kmax}) indicated by the vertical lines. To reduce scatter, we average the power spectra over 25 realizations, but show the errorbars for a single $8h^{-3}\mathrm{Gpc}^3$ volume. The colored bands in the bottom panels show the (subdominant) Monte Carlo error in the theory model. Note that the low-$k$ scatter in the hexadecapole measurements arises from discreteness effects rather than noise; as such, it is present in both the data and the (windowed) model. Across both real- and redshift-space, the two-loop model is able to fit the power spectrum at high precision across a wide range of scales.}
    \label{fig: dm-pk}
\end{figure}

\begin{figure}
    \centering
    \includegraphics[width=0.9\linewidth]{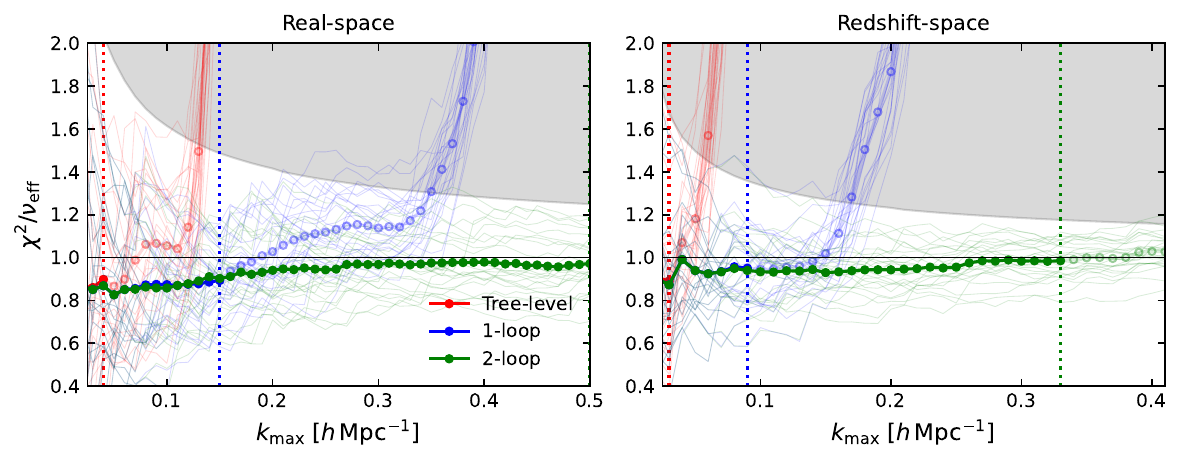}
    \caption{\textbf{Dark Matter Fit}. Goodness-of-fit of the dark matter power spectrum model in real-space (left) and redshift-space (right). We plot the best-fit $\chi^2$ values from \eqref{eq: chi2} as a function of $k_{\rm max}$ for the tree-level, one-loop and two-loop models, normalizing by the effective number of degrees of freedom, $\nu_{\rm eff}$ \citep{Raveri:2018wln}. This is computed for each of 25 independent realizations, with the means indicated by circles. The vertical lines indicate the scale-cuts given in Tab.\,\ref{tab: kmax}, and the gray band indicates regions excluded at $95\%$ confidence. On large-scales, we find a good match of theory and data, but on small-scales, only the two-loop theory matches the simulations.}
    \label{fig: dm-chi2}
\end{figure}

In Fig.\,\ref{fig: dm-pk}, we plot the best-fit models at $z=0.5$, fixing to the true cosmology. Across both real- and redshift-space, the two-loop model closely matches the data across a wide range of scales, capturing both the broadband shape of the power spectra and the BAO oscillations (as well as the discreteness artifacts in $P_4$). This can be quantified by the goodness-of-fit, which Fig.\,\ref{fig: dm-chi2} shows as a function of $k_{\rm max}$. Across all realizations, the two-loop model achieves $\chi^2/\nu_{\rm eff}$ of order-one down to small scales, while the one-loop theory provides a poor fit beyond $0.16\hMpc$ in redshift-space. This test clearly demonstrates that the two-loop model is flexible enough to capture non-linear physics on small scales. 

\begin{figure}
    \centering
    \includegraphics[width=0.9\linewidth]{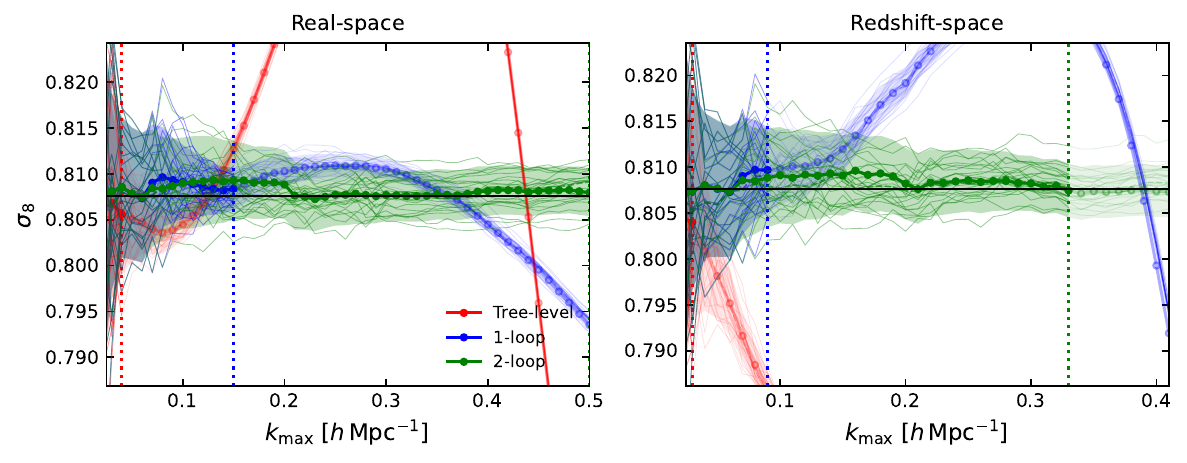}
    \caption{\textbf{Dark Matter Inference}. Constraints on $\sigma_8$ as a function of $k_{\rm max}$ from the real-space (left) and redshift-space (right) dark matter power spectrum. The circles show best-fit results obtained from the mean of 25 \textsc{AbacusSummit} realizations, with inferences from individual realizations shown as faint lines. The errorbars represent the error in a single realization (mimicking that of a realistic observational study). While the tree-level results quickly become biased, we find that the one-loop model is unbiased for $k_{\rm max}\leq 0.15\hMpc$ in real-space and $k_{\rm max}\leq 0.09\hMpc$ in redshift-space. Despite the high precision of the test, the two-loop model is able to fit the data down to small scales, obtaining errors below $0.5\sigma$ up to $k_{\rm max}=0.5\hMpc$ in real-space, or $0.33\hMpc$ in redshift-space. Notably, the individual inferences agree well with the mean; this indicates that prior volume shifts are subdominant.}
    \label{fig: dm-sigma8}
\end{figure}

Fig.\,\ref{fig: dm-sigma8} shows the constraints on $\sigma_8$ obtained from the dark matter power spectrum at $z=0.5$. While the results are strongly biased when using the tree-level model, we find tight constraints from both the one-loop and two-loop theories, reaching a precision of around $0.3\%$. Analyzing the mean of 25 realizations with the covariance of a single volume, we find $\sigma_8=0.8084 \pm 0.0023$ ($0.8094 \pm 0.0047$) at one-loop in real-space (redshift-space) or $0.8085 \pm 0.0026$ ($0.8077 \pm 0.0029$) at two-loop, using the scale cuts of Tab.\,\ref{tab: kmax}. These are in excellent agreement with the true value of $\sigma_8=0.8076$.\footnote{Using instead the covariance for the combined $200h^{-3}\mathrm{Gpc}^{3}$ volume, we find the real-space (redshift-space) results $\sigma_8=0.8084 \pm 0.0005$ and $0.8080\pm0.0006$ ($0.8097\pm0.0009$ and $0.8074\pm0.0006$) at one- and two-loop orders, which match the input value within $1.7\sigma$ and $0.6\sigma$ ($2.2\sigma$ and $0.3\sigma$) respectively.}
Extending beyond the nominal $k_{\rm max}$ values, the one-loop constraints tighten significantly; however, they quickly become biased, indicating a breakdown of the model. Despite the wider range of validity, the two-loop model does not give significant improvements in $\sigma_8$ precision due to the large number of additional Wilson coefficients, whose effects are degenerate with $\sigma_8$ variations. Indeed, at the maximal scale of the one-loop model, the two-loop redshift-space errorbar is $16\%$ larger than the one-loop error (indicating that two-loop effects are becoming important), though stronger constraints are found by $k_{\rm max}=0.33\hMpc$. Similarly, extending the analyses from $k_{\rm max}=0.4\hMpc$ to $k_{\rm max}=0.5\hMpc$ (where possible) improves $\sigma_8$ constraints by $\lesssim 5\%$. Physically, this implies that, though small-scales are highly constraining, their cosmological information content is difficult to access without a better understanding of ultraviolet physics. 

\subsubsection{Galaxies}
\noindent Having validated the dark matter components of the model, we now turn to galaxies. In this section, we will focus on the redshift-space results since the real-space galaxy power spectrum is limited by parameter degeneracies (with $\sigma_8$ information arising only through loop effects).\footnote{See \citep{Ivanov:2026zos} for a discussion of the real-space galaxy power spectrum and the cross-correlation of galaxies and dark matter.} As discussed in \S\ref{subsec: sim-data}, we will present results for two mock galaxy catalogs, which mimic the main DESI samples \citep{DESI:2024aax}: LRGs at $z=0.5$ and ELGs at $z=1.1$. \textit{A priori}, we expect that the former will be harder to model due to its lower redshift and enhanced velocity non-linearity (the fingers-of-God effect).

The scale cuts obtained for each analysis are given in Tab.\,\ref{tab: kmax}. We find a similar story to that of dark matter: the tree-level theory breaks down very quickly, while one-loop provides a good fit up to intermediate scales, which is extended further by two-loop corrections. As expected, both the one- and two-loop models perform better on the ELG sample than on LRGs; moreover, the one-loop reach is larger for both samples than for dark matter. Across all three of the metrics discussed in \S\ref{sec: sim-priors}, the two-loop galaxy model provides an accurate model up to $k_{\rm max}=0.34\hMpc$ for LRGs and $k_{\rm max}=0.42\hMpc$ for ELGs.

\begin{figure}
    \centering
    \includegraphics[width=\linewidth]{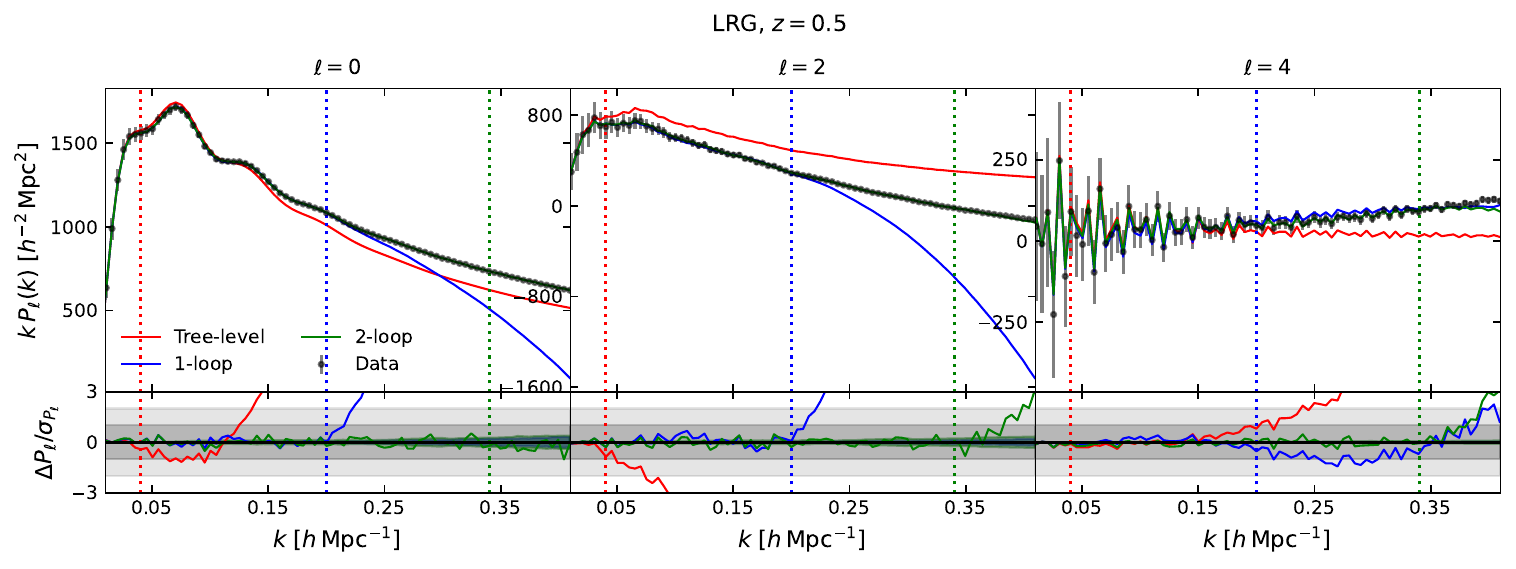}
    \includegraphics[width=\linewidth]{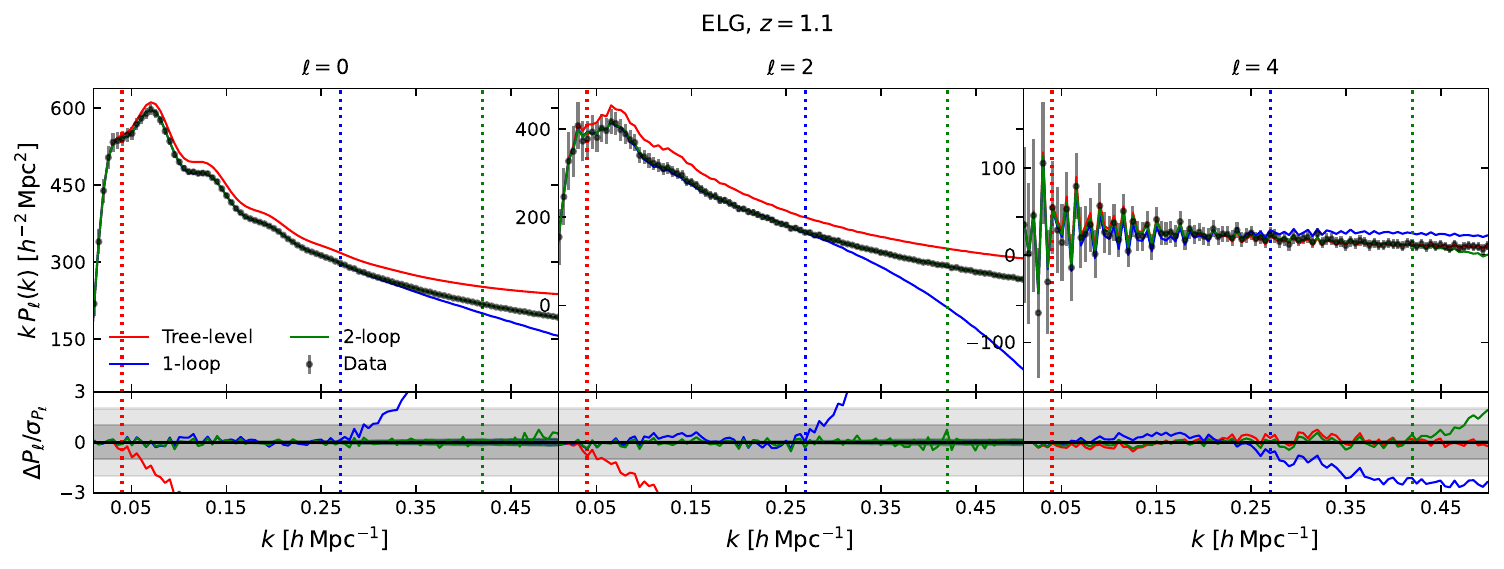}
    \caption{\textbf{Galaxy Power Spectra}. Comparison of theoretical models for the redshift-space galaxy power spectra (following Fig.\,\ref{fig: dm-pk}). We show results for two datasets constructed from \textsc{AbacusSummit} dark matter halos, emulating the DESI LRG (top) and ELG (bottom) samples at redshifts $z=0.5$ and $z=1.1$. As for dark matter, we find that the tree-level models break down on large-scales, while the one-loop and two-loop prescriptions provide much improved fits extending to high-$k$. Due to the reduced gravitational and velocity non-linearities, both the one- and two-loop models perform noticeably better for ELGs than for LRGs.}
    \label{fig: gal-pk}
\end{figure}

\begin{figure}
    \centering
    \includegraphics[width=0.95\linewidth]{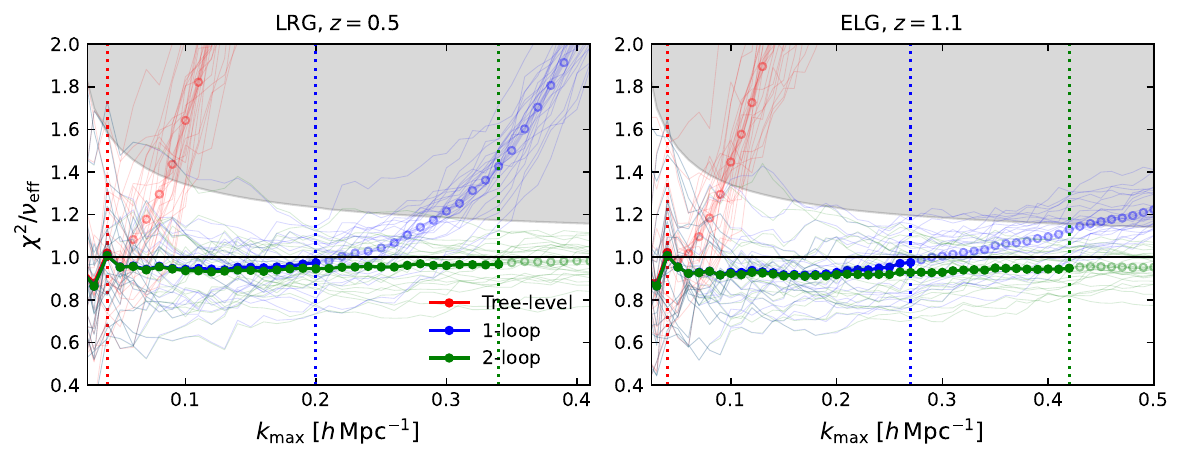}
    \caption{\textbf{Galaxy Fit}. Goodness-of-fit for the redshift-space galaxy power spectrum models as a function of scale (following Fig.\,\ref{fig: dm-chi2}). Similarly to the dark matter case, we find that the two-loop model provides an excellent fit to the data down to small scales, while the one-loop model breaks down at intermediate $k_{\rm max}$. As before, the low-redshift LRG sample shows a reduced regime of validity compared to the ELG sample.}
    \label{fig: gal-chi2}
\end{figure}

In Fig.\,\ref{fig: gal-pk} we compare the theoretical models, evaluated at their respective $k_{\rm max}$, to the simulated LRG and ELG power spectra. In all cases, we find excellent agreement in the spectra up to the truncation scale, including for the damped oscillatory components. This is confirmed by the goodness-of-fit statistics displayed in Fig.\,\ref{fig: gal-chi2}, which show $\chi^2/\nu_{\rm eff}$ of order one up to the fiducial scale-cuts (and growing deviations thereafter), both for the mean and for individual realizations. Consistent with the above discussion, the one-loop model breaks down significantly earlier for LRGs than for ELGs. 

\begin{figure}
    \centering
    \includegraphics[width=0.75\linewidth]{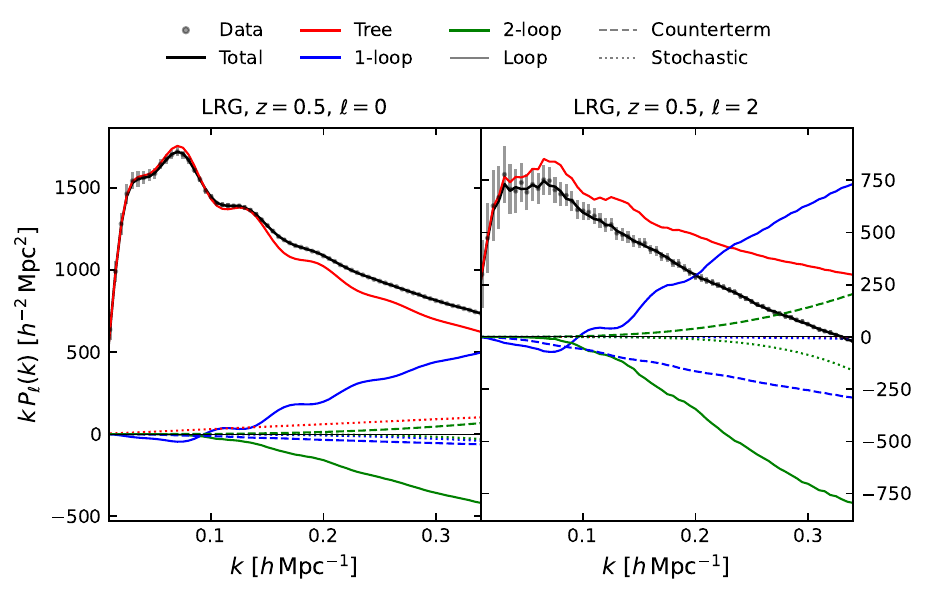}
    \includegraphics[width=0.75\linewidth]{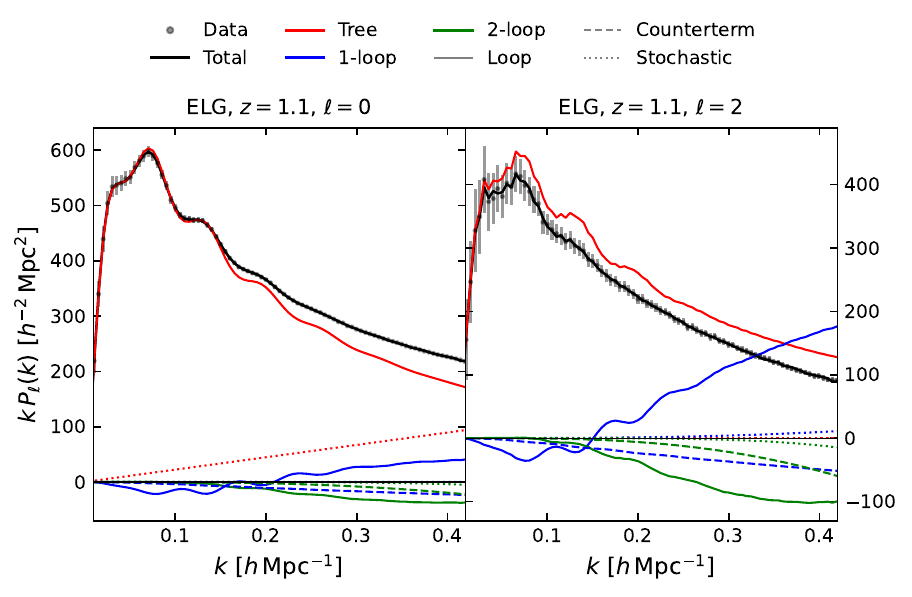}
    \caption{\textbf{Two-loop Breakdown}. Contributions to the LRG (top) and ELG (bottom) two-loop power spectrum, split by perturbative order. We additionally separate into deterministic (\textit{i.e.}\ leading-gradient), counterterm and stochastic pieces, following \S\ref{sec: theory}. For clarity, we omit the hexadecapole moment.. While the monopoles are dominated by the tree-level term (including residual shot-noise contributions), the quadrupoles receives large (and partly canceling) contributions from the deterministic loops, particularly for the LRG sample. This is characteristic of the fingers-of-God effect.}
    \label{fig: gal-breakdown}
\end{figure}

To further explore the differences between the LRG and ELG samples, we plot the best-fit two-loop power spectra for each sample in Fig.\,\ref{fig: gal-breakdown}, splitting the underlying contributions by perturbative order. In both cases, the monopole is dominated by the deterministic and stochastic tree-level pieces,\footnote{The two samples show a preference for super-Poissonian shot-noise, with $a_{1,0}> 0$ at $2.1\sigma$ ($3.0\sigma$) in the combined-volume LRG (ELG) fit.} with the higher-order corrections sourcing $10\%$ of the ELG signal at $k=0.3\hMpc$ and $9\%$ for LRGs (after large cancellations). In contrast, the quadrupole receives large contributions from each of the one- and two-loop pieces, which, despite their partial cancellation, dominate the spectra on small-scales. Here, much of the signal is carried by the deterministic loops (which are sourced by velocities), with rising contributions at large $k$ from the redshift-space stochasticity (generated by uncorrelated velocities). The large alternating contributions of the one- and two-loop spectra are indicative of fingers-of-God effects, which our two-loop model is clearly flexible enough to capture.
  
\begin{figure}
    \centering
    \includegraphics[width=0.95\linewidth]{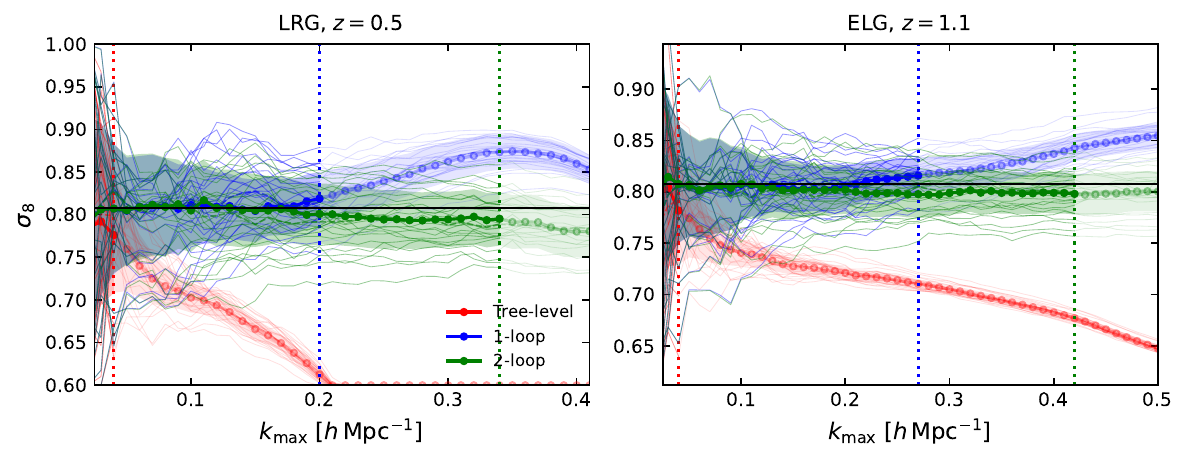}
    \caption{\textbf{Galaxy Inference}. Constraints on $\sigma_8$ from the redshift-space galaxy power spectrum as a function of scale (following Fig.\,\ref{fig: gal-pk}) for both the LRG (left) and ELG (right) samples. While the one-loop theory breaks down beyond $k_{\rm max}= 0.2\hMpc$ (with a somewhat larger reach for ELGs), the two-loop theory returns unbiased constraints on $\sigma_8$ up to $k_{\rm max}=0.42\hMpc$ for ELGs and $k_{\rm max}=0.34\hMpc$ for LRGs. These conclusions hold both for the mean of 25 realizations (points) and each simulation individually (lines), up to slight prior volume shifts ($\lesssim 0.4\sigma$) in the latter. This test indicates that the model is both accurate and applicable to contemporary datasets.}
    \label{fig: gal-sigma8}
\end{figure}

In addition to testing the model at fixed cosmology, it is important to validate whether we can use the models to correctly recover cosmological parameters in the presence of a large number of unknown Wilson coefficients. Constraints on $\sigma_8$ obtained using each model are shown in Fig.\,\ref{fig: gal-sigma8}, both for the individual \textsc{AbacusSummit} boxes and the mean of 25 realizations. We find clear similarities to Fig.\,\ref{fig: dm-sigma8}; the one-loop model obtains unbiased constraints up to intermediate scales before breaking down rapidly, while the two-loop model is accurate, though conservative, up to large $k_{\rm max}$. Numerically, the constraints are far weaker in this scenario, with the tightest galaxy bounds reaching $2\%$ precision, compared to around $0.2\%$ for dark matter; this information loss occurs due to the unknown physics of galaxy formation. Applying the one-loop model to the mean of $25$ simulations with the covariance of a single box, we find $\sigma_8=0.818\pm0.031$ ($0.816\pm0.017$) for LRGs (ELGs) at the $k_{\rm max}$ specified in Tab.\,\ref{tab: kmax}, whereas the two-loop theory gives $0.787\pm0.032$ ($0.796\pm0.020$).\footnote{The corresponding LRG (ELG) results with a $200h^{-3}\mathrm{Gpc}^{3}$ covariance are $0.8188\pm0.0069$ ($0.8155\pm0.0044$) at one-loop and $0.7950\pm0.0083$ ($0.7981\pm0.0052$) at two-loops, all of which lie within $1.8\sigma$ of the input values.} Notably, the two-loop constraints are no tighter than the one-loop results, indicating that the small-scale information primarily acts to calibrate the galaxy formation parameters. 

Lastly, we assess the dependence of our results on the analysis choices. Central to our analysis are power spectrum templates computed using Monte Carlo methods, which can induce biases in the cosmological parameters if insufficiently converged. Here, such biases are are negligible: at the $k_{\rm max}$ values given in Tab.\,\ref{tab: kmax}, sampling noise contributes at most $0.09\sigma$ to the covariance (or $0.16\sigma$ for dark matter) and shifts the best-fit parameters by less than $0.05\sigma$ (with the errors inflating by $<2\%$). Another potential uncertainty arises from our choice of priors (cf.\,\ref{tab: priors}): increasing the fiducial widths of all parameters by a factor of $2$ has minimal effect on the parameter constraints, with the best-fit values and widths shifting by at most $0.15\sigma$ and $13\%$ respectively. To test for potential prior volume shifts, we repeat our analysis on noiseless mock data; here, we recover the input $\sigma_8$ parameter to within $0.25\sigma$ at one-loop, or $0.3\sigma$ at two-loop (or $<0.1\sigma$ if using the full $200h^{-3}\mathrm{Gpc}^3$ covariance). The latter sources the slight deviations seen in Fig.\,\ref{fig: gal-sigma8} between the $\sigma_8$ constraints from individual realizations and those of the mean.

\section{Conclusions}\label{sec: conclusions}
\noindent In this work, we have derived the complete EFT model for the two-loop power spectrum of biased tracers in redshift-space. We include all ingredients required to compare theory and observational data: galaxy bias (at leading, next-to-leading and next-to-next-to-leading gradient order), ultraviolet counterterms, infrared resummation, stochasticity, and coordinate distortions. A notable feature of the model is the appearance of non-local-in-space bias operators in both galaxy density and velocity: despite our assumptions that galaxy formation is a local-in-space process (but non-local-in-time), these are sourced from dark matter renormalization (assuming the stress tensor, $\tau_{ij}$, to be local-in-space), and are required by renormalization (as discovered in \citep{Bakx:2026rmd} for the real-space power spectrum and \citep{DAmico:2022ukl} for the redshift-space bispectrum). The full model depends on $49$ Wilson coefficients (\textit{i.e.}\ biases, counterterms and stochasticity amplitudes) in redshift-space, or $29$ in real-space, with all but seven coefficients appearing also in the one-loop trispectrum. This model has been implemented numerically using a carefully calibrated numerical scheme, resulting in a likelihood that can be evaluated in milliseconds. 

Confronting our models with high-resolution simulations from the \textsc{AbacusSummit} suite, we find that the two-loop theory provides an excellent fit to both dark matter and galaxy power spectra. In both cases, the higher-loop corrections substantially increase the range of scales over which the power spectrum can be modeled, including in the presence of non-linear fingers-of-God effects. That said, the addition of small-scale information does not lead to great advances in parameter precision: we find similar constraints on $\sigma_8$ from the one-loop theory at $k_{\rm max}\approx 0.2\hMpc$ as we do from the two-loop result at $k_{\rm max}\approx 0.3$--$0.5\hMpc$. This is somewhat unsurprising: on small scales, variations of $\sigma_8$ primarily change the broadband shape of the power spectrum, which is degenerate with a variety of non-linear phenomena and thus the EFT coefficients. It is plausible that the two-loop power spectrum will yield tighter constraints on other parameters (as seen for one-loop bispectra \citep{Bakx:2025pop}) or when combined with higher-order statistics: this remains to be investigated.

Our analysis can be extended in a variety of ways. These include:
\begin{itemize}
    \item \textbf{Cosmological parameters}: In this work, we have varied only $\sigma_8$ and the Wilson coefficients. To fully assess the information content of the two-loop model, we must compute the predictions as a function of all relevant parameters (adding, for example, $H_0$, $\Omega_m$, and $n_s$) and validate the full cosmology fits using simulations. This can be efficiently realized by expanding the linear power spectra in a suitable basis (e.g., \textsc{cobra} \citep{Bakx:2024zgu,Bakx:2025jwa}), and adopting scaling symmetries for $H_0$ and the amplitude parameters \citep[e.g.,][]{Bakx:2025pop}. At the precision required for current datasets, this will require $\mathcal{O}(10\times)$ more loop integral templates, which remains computationally feasible.
    \item \textbf{Application to data}: Building on previous analyses of the one-loop power spectrum \citep[e.g.,][]{Chudaykin:2025aux,DESI:2024jxi}, we can apply the two-loop model to data. Though this is somewhat more expensive (since 12 nuisance parameters must be explicitly varied rather than 3), we expect it be a straightforward extension of current pipelines. This will allow a greater range of scales to be included in the analysis and reduce theoretical systematics.
    \item \textbf{Higher-order statistics}: We expect increased gains on cosmological parameter precision when combining the two-loop power spectrum with the one-loop bispectrum and/or trispectrum, which rest on the same deterministic model. Combination with the three-point function is straightforward (up to a redefinition of the bias parameters) and builds on previous one-loop analyses \citep{Bakx:2025pop,DAmico:2022osl,Ivanov:2026dvl}; far more work will be required to add with the four-point function, for which the first EFT analyses have only just been completed \citep{Tucci:2026wzg}.
    \item \textbf{New physics}: A wealth of non-$\Lambda$CDM physics can be probed in the power spectrum, particularly as one extends to smaller scales. It will be exciting to assess the utility of the two-loop model for constraining new dark matter physics, primordial non-Gaussianity, dark energy models, modified gravity and beyond.
\end{itemize}
We look forward to exploring these extensions, and many others, in future work. 

\acknowledgements
\vskip 8pt
{\small
\begingroup
\hypersetup{hidelinks}
\noindent 
OHEP thanks Samuel Goldstein, Mikhail Ivanov \& Beatriz Tucci for insightful discussions, as well as the \href{https://www.flickr.com/photos/198816819@N07/55577968911}{cariocas of Sugarlloaf mountain} for emotional support. 
OHEP is supported in part by the National Science Foundation under grant number PHY-2609396. Some of the computing for this project was performed on the Sherlock cluster. Additional computations were run at facilities supported by the Scientific Computing Core at the Flatiron Institute, a division of the Simons Foundation. We would like to thank Stanford University, Stanford Research Computing and the Simons Foundation for providing computational resources and support that contributed to these research results.  

\noindent \textit{AI usage statement}: This manuscript was entirely conceptualized and written by a human. Claude (Opus 4.8-5.5 and Fable 5-5.1) was used to perform numerical and analytic computations, interfacing with \textsc{python} and \textsc{mathematica}. The entire project involved around 50 Claude Code sessions with more than 8000 user prompts, with an equivalent API cost of around $\$20\,000$. The author takes full responsibility for all results presented herein.
\endgroup
}
\appendix

\section{Non-Locality in Time}\label{app: time-nonlocality}
\noindent Below, we explicitly demonstrate that the $\Pim{m}_{ij}$ basis from \citep{Mirbabayi:2014zca} incorporates non-locality in time induced by galaxy formation effects \citep[cf.][]{Donath:2023sav,Edison:2025bmj}. From \citep{Senatore:2014eva}, the contribution of a generic operator $\mathcal{O}$ to the galaxy overdensity at some time $t$ can be written as an integral over fluid trajectories:
\beq
    \delta_g^{\mathcal{O}}(\vx,t) = \int^t dt'K_{\mathcal{O}}(t,t')\mathcal{O}(\vx_{\rm fl}(\vx,t,t'),t'), \qquad \vx_{\rm fl}(\vx,t,t') = \vx - \int_{t'}^{t}\frac{dt''}{a(t'')}\vec v(\vx_{\rm fl}(\vx,t,t''),t'')
\eeq
where $K_{\mathcal{O}}$ kernel encodes the formation history and $\vx_{\rm fl}(\vx,t,t')$ describes the position of the fluid element at time $t'$. Noting that
\beq
    \frac{d}{dt'}\mathcal{O}(\vx_{\rm fl}(\vx,t,t'),t') = \left(\partial_{t'} + \frac{\vec v(t')}{a(t')}\cdot\nabla\right)\mathcal{O}(\vx_{\rm fl}(\vx,t,t'),t') = H(t')f(t')D_{\eta'}\mathcal{O}(\vx_{\rm fl}(\vx,t,t'),t')
\eeq
for convective time derivative $D_{\eta'}$ (with $\eta'=\log D(t')$), we can Taylor expand $\mathcal{O}$ about the final time $\eta'=\eta$ to find:
\beq\label{eq: time-nonlocal}
    \delta_g^{\mathcal{O}}(\vx, t) = \int^t dt'K_{\mathcal{O}}(t,t')\left[\sum_{N=0}^{\infty}\frac{(\eta'-\eta)^N}{N!}D_\eta^N\right]\mathcal{O}(\vx,t) = \int^t dt'K_{\mathcal{O}}(t,t')\exp\left[\log\frac{D(t')}{D(t)}D_\eta\right]\mathcal{O}(\vx,t).
\eeq
noting that $d\eta/dt = H(t)f(t)$. 

Next, we specialize to an operator of the form $\mathcal{O} = \Tr\Pim{m}$. From \eqref{eq: Pi-ij}, $D_\eta\Pim{m}_{ij} = m\,\Pim{m}_{ij}+m!\,\Pim{m+1}_{ij}$ thus
\beq\label{eq: nonlocal-Pim}
    \delta_g^{\Tr\Pim{m}}(\vx,t) = \sum_{k=0}^{\infty}\mathcal{C}_{m,k}(t)\Tr\Pim{m+k}(\vx,t), \quad \mathcal{C}_{m,k}(t) = \frac{1}{k!}\prod_{j=0}^{k-1}(m+j)!\int^t dt'K_{\Tr\Pim{m}}(t,t') \left[\frac{D(t')}{D(t)}\right]^m\left[\frac{D(t')}{D(t)}-1\right]^{k}
\eeq
with $\mathcal{C}_{m,0}(t) = \int^t dt'\,K_{\Tr\Pim{m}}(t,t') \left[D(t')/D(t)\right]^m$. This factorizes into evolution-dependent coefficients, $\mathcal{C}_{m,k}(t)$, and higher-order local-in-time operators $\Tr\Pim{m'}(\vx,t)$ with $m'\geq m$, implying that the $\{\Tr\Pim{m'}: m'\geq 1\}$ basis is closed under time evolution. Using the product rule, it is straightforward to show that this argument also extends to composite operators (e.g., $\Tr\Pim{2}\Pim{1}$), including those with multiple time dependencies. This confirms that the $\Pim{n}$ basis captures the effects of time non-locality, practically allowing us to omit the formation time integrals.\footnote{This is also evidenced by the fact that the $\Pim{m}_{ij}$ basis has 29 independent terms at fifth order, which matches the explicitly non-local-in-time basis of \citep{Donath:2023sav} (with three more operators than the local-in-time basis).}

A similar conclusion holds for operators formed from arbitrary products of $\Pim{m}$ and $\partial_k$. To demonstrate this, we start from the commutator
\beq
    \left[D_\eta,\partial_i\right]\mathcal{O} = -\frac{1}{aHf}(\partial_iv^j)(\partial_j \mathcal{O})
\eeq
for arbitrary $\mathcal{O}$, which can be derived using the definition of the convective time derivative. Applied to $\Pim{m}_{ij}$, this gives
\beq\label{eq: D-eta-delPi}
    D_\eta\left(\partial_k\Pim{m}_{ij}\right) = m\,\partial_k\Pim{m}_{ij} + m!\,\partial_k\Pim{m+1}_{ij}-\frac{1}{aHf}\partial_kv^l \partial_l \Pim{m}_{ij}.
\eeq
As demonstrated in Appendix A of \citep{Mirbabayi:2014zca}, $\partial_k v_l$ can be written in terms of convective derivatives of the tidal tensor (via the Euler equation, up to EFT corrections), thus the right-hand side can be written in terms of products of $\Pim{m'}$ tensors and a single spatial derivative. Noting that the extension to higher-gradients holds by induction, we find that the $\Pim{m}$-plus-gradients basis is closed under $D_\eta$,\footnote{This is straightforward to verify numerically by computing $D_\eta \mathcal{O}$ for each gradient operator $\mathcal{O}$, and checking that it can be expressed in terms of the operators of Tab.\,\ref{tab: operators} (up to higher-order corrections).} facilitating factorization of the time-integrals as in \eqref{eq: nonlocal-Pim}. 

Notably, the above conclusions do not hold if one extends the basis to include non-local operators involving $(\partial^{kl}/\nabla^2)\Pim{m}_{kl}$ (which arise at fifth order through the counterterm sector, cf.\,\S\ref{sec: renorm-real}). In this case, the time evolution induces new higher-order terms such as $(\partial^{kl}/\nabla^2)\Pim{m+1}_{kl}$ and $\widehat{\Pi}^{[m+1]}$ (cf.\,\eqref{eq: bakx-nonlocal-ops}). Since the new contributions arise only at sixth-order and beyond, we may ignore the effect in this work.

\section{Comparison with CLASS-PT}\label{app: class-pt}
\newcommand\cpt{\textsc{class-pt}\xspace}
\noindent Here, we validate our numerical pipeline against the one-loop power spectrum predictions of the established \textsc{class-pt} code \citep{Chudaykin:2020aoj}. All comparisons are performed at the \textsc{AbacusSummit} fiducial cosmology (matching \S\ref{sec: simulations}, without Alcock--Paczy\'{n}ski rescalings), using enhanced \textsc{class} precision settings: \texttt{k\_min\_tau0 = 0.002}, \texttt{k\_per\_decade\_for\_pk = 30}, \texttt{k\_per\_decade\_for\_bao = 80}. The \cpt loop integrals are computed using $512$ \textsc{FFTLog} frequencies \citep{Simonovic:2017mhp}, reducing to $256$ to assess numerical convergence. We fix the Wilson coefficients to the best-fit values from the $z=0.5$ \textsc{AbacusSummit} redshift-space galaxy clustering analysis of \S\ref{sec: simulations} at $k_{\rm max}=0.15\hMpc$, and use the same \textsc{vegas} precision settings as in \S\ref{sec: simulations}.

\begin{figure}
    \centering
    \includegraphics[width=0.95\linewidth]{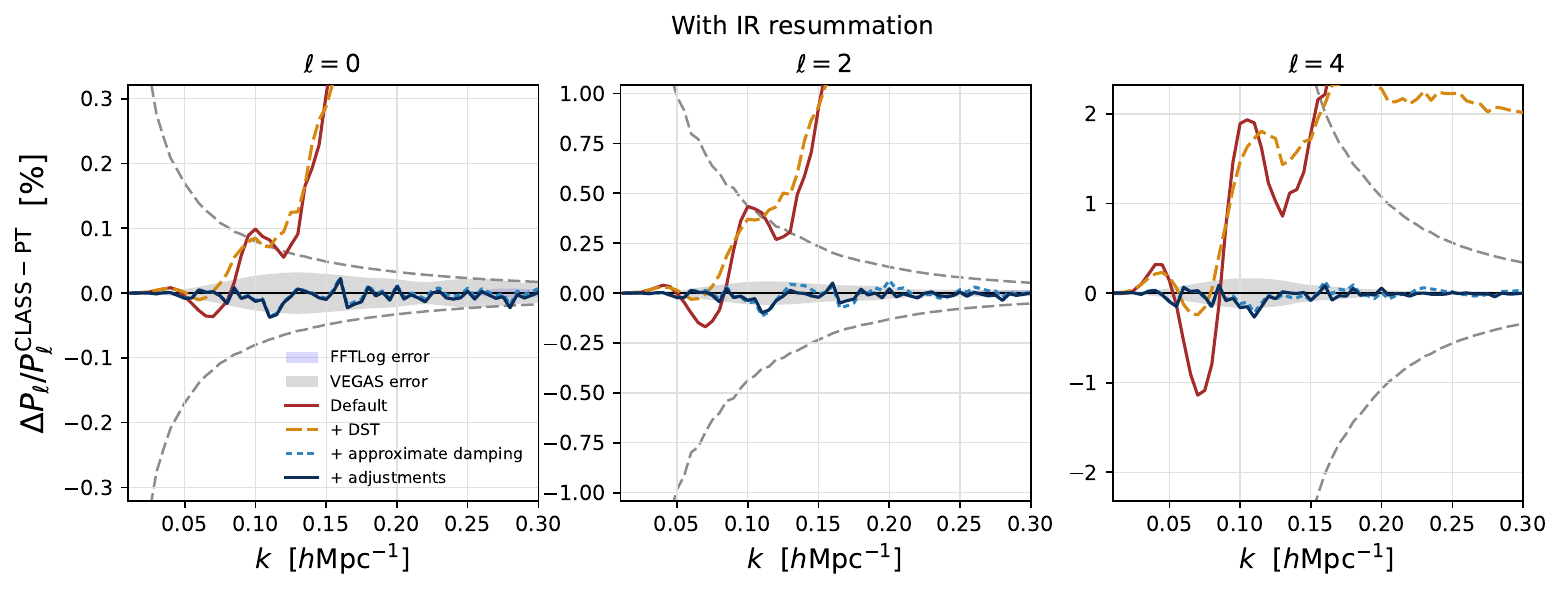}
    \includegraphics[width=0.95\linewidth]{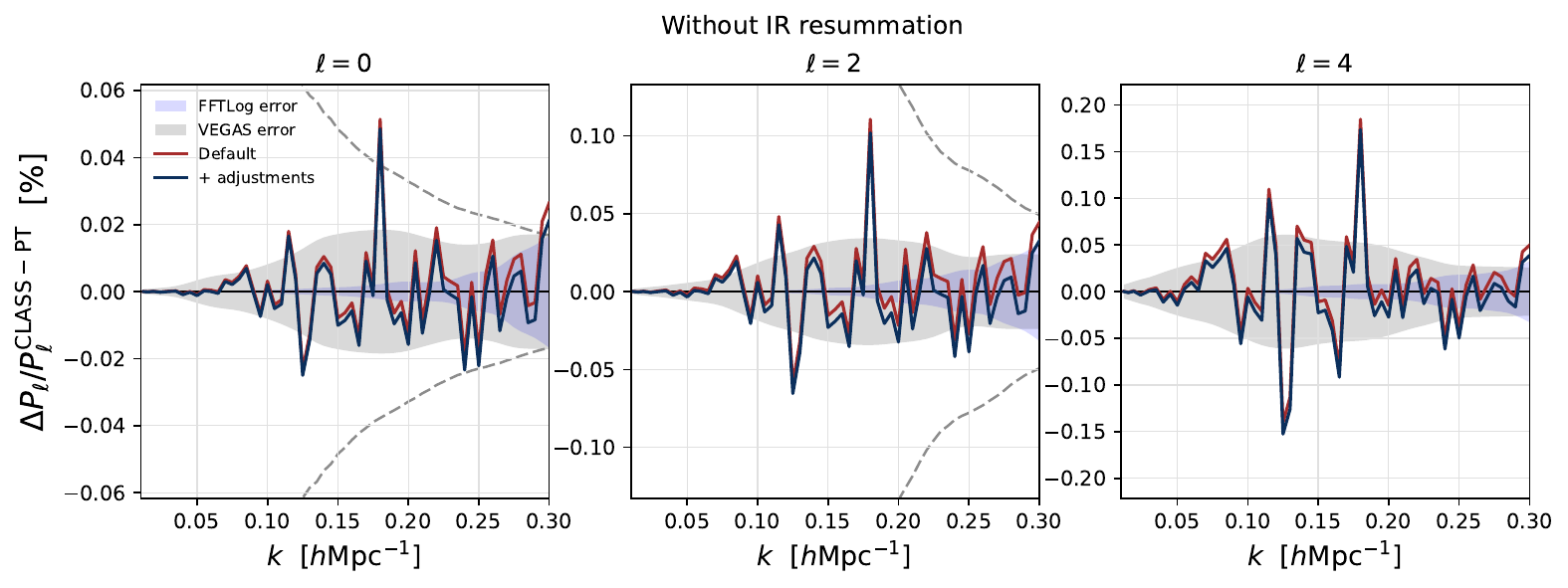}
    \caption{Comparison of the one-loop galaxy power spectra computed with \textsc{class-pt} and the numerical approach of this work. We plot the fractional error relative to \cpt for the first three multipoles (left-to-right), optionally excluding IR resummation (bottom panel). The raw multipoles agree to within $4\%$ up to $k=0.3\hMpc$; this tightens to $0.2\%$ when IR damping is removed. Improved agreement between spectra is obtained by using the same no-wiggle split for both methods (orange), matching the IR damping approximations in the loop integrals (light blue), and correcting for three further \cpt subtleties discussed in the text (dark blue). The resulting deviations are fully consistent with the Monte Carlo errors (gray) and FFTLog accuracy (blue). The dashed lines indicate the $1\sigma$ error on the power spectrum from the combined $200h^{-3}\mathrm{Gpc}^{3}$ simulation volume analyzed in \S\ref{sec: simulations}.}
    \label{fig: class-pt-comparison}
\end{figure}

There are several differences between the \cpt methodology and that of this work. First, the two approaches adopt different bias parameter conventions. These are related by
\beq\label{eq: class-pt-basis-redef}
   &&\begin{pmatrix} b_1 \\[2pt] b_2 \\[2pt] b_{\mathcal{G}_2} \\[2pt] b_{\Gamma_3} \end{pmatrix}
  = \begin{pmatrix} 1 & 0 & 0 & 0 \\[2pt] 0 & 2 & 2 & 0 \\[2pt] 0 & 1 & 0 & 0 \\[2pt] 0 & 0 & 0 & \tfrac54 \end{pmatrix}
  \begin{pmatrix} c_{1,0} \\[2pt] c_{2,0} \\[2pt] c_{3,0} \\[2pt] c_{4,0} \end{pmatrix}\,,
  \qquad
  \begin{pmatrix} P_{\rm shot} \\[2pt] a_0 \\[2pt] a_2 \end{pmatrix}
  = \begin{pmatrix} \bar{n} & 0 & 0 \\[2pt] 0 & k_{\rm NL}^{2}\bar{n} & 0 \\[2pt] 0 & 0 & k_{\rm NL}^{2}\bar{n} \end{pmatrix}
  \begin{pmatrix} a_{1,0} \\[2pt] a_{1,1} \\[2pt] a_{2,1} \end{pmatrix}\\\nonumber
  &&\begin{pmatrix} c_{s,0} \\[3pt] c_{s,2} \\[3pt] c_{s,4} \end{pmatrix}
  =
  \begin{pmatrix}
    \frac{3 c_{1,0} + f}{3} & -\frac{f \left(5 c_{1,0} + 3 f\right)}{15} & \frac{f^{2} \left(7 c_{1,0} + 5 f\right)}{70} \\[3pt]
    1 & -\frac{7 c_{1,0} + 6 f}{7} & \frac{f \left(6 c_{1,0} + 5 f\right)}{14} \\[3pt]
    0 & -1 & \frac{11 c_{1,0} + 15 f}{22}
  \end{pmatrix}
  \begin{pmatrix} c_{1,1} \\[3pt] c^{(1)}_{1,1} - \tfrac{f}{2}\,c^{(2)}_{2,1} \\[3pt] c^{(2)}_{3,1} + f\,c^{(2)}_{1,1}\end{pmatrix},
\eeq
defining $k_{\rm NL}=0.45\hMpc$ (in the \cpt convention). We do not include the $\tilde{c}$ bias (sometimes labeled as $b_4$), since this arises at higher-order in perturbation theory. Next, our implementation defines the no-wiggle power spectrum, $P_{\rm nw}$, by smoothing the ratio of the linear power spectrum to the Eisenstein-Hu no-wiggle model \citep[e.g.,][]{Vlah:2015zda} (see \S\ref{subsec: impl-ir-res}), in contrast to the discrete sine transform (DST) approach used in \cpt \citep{Chudaykin:2020aoj}. This slightly changes the amplitude of $P_{\rm nw}$, leading to percent-level variations in the output power spectra. Furthermore, we implement a fully-anisotropic IR-resummation scheme (\S\ref{sec: ir-resum}), whereas \cpt makes a number of approximations: replacing the anisotropic damping exponent, $\Sigma^2_{\rm IR}(\mu)$, with an isotropic limit inside the loop integrals (keeping the full form outside); dropping higher powers of $P_{\rm w}$; omitting the damping term in the amplitude, $\sigma_v^2$, of the UV contributions to `13'-type power spectra. While the error of the approximate resummation schemes is typically absorbed by the power spectrum counterterms \citep{Chudaykin:2020aoj,Ivanov:2018gjr}, it leads to significant deviations in the power spectra at fixed parameters.

The two sets of power spectrum multipoles are compared in Fig.\,\ref{fig: class-pt-comparison}. We find fair agreement between methods, with deviations below $1\%$ for the monopole and $4\%$ for the quadrupole and hexadecapole up to $k=0.3\hMpc$. While promising, the remaining differences are large compared to the expected error of the \textsc{FFTLog} method and the \textsc{vegas} integration noise ($\lesssim 0.05\%$). Much of this lies in the IR-resummation scheme: switching the no-wiggle decomposition from the Eisenstein-Hu approach to the DST algorithm reduces the oscillatory part of the error, and adopting the damping approximations discussed above reduces the deviations to $\lesssim 0.3\%$. Indeed, if one turns off IR-resummation (bottom panel of Fig.\,\ref{fig: class-pt-comparison}), the two spectra agree to within $0.2\%$.

We identify a number of additional subtle differences between the codes:
\begin{itemize}
    \item The mapping between our counterterms and those of \cpt \eqref{eq: class-pt-basis-redef} does not account for the anisotropic IR-damping exponent, $\Sigma^2_{\rm IR}(\mu)$.
    \item Due to the finite $k$-range used in the \textsc{FFTLog} algorithm, the \cpt results suffer slightly from aliasing. This arises primarily in the `13'-type integrals involving the $\mathcal{G}_2$ and $\Gamma_3$ operators (whose UV contributions are captured by \textsc{FFTLog} rather than added by hand), and leads to a slight enhancement due to a larger value of $\sigma_v^2$. 
    \item Aliasing also impacts the calculation of the tadpole $k\to 0$ term (for $\ell=0$ only), sourced by the $\delta^2$ `22'-type integral. Moreover, \cpt calculates the tadpole at $k=k_{\rm min}$, whereas our code computes it at $k=0$.
\end{itemize}
Each of these deviations (which are percent-level for some choices of bias parameters) can be calculated analytically and removed from the power spectrum. The dark blue lines in Fig.\,\ref{fig: class-pt-comparison} show the corrected result: across all scales and multipoles, the residual differences are consistent with the Monte Carlo noise. 
The conclusion of this exercise is that our one-loop power spectrum (which uses exactly the same framework as the two-loop spectrum) provides highly accurate power spectrum predictions with an accuracy meeting (or exceeding) that of \cpt. 

\section{Dark Matter Renormalization}\label{app: dm-renorm}
\noindent In this appendix, we derive the renormalization of the dark matter overdensity, $\delta\equiv \rho/\bar\rho-1$, and velocity, $v^i\equiv \pi^i/\rho$ (for momentum $\pi^i$) from the stress-energy tensor, $\tau^{ij}$. This extends a number of previous works \citep{DAmico:2022ukl,Anastasiou:2025jsy,Raikman:2026gcx} by (a) working at third-order in $\delta_{\rm L}$ (as required for the two-loop power spectrum and one-loop trispectrum), (b) allowing for arbitrary time dependence in the Wilson coefficients, and (c) including both density and velocity contributions. We start from the equations of motion for density and momentum, written in terms of $\eta \equiv \log D$:
\beq
    &&-\frac{\pi_S}{f\mathcal{H}\bar\rho}=\partial_\eta\delta\\\nonumber    
    &&\left[\partial_\eta+\left(\frac{3}{2}\frac{\Omega_m}{f^2}-1\right)\right]\frac{\pi_V^i}{\mathcal{H}f\bar\rho} = -\epsilon_{ijk}\partial^j\partial_l\left[S^{kl}+\frac{\tau^{kl}}{f^2\mathcal{H}^2\bar\rho}\right]\\\nonumber
    &&\left[\partial_\eta^2+\left(\frac{3}{2}\frac{\Omega_m}{f^2}-1\right)\partial_\eta -\frac{3}{2}\frac{\Omega_m}{f^2}\right]\delta = \partial_{ij}\left[S^{ij}+\frac{\tau^{ij}}{f^2\mathcal{H}^2\bar\rho}\right]
\eeq
for $f \equiv d\log D/d\log a$ and $\mathcal{H} \equiv aH$ \citep[e.g.,][]{2002PhR...367....1B}. Here we have decomposed the momentum into scalar and vector components via 
\beq\label{eq: pi-SV}
    \pi_i = \frac{\partial_i}{\nabla^2}\pi_S - \epsilon_{ijk}\frac{\partial^j}{\nabla^2}\pi_V^k,
\eeq
and defined the source term
\beq\label{eq: source}
	S^{ij} = \frac{3}{2}\frac{\Omega_m}{f^2}\left(\frac{\partial^i}{\nabla^2}\delta\frac{\partial^j}{\nabla^2}\delta-\frac{1}{2}\delta_{\rm K}^{ij}\frac{\partial_k}{\nabla^2}\delta\frac{\partial^k}{\nabla^2}\delta\right)+\frac{1}{f^2\mathcal{H}^2\bar\rho^2}\frac{\pi^i\pi^j}{1+\delta}
\eeq
\citep{Anastasiou:2025jsy}. In the Einstein-de-Sitter limit ($\Omega_m/f^2\to 1$), these have the formal solutions:
\beq\label{eq: green-soln}
    \delta(\vx,\eta) &=& \delta_{\rm L}(\vx,\eta)+\frac{2}{5}\int^{\eta} d\eta'\left(e^{\eta-\eta'}-e^{-3(\eta-\eta')/2}\right)\partial_{ij}\left[S^{ij}+\frac{1}{f^2\mathcal{H}^2\bar\rho}\tau^{ij}\right](\vx,\eta')\\\nonumber
    -\frac{\pi_S}{f\mathcal{H}\bar\rho}(\vx,\eta) &=& \delta_{\rm L}(\vx,\eta)+\frac{2}{5}\int^{\eta} d\eta'\left(e^{\eta-\eta'}+\frac{3}{2}e^{-3(\eta-\eta')/2}\right)\partial_{ij}\left[S^{ij}+\frac{1}{f^2\mathcal{H}^2\bar\rho}\tau^{ij}\right](\vx,\eta')\\\nonumber
    -\frac{\pi_V^i}{\mathcal{H}f\bar\rho}(\vx,\eta) &=& \int^{\eta}d\eta'e^{-(\eta-\eta')/2}\epsilon^{ijk}\partial_j\partial^l\left[S_{kl}+\frac{1}{f^2\mathcal{H}^2\bar\rho}\tau_{kl}\right](\vx,\eta'),
\eeq
dropping the decaying mode and assuming vanishing primordial vorticity. Setting $\tau^{ij}=0$ and expressing the source function in powers of $\delta_{\rm L}$, we find the familiar SPT solutions for density and velocity.

\subsection{Generation of Counterterms}
\noindent Here, we are interested in the non-ideal terms sourced by the stress tensor. In general, $\tau^{ij}$ can be written as a local-in-space function of the gravitational potential, $\Phi$, and velocity shear, $\partial_iv_j$, possibly integrated over the fluid trajectory \citep{Baumann:2010tm,Carrasco:2012cv,Carroll:2013oxa,Carrasco:2013mua,Senatore:2014eva,Baldauf:2014qfa,DAmico:2022ukl,Anastasiou:2025jsy}; as discussed in Appendix \ref{app: time-nonlocality}, this implies that it can be fully expressed in the $\Pim{n}_{ij}$ basis (up to stochastic contributions, which we omit in this discussion). Up to fifth-order in fields and derivatives, the deterministic part of $\tau^{ij}$ has the general expansion
\beq\label{eq: tauij-general}
    \frac{1}{f^2\mathcal{H}^2\bar\rho}\tau_{ij}^{\rm det}(\vx,\eta) &=& D^{2}(\eta)\sum_{a=1}^{18}t_{a,1}(\eta)\mathcal{O}^{ij}_{a,1}(\vx,\eta)+D^{4}(\eta)\sum_{a=1}^{2}t_{a,2}(\eta)\mathcal{O}^{ij}_{a,2}(\vx,\eta)\\\nonumber
    &=& D^2(\eta)\left[t_{1,1}\delta^{\rm K}_{ij}+t_{2,1}\delta^{\rm K}_{ij}\delta +t_{3,1}\Pim{1}_{ij}+\cdots\right]+D^4(\eta)\left[t_{1,2}\nabla^2\Pim{1}_{ij}+t_{2,2}\delta^{\rm K}_{ij}\nabla^2\delta+\cdots\right],
\eeq
where $\mathcal{O}^{ij}_{a,b}$ are the rank-two operators listed in Tab.\,\ref{tab: rsd-table} (which provide a complete basis for the local-in-space tensor, given the assumed symmetries) and $t_{a,b}(\eta)$ are unknown coefficients, encoding the fluid's microphysical properties such as sound-speed and viscosity.\footnote{Dropping the order-zero term $\mathcal{O}_{1,1}^{ij}=\delta^{\rm K}_{ij}$, our $b=1$ expansion matches that of \citep{Anastasiou:2025jsy}, though we include the additional $b=2$ term $\mathcal{O}_{1,2}^{ij}=\nabla^2\Pim{1}_{ij}$. The coupling parameters of this work are related to those of \citep{Anastasiou:2025jsy} by 
\beq\label{eq: anastasiou-dict}
    t_{2,1}&=&c_{\delta,1}, \quad 
    t_{3,1}=c_{r,1}, \quad 
    t_{4,1}=-c_{r,1}+c_{r,2}, \quad 
    t_{5,1}=c_{r\delta,1}, \quad 
    t_{6,1}=c_{r^2,1}, \quad 
    t_{7,1}=-\tfrac{2}{7}c_{\delta,1}+\tfrac{2}{7}c_{\delta,2}\\\nonumber
    t_{8,1}&=&-\tfrac{5}{7}c_{\delta,1}+\tfrac{5}{7}c_{\delta,2}+c_{\delta^2,1}, \quad 
    t_{9,1}=-\tfrac{101}{30}c_{r,1}+\tfrac{101}{15}c_{r,2}-\tfrac{101}{30}c_{r,3}-2c_{r^2,1}+2c_{r^2,2}, \quad 
    t_{10,1}=\tfrac{7}{3}c_{r,1}-\tfrac{14}{3}c_{r,2}+\tfrac{7}{3}c_{r,3}-c_{r\delta,1}+c_{r\delta,2}\\\nonumber
    t_{11,1}&=&\tfrac{7}{6}c_{r,1}-\tfrac{7}{3}c_{r,2}+\tfrac{7}{6}c_{r,3}+\tfrac{46}{105}c_{\delta,1}-\tfrac{4}{7}c_{\delta,2}+\tfrac{2}{15}c_{\delta,3}, \quad 
    t_{12,1}=c_{r,1}-2c_{r,2}+c_{r,3}, \quad 
    t_{13,1}=c_{r^3,1}+c_{r^2\delta,1}\\\nonumber
    t_{14,1}&=&\tfrac{254}{441}c_{\delta,1}-\tfrac{50}{49}c_{\delta,2}+\tfrac{4}{9}c_{\delta,3}-\tfrac{10}{7}c_{\delta^2,1}+\tfrac{10}{7}c_{\delta^2,2}+\tfrac{1}{6}c_{r^3,1}+c_{\delta^3,1}, \quad 
    t_{15,1}=\tfrac{11}{147}c_{\delta,1}-\tfrac{20}{49}c_{\delta,2}+\tfrac{1}{3}c_{\delta,3}-\tfrac{4}{7}c_{\delta^2,1}+\tfrac{4}{7}c_{\delta^2,2}-\tfrac{1}{2}c_{r^3,1}\\\nonumber
    t_{16,1}&=&-\tfrac{4}{45}c_{\delta,1}+\tfrac{4}{45}c_{\delta,3}+\tfrac{1}{3}c_{r^3,1}, \quad 
    t_{17,1}=-\tfrac{2}{7}c_{r\delta,1}+\tfrac{2}{7}c_{r\delta,2}+\tfrac{1}{2}c_{r^3,1}+c_{(r^2)r,1}, \quad 
    t_{18,1}=-\tfrac{5}{7}c_{r\delta,1}+\tfrac{5}{7}c_{r\delta,2}-\tfrac{1}{2}c_{r^3,1}+c_{r\delta^2,1}\\\nonumber
    t_{1,2}&=&0, \quad 
    t_{2,2}=c_{\partial^2\delta,1}.
\eeq
} Here, $b\geq 1$ labels the gradient order of the eventual density counterterms (\textit{i.e.}\ $b=1$ and $b=2$ give NLG and NNLG corrections to $\delta$). If $t_{a,b}(\eta)$ are independent of $\eta$, \eqref{eq: tauij-general} has the correct time dependence to capture the cut-off dependence of the loop integrals; here, we will assume general time dependence, which allows for both finite and divergent contributions to the counterterms.

The stress tensor sources corrections to $\delta$ and $v^i$ both directly, due to the $\tau^{ij}$ term on the right-hand-side of \eqref{eq: green-soln}, and through backreaction, sourced by modifications to $S^{ij}$ and $\tau^{ij}$ from lower-order counterterms \citep{Baldauf:2014qfa,DAmico:2022ukl}. The former terms can be computed from \eqref{eq: tauij-general}:
\newcommand{\Gd}[1]{\mathcal{G}^{(#1)}_{\delta}}
\newcommand{\Gs}[1]{\mathcal{G}^{(#1)}_{S}}
\newcommand{\Gv}[1]{\mathcal{G}^{(#1)}_{V}}
\beq\label{eq: direct-ct}
    \left.\delta^{(m)}_{{\rm ct}, ab}(\vx,\eta)\right|_{\rm direct} &=& \frac{2D^{2b}(\eta)}{(2m+4b+3)(m+2b-1)}\Gd{m+2b}[t_{a,b}](\eta)\left[\partial_{ij}\mathcal{O}^{ij}_{a,b}(\vx,\eta)\right]_{\mathcal{O}(\delta_{\rm L}^m)}\\\nonumber
    \left.-\frac{\pi_{S,{\rm ct},ab}}{f\mathcal{H}\bar\rho}(\vx,\eta)\right|_{\rm direct} &=& \frac{2(m+2b)D^{2b}(\eta)}{(2m+4b+3)(m+2b-1)}\Gs{m+2b}[t_{a,b}](\eta)\left[\partial_{ij}\mathcal{O}^{ij}_{a,b}(\vx,\eta)\right]_{\mathcal{O}(\delta_{\rm L}^m)}\\\nonumber
    \left.-\frac{\pi_{V, {\rm ct}, ab}^i}{\mathcal{H}f\bar\rho}(\vx,\eta)\right|_{\rm direct} &=& \frac{2D^{2b}(\eta)}{2m+4b+1}\Gv{m+2b}[t_{a,b}](\eta)\left[\epsilon^{i}_{\,\,jk}\partial^j\partial_l \mathcal{O}_{a,b}^{kl}(\vx,\eta)\right]_{\mathcal{O}(\delta_{\rm L}^m)},
\eeq
at $m$-th order in the linear density field, defining the Green's function integrals
\beq\label{eq: Gdefs}
    \Gd{p}[s](\eta) &\equiv& \frac{(2p+3)(p-1)}{5}\int^{\eta}d\eta'
      \left(e^{\eta-\eta'}-e^{-3(\eta-\eta')/2}\right)s(\eta')
      \left(\frac{D(\eta')}{D(\eta)}\right)^{p},\\\nonumber
    \Gs{p}[s](\eta) &\equiv& \frac{(2p+3)(p-1)}{5p}\int^{\eta}d\eta'
      \left(e^{\eta-\eta'}+\frac32 e^{-3(\eta-\eta')/2}\right)s(\eta')
      \left(\frac{D(\eta')}{D(\eta)}\right)^{p},\\\nonumber
    \Gv{p}[s](\eta) &\equiv& \frac{2p+1}{2}\int^{\eta}d\eta'\,
      e^{-(\eta-\eta')/2}\,s(\eta')\left(\frac{D(\eta')}{D(\eta)}\right)^{p},
\eeq
such that $\mathcal{G}[t] = t$ if $t$ is time-independent. At first-order in fields, we find the familiar NLG counterterm, proportional to $\nabla^2\delta$:
\beq\label{eq: ct1}
    \delta^{(1)}_{\rm ct, NLG}(\vx,\eta) &=&\frac{1}{9}D^2(\eta)
      \Gd{3}[t_{2,1}+t_{3,1}](\eta)\nabla^2\delta^{(1)}(\vx,\eta)\\\nonumber
    -\frac{\pi_{S,\rm ct, NLG}^{(1)}}{f\mathcal{H}\bar\rho}(\vx,\eta)
      &=&\frac{1}{3}D^2(\eta)\Gs{3}[t_{2,1}+t_{3,1}](\eta)\nabla^2\delta^{(1)}(\vx,\eta)\\\nonumber
    \Rightarrow\quad -\frac{v^{i,(1)}_{\rm ct, NLG}}{f\mathcal{H}}(\vx,\eta)
      &=&\frac{1}{3}D^2(\eta)\Gs{3}[t_{2,1}+t_{3,1}](\eta)\partial^i\delta^{(1)}(\vx,\eta),
\eeq
with $\pi_{V,\rm ct, NLG}^{i,(1)}=0$. Note that the ratio of the density and velocity counterterms is fixed to $1/3$ if $t_{a,b}$ are time-independent.

To compute the backreaction terms, we proceed order-by-order, substituting $\delta_{\rm ct}^{(1)}$ and $\pi_{\rm ct}^{(1)}$ into $S^{ij}$ and $\tau^{ij}$ to compute $\delta_{\rm ct}^{(2)}$ \textit{et cetera}. For example, the leading counterterm contribution to $S^{ij}$ is given by
\beq\label{eq: Sct1}
    S^{ij,(2)}_{\rm back}&=&\frac32\frac{\Omega_m}{f^2}\left[\Psi^i_{(1)}\Psi^j_{\rm ct,(1)}
      +(i\!\leftrightarrow\!j)-\dK_{ij}\Psi^k_{(1)}\Psi^k_{\rm ct,(1)}\right]
      +\frac{1}{f^2\mathcal{H}^2\bar\rho}
      \frac{\pi^i_{(1)}\pi^j_{\rm ct,(1)}+(i\!\leftrightarrow\!j)}{\bar\rho}\\\nonumber
    &=&\frac16 D^2(\eta)\,\Gd{3}[t_{2,1}+t_{3,1}]
      \left(\Psi^i_{(1)}\partial^j\delta^{(1)}+\partial^i\delta^{(1)}\Psi^j_{(1)}-\dK_{ij}\Psi^k_{(1)}\partial_k\delta^{(1)}\right)\\\nonumber
      &&\,+\,\frac13 D^2(\eta)\,\Gs{3}[t_{2,1}+t_{3,1}]\left(\Psi^i_{(1)}\partial^j\delta^{(1)}+\partial^i\delta^{(1)}\Psi^j_{(1)}\right)
\eeq
for $\Psi^i\equiv (\partial^i/\nabla^2)\delta$. Note that the density and velocity pieces have different dependence on $\eta$ if $t_{a,b}$ are not constant. In combination with the direct contributions, we find the following second-order counterterms for density and momentum:
\beq\label{eq: ct2}
    \delta^{(2)}_{\rm ct, NLG}(\vx,\eta) &=& \frac{2}{33}D^2(\eta)\Big(
      \Gd{4}\big[t_{2,1}+t_{3,1}+s_1-\tfrac16 s_2\big]\nabla^2\delta+\Gd{4}\big[t_{7,1}-\tfrac{3}{14}t_{4,1}+\tfrac{1}{2}t_{6,1}-\tfrac{2}{7}s_1+\tfrac{1}{21}s_2\big]\nabla^2\Tr[(\Pim{1})^2]\nonumber\\
    &&\qquad\quad\quad\,+\,\Gd{4}\big[t_{8,1}+\tfrac{3}{14}t_{4,1}-\tfrac{1}{2}t_{6,1}+\tfrac{2}{7}s_1+\tfrac{5}{42}s_2\big]\nabla^2\delta^2\\\nonumber
    &&\qquad\quad\quad\,+\,\Gd{4}\big[t_{4,1}+t_{5,1}+t_{6,1}-s_1\big]\partial^{kl}\big[\Pim{1}_{kl}\,\delta\big]\Big)_{\mathcal{O}(\delta_{\rm L}^2)}\\\nonumber
    -\frac{\pi^{(2)}_{S,\rm ct, NLG}}{f\mathcal{H}\bar\rho}&=&\partial_\eta\delta^{(2)}_{\rm ct, NLG}\\\nonumber
    -\frac{\pi^{i,(2)}_{V,\rm ct, NLG}}{f\mathcal{H}\bar\rho} &=&  \frac{1}{9}D^2(\eta)
      \Big(2\,\Gv{4}\big[t_{4,1}+t_{5,1}+t_{6,1}\big]\epsilon^{ijk}\partial_j\partial^{l}\Pim{2}_{kl}+s_2\epsilon^{ijk}\partial_j[\Psi_k\nabla^2\delta]\Big)_{\mathcal{O}(\delta_{\rm L}^2)},
\eeq
defining the backreaction amplitudes $s_1\equiv\tfrac13\Gd{3}[t_{2,1}+t_{3,1}]+\tfrac23\Gs{3}[t_{2,1}+t_{3,1}]$ and $s_2\equiv\Gd{3}[t_{2,1}+t_{3,1}]$.\footnote{Using the relation $\Gd{N+1}[\Gd{N}]=\frac{2}{21}[A_{N+1}(4N\Gs{N}-\Gd{N})-A_N(3\,\Gd{N+1}+4(N{+}1)\,\Gs{N+1})]$ for $A_N=\tfrac12(2N+3)(N-1)$ (which can be derived from the fluid equations), the coefficient of the $\nabla^2\delta$ term can be written as $\frac{1}{9}D^2(\eta)\Gd{3}[t_{2,1}+t_{3,1}]$, matching \eqref{eq: ct1}.} Combining equations according to \eqref{eq: pi-SV} yields the velocity counterterm at second-order in fields:
\beq\label{eq: ct2-v}
    -\frac{v^{i,(2)}_{\rm ct, NLG}}{f\mathcal{H}}(\vx,\eta)
    &=& D^2(\eta)\Big(\tfrac13\Gs{3}[t_{2,1}+t_{3,1}]\partial^{i}\delta
     +(\tfrac{8}{33}\Gs{4}[t_{7,1}+\tfrac{3}{14}t_{5,1}+\tfrac{5}{7}t_{6,1}-\tfrac{1}{2}s_1+\tfrac{1}{21}s_2]+\tfrac{5}{63}s_2)\partial^{i}\Tr[(\Pim{1})^2]\nonumber\\
    &&\qquad\,+\,(\tfrac{8}{33}\Gs{4}[t_{8,1}-\tfrac{3}{14}t_{5,1}-\tfrac{5}{7}t_{6,1}+\tfrac{1}{2}s_1+\tfrac{5}{42}s_2]-\tfrac{1}{6}\Gs{3}[t_{2,1}+t_{3,1}]-\tfrac{17}{126}s_2)\partial^{i}\delta^2\nonumber\\
    &&\qquad\,+\,\tfrac{2}{9}\Gv{4}[t_{4,1}+t_{5,1}+t_{6,1}]\partial_{a}\Pim{2}_{ia}\nonumber\\
    &&\qquad\,+\,(\tfrac8{33}\Gs{4}[t_{4,1}+t_{5,1}+t_{6,1}-s_1]+\tfrac29 s_2
          -\tfrac29\Gv{4}[t_{4,1}+t_{5,1}+t_{6,1}])
        \partial^{i}\frac{\partial_{ab}}{\nabla^2}\Pim{2}_{ab}\Big)_{\mathcal{O}(\delta_{\rm L}^2)}.
\eeq
Keeping the time-dependence of $t_{a,b}$ free, we find four scalar counterterms and five velocity counterterms up to second-order in fields; at fixed $t_{a,b}$, the amplitudes of the velocity counterterms are fixed by the four scalar counterterms (matching \citep{DAmico:2022ukl}).\footnote{Explicitly, the independent degrees of freedom are $t_{2,1}+t_{3,1}$, $t_{4,1}+t_{5,1}+t_{6,1}$, $t_{7,1}+\tfrac3{14}t_{5,1}+\tfrac57t_{6,1}$, and $t_{8,1}-\tfrac3{14}t_{5,1}-\tfrac57t_{6,1}$.}

\subsection{Non-Local Contributions}\label{subsec: nonlocal-explain}
\noindent Up to second-order in fields, the density counterterms specified by \eqref{eq: ct1}\,\&\,\eqref{eq: ct2} can be fully expressed in terms of the local-in-space NLG operators given in Tab.\,\ref{tab: operators} (specifically, $\mathcal{O}_{1,1}$ through $\mathcal{O}_{4,1}$). In contrast, the velocity sector of \eqref{eq: ct2-v} acquires a \textit{non-local-in-space} contribution proportional to $(\partial^{iab}/\nabla^2)\Pim{2}_{ab}$, in addition to the four local-in-space operators of Tab.\,\ref{tab: rsd-table}, as noted in \citep{DAmico:2022ukl}.\footnote{The analogous result for $\pi^i_{\rm ct}$ includes two terms with inverse Laplacians; the second arises from a product of $\delta_{\rm ct}$ and $v^i$, and is not a true signature of non-locality-in-space.} To see how this arises, consider the formal definition of $\pi_{\rm ct}^i$ obtained from \eqref{eq: pi-SV}\,\&\,\eqref{eq: green-soln}:
\beq
    -\frac{\pi^i_{\rm ct}}{f\mathcal{H}\bar\rho}(\vx,\eta) &=& \int^{\eta} d\eta'\left\{\left(\frac{2}{5}e^{\eta-\eta'}+\frac{3}{5}e^{-3(\eta-\eta')/2}\right)\frac{\partial^{ij}}{\nabla^2}\,+\,\left(e^{-(\eta-\eta')/2}\right)\left[\delta_{\rm K}^{ij}-\frac{\partial^{ij}}{\nabla^2}\right]\right\}\\\nonumber
    &&\qquad\,\times\,\partial^k\big[S^{\rm back}_{jk}+\frac{1}{f^2\mathcal{H}^2\bar\rho}\tau_{jk}\big](\vx,\eta').
\eeq
If the two parentheses in the first line are equal (\textit{i.e.}\ if the transverse and longitudinal components have the same Green's functions), the local-in-space contributions from $\tau_{jk}$ will induce local-in-space contributions; if they differ (which occurs in both $\Lambda$CDM and the Einstein-de-Sitter approximation), non-locality is induced by transverse contributions to $\partial^k\tau_{jk}$, starting at second order in fields \citep{DAmico:2022ukl}. Backreaction through $S_{jk}$ also sources inverse Laplacian contributions; these do not induce new terms in $v^i\propto \pi^i/(1+\delta)$ once we subtract the contributions from $v^i\delta_{\rm ct}$. The net effect is a single additional term in $v_{\rm ct}^{i,(2)}$, which induces a characteristic shape in the redshift-space bispectrum \citep{DAmico:2022ukl}.

At next order, non-locality enters the density sector (as predicted in \citep{Bakx:2026rmd}). Since $\tau^{ij}$ is local-in-space (and thus so are the direct contributions to $\delta_{\rm ct}$, cf.\,\eqref{eq: green-soln}), this must arise from backreaction via the source term $S_{ij}$ \eqref{eq: source}. To understand this, consider the third-order contributions of $\delta_{\rm ct}^{(2)}$ and $v_{\rm ct}^{i,(2)}$ to the NLG source function:
\beq\label{eq: o3-Sij}
    S^{ij,(3)}_{\rm back} &\supset&
      \frac{3}{2}\left(\Psi_{\rm ct}^{i,(2)}\Psi^{j,(1)}+(i\leftrightarrow j)
      -\delta_{\rm K}^{ij}\Psi^{k,(2)}_{\rm ct}\Psi^{(1)}_k\right)+\frac{1}{f^2\mathcal{H}^2}\left(v_{\rm ct}^{i,(2)}v^{j,(1)}+(i\leftrightarrow j)\right),
\eeq
for $\Psi^i\equiv (\partial^i/\nabla^2)\delta$. Non-locality is induced in two manners: (1) from the non-local second-order velocity $-v_{\rm ct}^{i,(2)}/f\mathcal{H}\supset D^2\gamma_v(\partial^{iab}/\nabla^2)\Pim{2}_{ab}$ \eqref{eq: ct2-v}; (2) from the non-Laplacian contribution to the second-order density: $\delta_{\rm ct}^{(2)}\supset D^2\gamma_\delta\partial^{kl}[\Pim{1}_{kl}\delta]$ \eqref{eq: ct2} (denoting the amplitudes by $\gamma_{v,\delta}$). Since $(\partial^{kl}/\nabla^2)[\Pim{1}_{kl}\delta]=(\partial^{kl}/\nabla^2)\Pim{2}_{kl}$ up locally-contributing operators, the two contributions have the same spatial structure. Recalling that $v^{i,(1)}=-f\mathcal{H}\Psi^{i,(1)}$, we obtain the non-local source term
\beq\label{eq: o3-Sij-nl}
    \left.S^{ij,(3)}_{\rm back}\right|_{\rm non-local} &=&
      D^2\left[\frac{3}{2}\gamma_{\delta}\left(\Psi^{j}\frac{\partial^{iab}}{\nabla^2}\Pim{2}_{ab}+(i\leftrightarrow j)
      -\delta_{\rm K}^{ij}\Psi_k\frac{\partial^{kab}}{\nabla^2}\Pim{2}_{ab}\right)+\gamma_v\left(\Psi^{j}\frac{\partial^{iab}}{\nabla^2}\Pim{2}_{ab}+(i\leftrightarrow j)\right)\right]_{\mathcal{O}(\delta_{\rm L}^3)},
\eeq
from which the backreaction counterterms can be derived.

Combining \eqref{eq: green-soln}\,\&\,\eqref{eq: o3-Sij-nl} yields the non-local contributions to the density counterterm at third-order in fields. Using the relations
\beq\label{eq: o3-ids}
    \partial_{ij}\bigg[\frac{\partial^{iab}}{\nabla^2}\Pim{2}_{ab}
      \Psi^j\bigg]
    =-2\widehat\Pi^{[3]}+\Pi_{B1}-\Pi_{B2}+({\rm local}) ,
    \qquad
    \nabla^2\bigg[\frac{\partial^{kab}}{\nabla^2}\Pim{2}_{ab}\Psi_k\bigg]
    =-2\widehat\Pi^{[3]}+({\rm local}),
\eeq
where $\Pi_{B1}$, $\Pi_{B2}, \widehat\Pi^{[3]}$ are the non-local operators of \eqref{eq: bakx-nonlocal-ops} (introduced in \citep{Bakx:2026rmd}), we find:
\beq
    \left.\delta_{\rm ct}^{(3)}(\vx,\eta)\right|_{\rm non-local} &=& -\frac{1}{26}D^2(\eta)\left(\mathcal{G}_{\delta}^{(5)}[3\gamma_{\delta}+4\gamma_v](\eta)\widehat\Pi^{[3]}(\vx,\eta)+\mathcal{G}_{\delta}^{(5)}[3\gamma_{\delta}+2\gamma_v](\eta)\left(\Pi_{B2}-\Pi_{B1}\right)(\vx,\eta)\right)_{\mathcal{O}(\delta_{\rm L}^3)},
\eeq
where the time-dependent coefficients are given explicitly by 
\beq\label{eq: gamma-v-delta}
    \gamma_v(\eta) &=& \frac8{33}\Gs{4}[t_{4,1}+t_{5,1}+t_{6,1}-s_1]+\frac29 s_2
          -\frac29\Gv{4}[t_{4,1}+t_{5,1}+t_{6,1}]          \\\nonumber
    \gamma_\delta(\eta) &=& \frac{2}{33}\Gd{4}[t_{4,1}+t_{5,1}+t_{6,1}-s_1].    .
\eeq
In the limit of time-independent $t_{a,b}$ (omitting the finite parts of the counterterms), we find $\gamma_{\delta}=3\gamma_v$ and thus
\beq\label{eq: delta-ct-nonlocal-fixt}
    \left.\delta_{\rm ct}^{(3)}(\vx,\eta)\right|_{\rm non-local} &\to & -\frac{1}{1287}D^2(\eta)\left[\left(t_{4,1}+t_{5,1}+t_{6,1}-t_{2,1}-t_{3,1}\right)\left(13\,\widehat\Pi^{[3]}-11(\Pi_{B1}-\Pi_{B2})\right)(\vx,\eta)\right]_{\mathcal{O}(\delta_{\rm L}^3)}.
\eeq
Notably, $\delta_{\rm ct}^{(3)}$ depends on only two of the three permissible non-local structures: $\Pi_{B1}-\Pi_{B2}$ and $\widehat{\Pi}^{[3]}$, due the structure of the source term; moreover, only a single combination arises if $t_{a,b}$ are time-independent. As discussed in \S\ref{sec: renorm-real}, both terms are sourced for biased tracers due to non-linear effects, though only the latter yields a non-degenerate signature in the two-loop power spectrum. 

\subsection{Explicit Results}
\noindent Finally, we present the complete NLG (NNLG) density and velocity counterterms up to third-order (first-order) in fields, generalizing \citep{DAmico:2022ukl,Raikman:2026gcx}. These are obtained analogously to the lower-order pieces, though the computational complexity quickly becomes large due to the nested time integrals sourced by backreaction. The (real-space) dark matter density counterterm reads:
\beq\label{eq: dm-density-ct}
    \delta_{\rm ct}(\vx,\eta) &=&D^2(\eta)\bigg[\sum_{a=1}^{13}T^{(\delta)}_{a,1}(\eta)
      \mathcal{O}_{a,1}(\vx,\eta)
      +\sum_{a=1}^{2}\tilde{T}^{(\delta)}_{a,1}(\eta)\tilde{\mathcal{O}}_{a,1}(\vx,\eta)\bigg]+D^4(\eta)T^{(\delta)}_{1,2}(\eta)\mathcal{O}_{1,2}(\vx,\eta)+\cdots,
\eeq
dropping stochastic terms. Here, $\mathcal{O}_{a,b}$ are the scalar operators listed in Tab.\,\ref{tab: operators} and $ \tilde{\mathcal{O}}_{1,1}\equiv \widehat{\Pi}^{[3]}$ and $\tilde{\mathcal{O}}_{2,1}\equiv \Pi_{B1}-\Pi_{B2}-\tfrac75\mathcal{O}_{7,1}$ are the non-local operators discussed above (subtracting a degenerate local contribution from $\mathcal{O}_{7,1}$). The time-dependent amplitudes, $T_{a,b}^{(\delta)}$, take the explicit form (simplifying products of $\mathcal{G}$ where possible):
{\small
\beq\label{eq: dm-density-T}
    T^{(\delta)}_{1,1}&=&\tfrac{1}{9}\Gd{3}[t_{23}]\\\nonumber
    T^{(\delta)}_{2,1}&=&\tfrac{2}{33}\Gd{4}[t_{7,1}-\tfrac{3}{14}t_{4,1}+\tfrac{1}{2}t_{6,1}-\tfrac{2}{7}s_1+\tfrac{1}{21}s_2]\\\nonumber
    T^{(\delta)}_{3,1}&=&\tfrac{2}{33}\Gd{4}[t_{8,1}+\tfrac{3}{14}t_{4,1}-\tfrac{1}{2}t_{6,1}+\tfrac{2}{7}s_1+\tfrac{5}{42}s_2]\\\nonumber
    T^{(\delta)}_{4,1}&=&\tfrac{2}{33}\Gd{4}[t_{456}-s_1]\\\nonumber
    T^{(\delta)}_{5,1}&=&0\\\nonumber
    T^{(\delta)}_{6,1}&=&\tfrac{1}{26}\Gd{5}[t_{11,1}]+\tfrac{16}{21}s_1-\tfrac{100}{189}s_2+\tfrac{4}{9}s_3\\\nonumber
      &&+\tfrac{1}{11}\Gd{4}[\tfrac{32}{21}t_{23}+\tfrac{34}{21}t_{4,1}+\tfrac{4}{3}t_{5,1}+\tfrac{2}{3}t_{6,1}-\tfrac{4}{3}t_{7,1}]+\tfrac{64}{231}\Gs{4}[t_{23}]\\\nonumber
      &&+\Gd{5}[-\tfrac{2}{13}t_{23}-\tfrac{1}{6}t_{4,1}-\tfrac{2}{13}t_{5,1}-\tfrac{3}{26}t_{6,1}+\tfrac{1}{13}t_{7,1}-\tfrac{1}{26}t_{10,1}]+\tfrac{1}{13}\Gs{5}[-\tfrac{20}{3}t_{23}-5t_{456}]\\\nonumber
    T^{(\delta)}_{7,1}&=&\tfrac{1}{26}\Gd{5}[t_{9,1}]+\tfrac{13}{105}s_1+\tfrac{4}{45}s_2+\tfrac{8}{45}s_3\\\nonumber
      &&+\Gd{4}[-\tfrac{4}{35}t_{23}-\tfrac{6}{55}t_{456}]+\tfrac{1}{11}\Gs{4}[-\tfrac{32}{105}t_{23}+\tfrac{32}{15}t_{456}]\\\nonumber
      &&+\tfrac{1}{13}\Gd{5}[\tfrac{1}{6}t_{23}-\tfrac{1}{10}t_{4,1}+\tfrac{1}{10}t_{5,1}+\tfrac{1}{10}t_{6,1}+t_{10,1}]+\tfrac{1}{13}\Gs{5}[-\tfrac{1}{3}t_{23}-4t_{456}]\\\nonumber
    T^{(\delta)}_{8,1}&=&\tfrac{1}{26}\Gd{5}[t_{14,1}]-\tfrac{397}{882}s_1+\tfrac{5953}{19845}s_2-\tfrac{32}{189}s_3\\\nonumber
      &&+\tfrac{1}{11}\Gd{4}[-\tfrac{362}{441}t_{23}-\tfrac{251}{441}t_{4,1}-\tfrac{215}{441}t_{5,1}-\tfrac{131}{441}t_{6,1}-\tfrac{8}{21}t_{8,1}]+\tfrac{1}{11}\Gs{4}[-\tfrac{80}{63}t_{23}-\tfrac{16}{49}t_{5,1}-\tfrac{160}{147}t_{6,1}+\tfrac{32}{21}t_{8,1}]\\\nonumber
      &&+\tfrac{1}{13}\Gd{5}[\tfrac{3173}{2940}t_{23}+\tfrac{977}{1260}t_{4,1}+\tfrac{475}{588}t_{5,1}+\tfrac{475}{588}t_{6,1}+\tfrac{4}{21}t_{10,1}-\tfrac{89}{360}t_{12,1}]\\\nonumber
      &&+\Gs{5}[\tfrac{3035}{11466}t_{23}+\tfrac{40}{273}t_{4,1}+\tfrac{25}{147}t_{5,1}+\tfrac{430}{1911}t_{6,1}-\tfrac{10}{91}t_{8,1}]\\\nonumber
    T^{(\delta)}_{9,1}&=&\tfrac{1}{26}\Gd{5}[t_{18,1}]+\tfrac{1}{42}s_1-\tfrac{2}{35}s_2-\tfrac{8}{63}s_3\\\nonumber
      &&+\Gd{4}[\tfrac{74}{1617}t_{23}-\tfrac{1}{33}t_{4,1}-\tfrac{31}{1617}t_{5,1}+\tfrac{1}{147}t_{6,1}-\tfrac{4}{77}t_{8,1}]+\tfrac{1}{11}\Gs{4}[\tfrac{272}{147}t_{23}+\tfrac{16}{49}t_{5,1}+\tfrac{160}{147}t_{6,1}-\tfrac{32}{21}t_{8,1}]\\\nonumber
      &&+\tfrac{1}{13}\Gd{5}[-\tfrac{1619}{2940}t_{23}+\tfrac{61}{140}t_{4,1}+\tfrac{75}{196}t_{5,1}+\tfrac{5}{196}t_{6,1}+\tfrac{5}{7}t_{8,1}+\tfrac{1}{7}t_{10,1}-\tfrac{31}{120}t_{12,1}]\\\nonumber
      &&+\tfrac{1}{13}\Gs{5}[-\tfrac{545}{294}t_{23}+\tfrac{10}{7}t_{4,1}+\tfrac{55}{49}t_{5,1}+\tfrac{20}{49}t_{6,1}+\tfrac{10}{7}t_{8,1}]\\\nonumber
    T^{(\delta)}_{10,1}&=&\tfrac{1}{26}\Gd{5}[t_{15,1}]-\tfrac{29}{98}s_1+\tfrac{1042}{6615}s_2-\tfrac{8}{63}s_3\\\nonumber
      &&+\Gd{4}[-\tfrac{10}{1617}t_{23}-\tfrac{89}{1617}t_{4,1}-\tfrac{59}{1617}t_{5,1}+\tfrac{1}{147}t_{6,1}+\tfrac{4}{77}t_{7,1}-\tfrac{8}{231}t_{8,1}]\\\nonumber
      &&+\tfrac{1}{11}\Gs{4}[\tfrac{16}{21}t_{23}+\tfrac{16}{49}t_{5,1}+\tfrac{160}{147}t_{6,1}+\tfrac{32}{21}t_{7,1}]\\\nonumber
      &&+\tfrac{1}{13}\Gd{5}[\tfrac{691}{2940}t_{23}+\tfrac{323}{420}t_{4,1}+\tfrac{103}{196}t_{5,1}+\tfrac{5}{196}t_{6,1}-\tfrac{5}{7}t_{7,1}+\tfrac{2}{7}t_{8,1}+\tfrac{1}{7}t_{10,1}-\tfrac{11}{120}t_{12,1}]\\\nonumber
      &&+\tfrac{1}{13}\Gs{5}[\tfrac{75}{98}t_{23}+\tfrac{10}{7}t_{4,1}+\tfrac{55}{49}t_{5,1}+\tfrac{20}{49}t_{6,1}-\tfrac{10}{7}t_{7,1}]\\\nonumber
    T^{(\delta)}_{11,1}&=&\tfrac{1}{26}\Gd{5}[t_{17,1}]+\tfrac{13}{42}s_1-\tfrac{52}{315}s_2+\tfrac{8}{63}s_3\\\nonumber
      &&+\Gd{4}[\tfrac{10}{1617}t_{23}+\tfrac{1}{33}t_{4,1}+\tfrac{31}{1617}t_{5,1}-\tfrac{1}{147}t_{6,1}-\tfrac{4}{77}t_{7,1}]+\tfrac{1}{11}\Gs{4}[-\tfrac{16}{49}t_{23}-\tfrac{16}{49}t_{5,1}-\tfrac{160}{147}t_{6,1}-\tfrac{32}{21}t_{7,1}]\\\nonumber
      &&+\tfrac{1}{13}\Gd{5}[-\tfrac{971}{2940}t_{23}-\tfrac{61}{140}t_{4,1}-\tfrac{75}{196}t_{5,1}-\tfrac{5}{196}t_{6,1}+\tfrac{5}{7}t_{7,1}-\tfrac{1}{7}t_{10,1}+\tfrac{31}{120}t_{12,1}]\\\nonumber
      &&+\tfrac{1}{13}\Gs{5}[-\tfrac{365}{294}t_{23}-\tfrac{10}{7}t_{4,1}-\tfrac{55}{49}t_{5,1}-\tfrac{20}{49}t_{6,1}+\tfrac{10}{7}t_{7,1}]\\\nonumber
    T^{(\delta)}_{12,1}&=&\tfrac{1}{26}\Gd{5}[t_{16,1}]-\tfrac{16}{63}s_1+\tfrac{74}{405}s_2-\tfrac{4}{27}s_3\\\nonumber
      &&+\tfrac{1}{11}\Gd{4}[-\tfrac{40}{63}t_{23}-\tfrac{4}{9}t_{456}]-\tfrac{64}{693}\Gs{4}[t_{23}]\\\nonumber
      &&+\tfrac{1}{13}\Gd{5}[\tfrac{11}{15}t_{23}+\tfrac{28}{45}t_{4,1}+\tfrac{2}{3}t_{5,1}+\tfrac{2}{3}t_{6,1}+\tfrac{1}{6}t_{10,1}-\tfrac{11}{45}t_{12,1}]+\tfrac{1}{13}\Gs{5}[\tfrac{20}{9}t_{23}+\tfrac{5}{3}t_{456}]\\\nonumber
    T^{(\delta)}_{13,1}&=&\tfrac{1}{26}\Gd{5}[t_{13,1}]+\tfrac{5}{7}s_1-\tfrac{22}{45}s_2+\tfrac{4}{9}s_3\\\nonumber
      &&+\tfrac{1}{11}\Gd{4}[\tfrac{20}{21}t_{23}+\tfrac{2}{3}t_{456}]+\tfrac{32}{231}\Gs{4}[t_{23}]\\\nonumber
      &&+\Gd{5}[-\tfrac{1}{10}t_{23}-\tfrac{1}{10}t_{4,1}-\tfrac{3}{26}t_{5,1}-\tfrac{3}{26}t_{6,1}-\tfrac{1}{26}t_{10,1}+\tfrac{61}{780}t_{12,1}]+\tfrac{1}{13}\Gs{5}[-5t_{23}-5t_{456}]\\\nonumber
    \tilde{T}^{(\delta)}_{1,1}&=&-\tfrac{1}{13}\Gd{5}[\tfrac{3}{2}T^{(\delta)}_{4,1}+\tfrac{16}{33}\Gs{4}[t_{456}-s_1]+\tfrac{4}{9}s_2-\tfrac{4}{9}s_3]\\\nonumber
    \tilde{T}^{(\delta)}_{2,1}&=&\tfrac{1}{13}\Gd{5}[\tfrac{3}{2}T^{(\delta)}_{4,1}+\tfrac{8}{33}\Gs{4}[t_{456}-s_1]+\tfrac{2}{9}s_2-\tfrac{2}{9}s_3]\\\nonumber
    T^{(\delta)}_{1,2}&=&\tfrac{1}{26}\Gd{5}[t_{1,2}+t_{2,2}+\tfrac{1}{9}t_{23}s_2+\tfrac{2}{9}t_{4,1}\Gv{4}[t_{23}]+t_{12,1}(\tfrac{1}{2}t_{23}-\tfrac{7}{18}\Gv{4}[t_{23}])]
\eeq
}\noindent
defining $t_{23}\equiv t_{2,1}+t_{3,1}$, $t_{456}\equiv t_{4,1}+t_{5,1}+t_{6,1}$ as well as the backreaction functions $s_1\equiv\tfrac13\Gd{3}[t_{23}]+\tfrac23\Gs{3}[t_{23}]$, $s_2\equiv\Gd{3}[t_{23}]$ and $s_3\equiv\Gv{4}[t_{456}]$. For time-independent $t_{a,b}$, these can be further simplified with the relations $\mathcal{G}[t]\to t$, $s_1,s_2\to t_{23}$ and $s_3\to t_{456}$.\footnote{Using \eqref{eq: anastasiou-dict}, we can additionally relate the $T^{(\delta)}$ amplitudes to the counterterm coefficients of \citep{Anastasiou:2025jsy}.}

In total, $\delta_{\rm ct}$ depends on 12 local NLG amplitudes, $T_{a,1}^{(\delta)}$, two non-local coefficients, $\tilde{T}_{a,1}^{(\delta)}$, and one NNLG coefficient $T_{1,2}^{(\delta)}$ (which includes terms quadratic in $t_{a,1}$). Notably, \eqref{eq: dm-density-ct} involves only NLG operators with overall gradients; other terms (e.g., 
$\mathcal{O}_{5,1}=\delta\nabla^2\delta$) are not generated by dark matter renormalization.\footnote{Due to the non-linear connection between dark matter and tracers, such terms do arise in the galaxy counterterms.} As discussed above, the amplitudes of the non-local terms are set by the second-order-in-fields sector:
\beq
    \tilde{T}^{(\delta)}_{1,1}+2\tilde{T}^{(\delta)}_{2,1}=\tfrac{3}{26}\Gd{5}[T^{(\delta)}_{4,1}],
    \qquad
    \tilde{T}^{(\delta)}_{1,1}+\tilde{T}^{(\delta)}_{2,1}
      =-\tfrac{1}{13}\Gd{5}[\gamma_v],
\eeq
where $\gamma_v$ is the non-local velocity coefficient defined in \eqref{eq: gamma-v-delta}. If the time-dependence of $t_{a,b}$ is unknown, these do not restrict the dimensionality. In contrast, assuming time-independent $t_{a,b}$ fixes $\tilde{T}^{(\delta)}_{1,1}\to(t_{23}-t_{456})/99=-\tfrac16 T^{(\delta)}_{4,1}$ and $\tilde{T}^{(\delta)}_{2,1}=-\tfrac{11}{13}\tilde{T}^{(\delta)}_{1,1}$, such that the non-local sector is entirely fixed by the local sector with a total of $13$ degrees of freedom. As seen in Tab.\,\ref{tab: operators}, only seven of these contribute to the two-loop power spectrum (noting that $T^{(\delta)}_{5,1}=0$ and $\tilde{T}_{1,1}^{(\delta)}$ is fixed); this is consistent with the real-space results of \citep{Anastasiou:2025jsy} up to differences in the renormalization scheme.

We can similarly obtain the deterministic velocity counterterm, $v_{\rm ct}^i$ (combining both longitudinal and transverse components). This takes the explicit form:
\beq\label{eq: dm-velocity-ct}
    -\frac{v^i_{\rm ct}}{f\mathcal{H}}(\vx,\eta)=D^2(\eta)\left[\sum_{a=1}^{14}T^{(v)}_{a,1}(\eta)\mathcal{O}^{i}_{a,1}(\vx,\eta)+\sum_{a=1}^{7}\tilde{T}^{(v)}_{a,1}(\eta)\tilde{\mathcal{O}}^{i}_{a,1}(\vx,\eta)\right]+D^4(\eta)T^{(v)}_{1,2}(\eta)\mathcal{O}^{i}_{1,2}(\vx,\eta)+\cdots,
\eeq
where $\mathcal{O}_{a,b}^i$ and $\tilde{\mathcal{O}}_{a,b}^i$ are the local- and non-local-in-space operators displayed in Tab.\,\ref{tab: rsd-table} (excluding $\tilde{\mathcal{O}}^{i}_{8,1}$).\footnote{The momentum counterterm, $-\pi^i_{\rm ct}/(f\mathcal{H}\bar\rho)$, differs from \eqref{eq: dm-velocity-ct} in three ways: (1) the $(1+\delta)$ weighting induces shifts in the $T^{(v)}_{a,b}$ coefficients, (2) a new operator arises ($\tilde{\mathcal{O}}^{i}_{8,1}=\delta\,\tilde{\mathcal{O}}^{i}_{1,1}$) whose coefficient is fixed to $D^2(\eta)\tilde{T}^{(v)}_{1,1}(\eta)$; (3) there are additional contributions from $v^i\delta_{\rm ct}$, involving $\frac{\partial^i}{\nabla^2}\delta$ (since $\pi^i_{\rm ct}$ is not invariant under Galilean transformations).} Notably, this spans all fourteen of the possible local-in-space NLG operators, $\mathcal{O}_{a,1}^i$ (in contrast to the density, which uses only a subset). We additionally require seven non-local-in-space operators, which are generated both directly from the stress tensor and through backreaction. The time-dependent amplitudes are given by:
{\footnotesize
\beq\label{eq: dm-velocity-T}
    T^{(v)}_{1,1}&=&\tfrac{1}{3}\Gs{3}[t_{23}]\\\nonumber
    T^{(v)}_{2,1}&=&\tfrac{8}{33}\Gs{4}[t_{7,1}+\tfrac{3}{14}t_{5,1}+\tfrac{5}{7}t_{6,1}-\tfrac{1}{2}s_1+\tfrac{1}{21}s_2]+\tfrac{5}{63}s_2\\\nonumber
    T^{(v)}_{3,1}&=&\tfrac{8}{33}\Gs{4}[t_{8,1}-\tfrac{3}{14}t_{5,1}-\tfrac{5}{7}t_{6,1}+\tfrac{1}{2}s_1+\tfrac{5}{42}s_2]-\tfrac{1}{6}\Gs{3}[t_{23}]-\tfrac{17}{126}s_2\\\nonumber
    T^{(v)}_{4,1}&=&\tfrac{2}{9}\Gv{4}[t_{456}]\\\nonumber
    T^{(v)}_{5,1}&=&\tfrac{5}{26}\Gs{5}[t_{11,1}]+\tfrac{1420}{1911}s_1-\tfrac{3824}{12285}s_2+\tfrac{98}{351}s_3+\tfrac{1}{143}\Gd{4}[-\tfrac{368}{441}t_{23}+\tfrac{70}{9}t_{456}]\\\nonumber
      &&+\tfrac{1}{11}\Gs{4}[-\tfrac{9664}{1911}t_{23}-\tfrac{368}{117}t_{4,1}-\tfrac{3512}{819}t_{5,1}-\tfrac{5696}{819}t_{6,1}-\tfrac{16}{3}t_{7,1}]+\tfrac{1}{169}\Gd{5}[-\tfrac{2}{3}t_{23}-7t_{456}]\\\nonumber
      &&+\tfrac{1}{13}\Gs{5}[\tfrac{40}{39}t_{23}-\tfrac{5}{26}t_{4,1}+\tfrac{20}{13}t_{5,1}+\tfrac{105}{26}t_{6,1}+5t_{7,1}+\tfrac{175}{78}t_{9,1}+\tfrac{155}{78}t_{10,1}]\\\nonumber
    T^{(v)}_{6,1}&=&\tfrac{50}{91}s_1-\tfrac{8}{39}s_2+\tfrac{28}{117}s_3+\tfrac{1}{143}\Gd{4}[-\tfrac{136}{21}t_{23}+\tfrac{20}{3}t_{456}]+\tfrac{1}{143}\Gs{4}[-\tfrac{1472}{21}t_{23}-\tfrac{224}{3}t_{456}]+\tfrac{1}{169}\Gd{5}[5t_{23}-6t_{456}]\\\nonumber
      &&+\tfrac{1}{169}\Gs{5}[30t_{23}+35t_{4,1}+45t_{5,1}+45t_{6,1}+25t_{9,1}+50t_{10,1}]\\\nonumber
    T^{(v)}_{7,1}&=&\tfrac{5}{26}\Gs{5}[t_{14,1}]-\tfrac{64717}{27783}s_1+\tfrac{17293}{11907}s_2\\\nonumber
      &&+\tfrac{1}{11}\Gd{4}[-\tfrac{243328}{83349}t_{23}+\tfrac{10}{81}t_{5,1}+\tfrac{100}{243}t_{6,1}-\tfrac{140}{243}t_{8,1}+\tfrac{1060}{3969}t_{9,1}+\tfrac{2120}{3969}t_{10,1}-\tfrac{1378}{3969}t_{12,1}-\tfrac{8480}{11907}t_{13,1}-\tfrac{4240}{1701}t_{18,1}]\\\nonumber
      &&+\tfrac{1}{11}\Gs{4}[-\tfrac{232256}{83349}t_{23}-\tfrac{9136}{3969}t_{5,1}-\tfrac{91360}{11907}t_{6,1}+\tfrac{18272}{1701}t_{8,1}-\tfrac{4240}{3969}t_{9,1}-\tfrac{8480}{3969}t_{10,1}+\tfrac{5512}{3969}t_{12,1}+\tfrac{33920}{11907}t_{13,1}+\tfrac{16960}{1701}t_{18,1}]\\\nonumber
      &&+\tfrac{1}{13}\Gd{5}[\tfrac{5389}{1134}t_{23}+\tfrac{31}{63}t_{5,1}+\tfrac{310}{189}t_{6,1}-\tfrac{62}{27}t_{8,1}]\\\nonumber
      &&+\tfrac{1}{13}\Gs{5}[\tfrac{815}{54}t_{23}+\tfrac{515}{378}t_{5,1}+\tfrac{4100}{567}t_{6,1}-\tfrac{6655}{567}t_{8,1}-\tfrac{175}{81}t_{13,1}-\tfrac{1765}{162}t_{18,1}]\\\nonumber
    T^{(v)}_{8,1}&=&\tfrac{5}{26}\Gs{5}[t_{15,1}]-\tfrac{28291}{9261}s_1+\tfrac{7552}{3969}s_2\\\nonumber
      &&+\tfrac{1}{11}\Gd{4}[-\tfrac{118540}{27783}t_{23}+\tfrac{40}{189}t_{5,1}+\tfrac{400}{567}t_{6,1}-\tfrac{4}{21}t_{7,1}-\tfrac{668}{567}t_{8,1}+\tfrac{10}{27}t_{9,1}+\tfrac{20}{27}t_{10,1}-\tfrac{13}{27}t_{12,1}-\tfrac{80}{81}t_{13,1}-\tfrac{2}{3}t_{17,1}-\tfrac{334}{81}t_{18,1}]\\\nonumber
      &&+\Gs{4}[-\tfrac{81248}{305613}t_{23}-\tfrac{368}{1323}t_{5,1}-\tfrac{3680}{3969}t_{6,1}+\tfrac{64}{231}t_{7,1}+\tfrac{9824}{6237}t_{8,1}-\tfrac{40}{297}t_{9,1}-\tfrac{80}{297}t_{10,1}+\tfrac{52}{297}t_{12,1}+\tfrac{320}{891}t_{13,1}+\tfrac{8}{33}t_{17,1}+\tfrac{1336}{891}t_{18,1}]\\\nonumber
      &&+\tfrac{1}{13}\Gd{5}[\tfrac{2227}{378}t_{23}+\tfrac{10}{21}t_{5,1}+\tfrac{100}{63}t_{6,1}-\tfrac{3}{7}t_{7,1}-\tfrac{167}{63}t_{8,1}]\\\nonumber
      &&+\tfrac{1}{13}\Gs{5}[\tfrac{1135}{63}t_{23}+\tfrac{10}{9}t_{5,1}+\tfrac{1475}{189}t_{6,1}-\tfrac{20}{7}t_{7,1}-\tfrac{3070}{189}t_{8,1}-\tfrac{100}{27}t_{13,1}-\tfrac{835}{54}t_{18,1}]\\\nonumber
    T^{(v)}_{9,1}&=&\tfrac{5}{26}\Gs{5}[t_{16,1}]-\tfrac{836}{567}s_1+\tfrac{1126}{1215}s_2+\tfrac{2}{27}s_3\\\nonumber
      &&+\tfrac{1}{11}\Gd{4}[-\tfrac{3728}{1701}t_{23}+\tfrac{2}{9}t_{4,1}+\tfrac{194}{567}t_{5,1}+\tfrac{1058}{1701}t_{6,1}-\tfrac{136}{243}t_{8,1}+\tfrac{26}{81}t_{9,1}+\tfrac{34}{81}t_{10,1}+\tfrac{41}{405}t_{12,1}-\tfrac{82}{243}t_{13,1}-\tfrac{476}{243}t_{18,1}]\\\nonumber
      &&+\tfrac{1}{11}\Gs{4}[-\tfrac{4672}{1701}t_{23}-\tfrac{8}{9}t_{4,1}-\tfrac{1592}{567}t_{5,1}-\tfrac{12392}{1701}t_{6,1}+\tfrac{2176}{243}t_{8,1}-\tfrac{104}{81}t_{9,1}-\tfrac{136}{81}t_{10,1}-\tfrac{164}{405}t_{12,1}+\tfrac{328}{243}t_{13,1}+\tfrac{1904}{243}t_{18,1}]\\\nonumber
      &&+\tfrac{1}{13}\Gd{5}[\tfrac{1735}{567}t_{23}+\tfrac{17}{63}t_{5,1}+\tfrac{170}{189}t_{6,1}-\tfrac{34}{27}t_{8,1}]\\\nonumber
      &&+\tfrac{1}{13}\Gs{5}[\tfrac{1880}{189}t_{23}+\tfrac{200}{189}t_{5,1}+\tfrac{2980}{567}t_{6,1}-\tfrac{680}{81}t_{8,1}-\tfrac{205}{162}t_{13,1}-\tfrac{595}{81}t_{18,1}]\\\nonumber
    T^{(v)}_{10,1}&=&\tfrac{1}{11}\Gd{4}[\tfrac{1}{7}t_{5,1}+\tfrac{10}{21}t_{6,1}+\tfrac{2}{3}t_{7,1}+\tfrac{5}{49}t_{9,1}+\tfrac{10}{49}t_{10,1}-\tfrac{13}{98}t_{12,1}-\tfrac{40}{147}t_{13,1}+\tfrac{2}{3}t_{17,1}-\tfrac{2}{7}t_{18,1}]\\\nonumber
      &&+\tfrac{1}{11}\Gs{4}[-\tfrac{4}{7}t_{5,1}-\tfrac{40}{21}t_{6,1}-\tfrac{8}{3}t_{7,1}-\tfrac{20}{49}t_{9,1}-\tfrac{40}{49}t_{10,1}+\tfrac{26}{49}t_{12,1}+\tfrac{160}{147}t_{13,1}-\tfrac{8}{3}t_{17,1}+\tfrac{8}{7}t_{18,1}]\\\nonumber
    T^{(v)}_{11,1}&=&-\tfrac{1}{9}s_3+\tfrac{1}{11}\Gd{4}[-\tfrac{1}{3}t_{456}-\tfrac{5}{21}t_{9,1}-\tfrac{1}{7}t_{10,1}-\tfrac{59}{126}t_{12,1}-\tfrac{1}{7}t_{13,1}-\tfrac{2}{3}t_{17,1}]\\\nonumber
      &&+\tfrac{1}{11}\Gs{4}[\tfrac{4}{3}t_{456}+\tfrac{20}{21}t_{9,1}+\tfrac{4}{7}t_{10,1}+\tfrac{118}{63}t_{12,1}+\tfrac{4}{7}t_{13,1}+\tfrac{8}{3}t_{17,1}]\\\nonumber
    T^{(v)}_{12,1}&=&\tfrac{1}{11}\Gd{4}[\tfrac{2}{3}t_{456}+\tfrac{10}{21}t_{9,1}+\tfrac{2}{7}t_{10,1}+\tfrac{59}{63}t_{12,1}+\tfrac{2}{7}t_{13,1}-\tfrac{4}{3}t_{18,1}]\\\nonumber
      &&+\tfrac{1}{11}\Gs{4}[-\tfrac{8}{3}t_{456}-\tfrac{40}{21}t_{9,1}-\tfrac{8}{7}t_{10,1}-\tfrac{236}{63}t_{12,1}-\tfrac{8}{7}t_{13,1}+\tfrac{16}{3}t_{18,1}]\\\nonumber
    T^{(v)}_{13,1}&=&\tfrac{5}{26}\Gs{5}[t_{9,1}]-\tfrac{3116}{315}s_1+\tfrac{848}{135}s_2+\tfrac{14}{45}s_3\\\nonumber
      &&+\tfrac{1}{11}\Gd{4}[-\tfrac{14288}{945}t_{23}+\tfrac{2}{3}t_{4,1}+\tfrac{94}{63}t_{5,1}+\tfrac{646}{189}t_{6,1}-\tfrac{104}{27}t_{8,1}+\tfrac{13}{9}t_{9,1}+\tfrac{26}{9}t_{10,1}-\tfrac{169}{90}t_{12,1}-\tfrac{104}{27}t_{13,1}-\tfrac{364}{27}t_{18,1}]\\\nonumber
      &&+\tfrac{1}{11}\Gs{4}[-\tfrac{13696}{945}t_{23}-\tfrac{8}{15}t_{4,1}-\tfrac{4328}{315}t_{5,1}-\tfrac{42104}{945}t_{6,1}+\tfrac{1664}{27}t_{8,1}-\tfrac{52}{9}t_{9,1}-\tfrac{104}{9}t_{10,1}+\tfrac{338}{45}t_{12,1}+\tfrac{416}{27}t_{13,1}+\tfrac{1456}{27}t_{18,1}]\\\nonumber
      &&+\Gd{5}[\tfrac{1286}{819}t_{23}-\tfrac{3}{65}t_{4,1}+\tfrac{44}{455}t_{5,1}+\tfrac{587}{1365}t_{6,1}-\tfrac{2}{3}t_{8,1}]\\\nonumber
      &&+\Gs{5}[\tfrac{1324}{273}t_{23}-\tfrac{3}{13}t_{4,1}+\tfrac{127}{273}t_{5,1}+\tfrac{2201}{819}t_{6,1}-\tfrac{40}{9}t_{8,1}+\tfrac{5}{13}t_{10,1}-\tfrac{1}{4}t_{12,1}-\tfrac{10}{9}t_{13,1}-\tfrac{35}{9}t_{18,1}]\\\nonumber
    T^{(v)}_{14,1}&=&\tfrac{2}{9}s_3+\tfrac{1}{11}\Gd{4}[\tfrac{2}{3}t_{4,1}+\tfrac{14}{9}t_{5,1}+\tfrac{14}{9}t_{6,1}+\tfrac{4}{9}t_{9,1}+\tfrac{14}{9}t_{10,1}-\tfrac{32}{15}t_{12,1}-\tfrac{2}{3}t_{13,1}]\\\nonumber
      &&+\tfrac{1}{11}\Gs{4}[-\tfrac{8}{3}t_{4,1}-\tfrac{56}{9}t_{5,1}-\tfrac{56}{9}t_{6,1}-\tfrac{16}{9}t_{9,1}-\tfrac{56}{9}t_{10,1}+\tfrac{128}{15}t_{12,1}+\tfrac{8}{3}t_{13,1}]\\\nonumber
    \tilde{T}^{(v)}_{1,1}&=&\tfrac{8}{33}\Gs{4}[t_{456}-s_1]+\tfrac{2}{9}s_2-\tfrac{2}{9}\Gv{4}[t_{456}]\\\nonumber
    \tilde{T}^{(v)}_{2,1}&=&-\tfrac{5}{7}s_1+\tfrac{4}{9}s_2-\tfrac{4}{9}s_3+\tfrac{1}{11}\Gd{4}[-\tfrac{20}{21}t_{23}-\tfrac{2}{3}t_{456}]+\tfrac{1}{11}\Gs{4}[-\tfrac{32}{21}t_{23}]+\tfrac{1}{13}\Gd{5}[\tfrac{3}{2}t_{23}+\tfrac{3}{2}t_{456}]+\tfrac{1}{13}\Gs{5}[5t_{23}+5t_{456}]\\\nonumber
    \tilde{T}^{(v)}_{3,1}&=&\tfrac{5}{26}\Gs{5}[t_{12,1}]+\tfrac{514}{63}s_1-\tfrac{136}{27}s_2+\tfrac{4}{9}s_3\\\nonumber
      &&+\tfrac{1}{11}\Gd{4}[\tfrac{2104}{189}t_{23}+\tfrac{4}{3}t_{4,1}+\tfrac{100}{63}t_{5,1}+\tfrac{20}{189}t_{6,1}+\tfrac{80}{27}t_{8,1}+\tfrac{2}{3}t_{12,1}+\tfrac{80}{27}t_{13,1}+\tfrac{280}{27}t_{18,1}]\\\nonumber
      &&+\tfrac{1}{11}\Gs{4}[\tfrac{1088}{189}t_{23}-\tfrac{32}{3}t_{4,1}-\tfrac{256}{63}t_{5,1}+\tfrac{3712}{189}t_{6,1}-\tfrac{1280}{27}t_{8,1}-\tfrac{8}{3}t_{12,1}-\tfrac{320}{27}t_{13,1}-\tfrac{1120}{27}t_{18,1}]\\\nonumber
      &&+\tfrac{1}{13}\Gd{5}[-\tfrac{965}{63}t_{23}-\tfrac{10}{7}t_{5,1}-\tfrac{100}{21}t_{6,1}+\tfrac{20}{3}t_{8,1}]+\tfrac{1}{13}\Gs{5}[-\tfrac{970}{21}t_{23}+5t_{4,1}-\tfrac{25}{21}t_{5,1}-\tfrac{1475}{63}t_{6,1}+\tfrac{400}{9}t_{8,1}+\tfrac{100}{9}t_{13,1}+\tfrac{350}{9}t_{18,1}]\\\nonumber
    \tilde{T}^{(v)}_{4,1}&=&\tfrac{5}{26}\Gs{5}[t_{10,1}]-\tfrac{58}{21}s_1+\tfrac{5}{3}s_2-\tfrac{2}{9}s_3\\\nonumber
      &&+\Gd{4}[-\tfrac{20}{63}t_{23}-\tfrac{2}{33}t_{4,1}-\tfrac{10}{231}t_{5,1}-\tfrac{2}{693}t_{6,1}-\tfrac{8}{99}t_{8,1}+\tfrac{2}{33}t_{10,1}-\tfrac{14}{99}t_{12,1}-\tfrac{14}{99}t_{13,1}-\tfrac{28}{99}t_{18,1}]\\\nonumber
      &&+\tfrac{1}{11}\Gs{4}[\tfrac{64}{63}t_{23}+\tfrac{16}{3}t_{4,1}+\tfrac{16}{7}t_{5,1}-\tfrac{304}{63}t_{6,1}+\tfrac{128}{9}t_{8,1}-\tfrac{8}{3}t_{10,1}+\tfrac{56}{9}t_{12,1}+\tfrac{56}{9}t_{13,1}+\tfrac{112}{9}t_{18,1}]\\\nonumber
      &&+\tfrac{1}{13}\Gd{5}[\tfrac{193}{42}t_{23}+\tfrac{3}{7}t_{5,1}+\tfrac{10}{7}t_{6,1}-2t_{8,1}]\\\nonumber
      &&+\tfrac{1}{13}\Gs{5}[\tfrac{90}{7}t_{23}-\tfrac{5}{2}t_{4,1}+\tfrac{5}{14}t_{5,1}+\tfrac{295}{42}t_{6,1}-\tfrac{40}{3}t_{8,1}-\tfrac{35}{6}t_{12,1}-\tfrac{35}{6}t_{13,1}-\tfrac{35}{3}t_{18,1}]\\\nonumber
    \tilde{T}^{(v)}_{5,1}&=&\tfrac{5}{26}\Gs{5}[t_{9,1}]-\tfrac{313}{63}s_1+\tfrac{88}{27}s_2+\tfrac{2}{9}s_3\\\nonumber
      &&+\tfrac{1}{11}\Gd{4}[-\tfrac{1588}{189}t_{23}+\tfrac{2}{3}t_{4,1}+\tfrac{2}{9}t_{5,1}+\tfrac{34}{27}t_{6,1}-\tfrac{56}{27}t_{8,1}+\tfrac{2}{3}t_{9,1}+\tfrac{101}{45}t_{12,1}-\tfrac{20}{27}t_{13,1}-\tfrac{196}{27}t_{18,1}]\\\nonumber
      &&+\tfrac{1}{11}\Gs{4}[-\tfrac{2720}{189}t_{23}-\tfrac{16}{3}t_{4,1}-\tfrac{80}{9}t_{5,1}-\tfrac{688}{27}t_{6,1}+\tfrac{896}{27}t_{8,1}-\tfrac{8}{3}t_{9,1}-\tfrac{404}{45}t_{12,1}+\tfrac{80}{27}t_{13,1}+\tfrac{784}{27}t_{18,1}]\\\nonumber
      &&+\tfrac{1}{13}\Gd{5}[\tfrac{211}{18}t_{23}+t_{5,1}+\tfrac{10}{3}t_{6,1}-\tfrac{14}{3}t_{8,1}]\\\nonumber
      &&+\tfrac{1}{13}\Gs{5}[\tfrac{115}{3}t_{23}+\tfrac{5}{2}t_{4,1}+\tfrac{35}{6}t_{5,1}+\tfrac{385}{18}t_{6,1}-\tfrac{280}{9}t_{8,1}+\tfrac{101}{12}t_{12,1}-\tfrac{25}{9}t_{13,1}-\tfrac{245}{9}t_{18,1}]\\\nonumber
    \tilde{T}^{(v)}_{6,1}&=&\tfrac{4}{7}s_1-\tfrac{1}{3}s_2+\tfrac{1}{11}\Gd{4}[\tfrac{4}{7}t_{23}+\tfrac{4}{21}t_{7,1}+\tfrac{4}{21}t_{8,1}+\tfrac{2}{3}t_{17,1}+\tfrac{2}{3}t_{18,1}]+\tfrac{1}{11}\Gs{4}[-\tfrac{64}{21}t_{7,1}-\tfrac{64}{21}t_{8,1}-\tfrac{8}{3}t_{17,1}-\tfrac{8}{3}t_{18,1}]\\\nonumber
      &&+\Gd{5}[-\tfrac{1}{14}t_{23}+\tfrac{3}{91}t_{7,1}+\tfrac{3}{91}t_{8,1}]+\tfrac{1}{13}\Gs{5}[-\tfrac{20}{7}t_{23}+\tfrac{20}{7}t_{7,1}+\tfrac{20}{7}t_{8,1}+\tfrac{5}{2}t_{17,1}+\tfrac{5}{2}t_{18,1}]\\\nonumber
    \tilde{T}^{(v)}_{7,1}&=&\tfrac{3}{7}s_1-\tfrac{2}{9}s_2+\tfrac{2}{9}s_3+\tfrac{1}{11}\Gd{4}[\tfrac{4}{21}t_{23}+\tfrac{10}{21}t_{456}]+\tfrac{1}{11}\Gs{4}[-\tfrac{32}{21}t_{23}-\tfrac{16}{7}t_{456}]+\tfrac{1}{13}\Gd{5}[-\tfrac{5}{14}t_{23}-\tfrac{3}{7}t_{456}]\\\nonumber
      &&+\tfrac{1}{13}\Gs{5}[-\tfrac{5}{7}t_{23}-\tfrac{5}{14}t_{456}]\\\nonumber
    T^{(v)}_{1,2}&=&\tfrac{5}{26}\Gs{5}[t_{1,2}+t_{2,2}+\tfrac{1}{9}t_{23}s_2+\tfrac{2}{9}t_{4,1}\Gv{4}[t_{23}]+t_{12,1}(\tfrac{1}{2}t_{23}-\tfrac{7}{18}\Gv{4}[t_{23}])],
\eeq
}\noindent
where $\tilde{T}^{(v)}_{1,1}\equiv \gamma_v$ from \eqref{eq: gamma-v-delta}. 

Allowing for general time-dependent $t_{a,b}$, we find 14 local NLG coefficients, $T_{a,1}^{(v)}$, 7 non-local NLG amplitudes, $\tilde{T}_{a,1}^{(v)}$, and 1 NNLG amplitude, $T_{1,2}^{(v)}$, giving a total of 22 coefficients at a single redshift. Notably, these are not independent from the density coefficients: three of the non-local terms are fixed via
\beq\label{eq: dm-xsec-free}
    \tilde{T}^{(v)}_{1,1}&=&-2\tilde{T}^{(v)}_{4,1}-\tilde{T}^{(v)}_{5,1}+\tfrac{7}{5}T^{(v)}_{2,1}-\tfrac{3}{10}T^{(v)}_{4,1}+\tfrac{7}{5}T^{(v)}_{10,1}+\tfrac{7}{5}T^{(v)}_{11,1}-\tfrac{3}{10}T^{(v)}_{12,1}+T^{(v)}_{13,1}-T^{(v)}_{14,1}+\tfrac{7}{5}\tilde{T}^{(\delta)}_{1,1}\\\nonumber
      &&\,+\,\tfrac{21}{10}\tilde{T}^{(\delta)}_{2,1}+\tfrac{2}{5}T^{(\delta)}_{1,1}-\tfrac{7}{5}T^{(\delta)}_{2,1}+\tfrac{7}{10}T^{(\delta)}_{4,1}\\\nonumber
    \tilde{T}^{(v)}_{2,1}&=&\tfrac{1}{2}\tilde{T}^{(\delta)}_{1,1}, \qquad\tilde{T}^{(v)}_{7,1}=-\tilde{T}^{(\delta)}_{1,1}-\tfrac{3}{2}\tilde{T}^{(\delta)}_{2,1}
\eeq
(with additional terms constrained via integral relations, if the time-dependence is known). If $t_{a,b}$ are time-independent, the local NLG sector collapses to 13 degrees of freedom via the relation
\beq\label{eq: dm-xsec-fixed-local}
    T^{(v)}_{12,1}&\to&\tfrac{167}{33}T^{(v)}_{1,1}-\tfrac{1344}{71}T^{(v)}_{2,1}+\tfrac{455}{71}T^{(v)}_{3,1}-\tfrac{2293}{781}T^{(v)}_{4,1}-\tfrac{1792}{71}T^{(v)}_{10,1}-\tfrac{1792}{71}T^{(v)}_{11,1}-\tfrac{780}{71}T^{(v)}_{13,1}+\tfrac{825}{71}T^{(v)}_{14,1}
\eeq
and the non-local sector reduces to 5 coefficients, which are themselves fixed by the local sector, for example $\tilde{T}^{(v)}_{1,1}\to \tfrac{1}{33}(3T^{(v)}_{4,1}-2T^{(v)}_{1,1})$. Moreover, the density sector fixes all-but-one of the local NLG terms, the non-local NLG terms and the local NNLG. In conclusion, we find $22$ ($14$) degrees of freedom in the velocity sector with free $t_{a,b}$ (constant $t_{a,b}$), or $19$ ($1$) when analyzed jointly with the density sector.

\section{Perturbation Theory Kernels}\label{app: kernels}

\definecolor{opnlcol}{rgb}{0.00,0.50,0.10}
\definecolor{opnopcol}{rgb}{0.58,0.58,0.58}
\definecolor{opnlnopcol}{rgb}{0.45,0.64,0.47}
\definecolor{opstarcol}{rgb}{0.28,0.40,0.58}
\definecolor{opnlstarcol}{rgb}{0.00,0.55,0.60}
\newcommand{\opNL}[1]{\textcolor{opnlcol}{#1}}
\newcommand{\opNoP}[1]{\textcolor{opnopcol}{#1}}
\newcommand{\opNLNoP}[1]{\textcolor{opnlnopcol}{#1}}
\newcommand{\opStar}[1]{\textcolor{opstarcol}{#1}}
\newcommand{\opNLStar}[1]{\textcolor{opnlstarcol}{#1}}

\subsection{Real-Space}\label{app: kernels-real}
\noindent Below, we list the full set of LG kernels up to fifth order using the operators defined
in Tab.\,\ref{tab: operators}. To aid interpretation, we show terms that do not contribute to the two-loop power spectrum in \opNoP{gray}, non-local terms in \opNL{green}, and the combination in \opNLNoP{light green}. We adopt the shorthands
$\vq_{i_1\ldots i_n}\equiv \vq_{i_1}+\cdots+\vq_{i_n}$, $\mu_{i,j} \equiv \hq_i\cdot\hq_j$,
$\beta_{i,j} \equiv \vq_i\cdot\vq_j/q_i^2$, $F_{i_1\cdots i_n} = F_n(\vq_{i_1},\ldots,\vq_{i_n})$ and $G_{i_1\cdots i_n} = G_n(\vq_{i_1},\ldots,\vq_{i_n})$.
{\footnotesize
\beq
    &&K^{\rm LG}_1[c_{1,0}](\vk) = c_{1,0}\\\nonumber
    &&K^{\rm LG}_2[c_{1,0},\ldots,c_{3,0}](\vq_1,\vq_2) = c_{1,0}F_{12}\,+\,c_{2,0}\mu_{1,2}^2\,+\,c_{3,0}\\\nonumber
    &&K^{\rm LG}_3[c_{1,0},\ldots,c_{7,0}](\vq_1,\vq_2,\vq_3) = c_{1,0}F_{123}\,+\,2c_{2,0}\mu_{1,23}^2F_{23}\,+\,2c_{3,0}F_{23}\,+\,c_{4,0}\big(\mu_{12,3}^2F_{12}-\mu_{2,3}^2\beta_{1,2}\big)\,+\,c_{5,0}\,+\,c_{6,0}\mu_{1,2}^2\\\nonumber
    &&\spacer\,+\,c_{7,0}\mu_{1,2}\mu_{1,3}\mu_{2,3}\\\nonumber
    &&K^{\rm LG}_4[c_{1,0},\ldots,c_{11,0},c_{18,0},\ldots,c_{21,0}](\vq_1,\vq_2,\vq_3,\vq_4) = c_{1,0}F_{1234}\,+\,c_{2,0}\big(2\mu_{1,234}^2F_{234}+\mu_{12,34}^2F_{12}F_{34}\big)\,+\,c_{3,0}\big(2F_{234}+F_{12}F_{34}\big)\\\nonumber
    &&\spacer\,+\,c_{4,0}\big(\mu_{12,34}^2F_{12}F_{34}-\mu_{2,34}^2F_{34}\beta_{1,2}+2\mu_{123,4}^2F_{123}-\mu_{23,4}^2F_{23}\beta_{1,23}-\mu_{3,4}^2G_{12}\beta_{12,3}\big)\,+\,3c_{5,0}F_{34}\\\nonumber
    &&\spacer\,+\,c_{6,0}\big(\mu_{1,2}^2F_{34}+2\mu_{1,23}^2F_{23}\big)\,+\,3c_{7,0}\mu_{1,2}\mu_{1,34}\mu_{2,34}F_{34}\,+\,c_{8,0}\big(\mu_{12,3}\mu_{12,4}\mu_{3,4}F_{12}-\mu_{2,3}\mu_{2,4}\mu_{3,4}\beta_{1,2}\big)\\\nonumber
    &&\spacer\,+\,c_{9,0}\big(\mu_{12,3}^2F_{12}-\mu_{2,3}^2\beta_{1,2}\big)\\\nonumber
    &&\spacer\,+\,c_{10,0}\big(-\frac{1}{2}\mu_{3,4}^2G_{12}\beta_{12,3}+\frac{1}{2}\mu_{3,4}^2\beta_{1,23}\beta_{2,3}+\mu_{123,4}^2F_{123}-\mu_{23,4}^2F_{23}\beta_{1,23}\big)\\\nonumber
    &&\spacer\,+\,c_{11,0}\big(\mu_{12,34}^2F_{12}F_{34}-\mu_{12,4}^2F_{12}\beta_{3,4}-\mu_{2,34}^2F_{34}\beta_{1,2}+\mu_{2,4}^2\beta_{1,2}\beta_{3,4}\big)\opNoP{\,+\,c_{18,0}}\opNoP{\,+\,c_{19,0}\mu_{1,2}\mu_{1,3}\mu_{2,3}}\\\nonumber
    &&\spacer\opNoP{\,+\,c_{20,0}\mu_{1,2}^2}\opNoP{\,+\,c_{21,0}\mu_{1,2}^2\mu_{3,4}^2}\\\nonumber
    &&K^{\rm LG}_5[c_{1,0},\ldots,c_{29,0}](\vq_1,\vq_2,\vq_3,\vq_4,\vq_5) = c_{1,0}F_{12345}\,+\,c_{2,0}\big(2\mu_{1,2345}^2F_{2345}+2\mu_{12,345}^2F_{12}F_{345}\big)\,+\,c_{3,0}\big(2F_{2345}+2F_{12}F_{345}\big)\\\nonumber
    &&\spacer\,+\,c_{4,0}\big(\mu_{12,345}^2F_{12}F_{345}-\mu_{2,345}^2F_{345}\beta_{1,2}+2\mu_{123,45}^2F_{123}F_{45}-\mu_{23,45}^2F_{23}F_{45}\beta_{1,23}-\mu_{3,45}^2F_{45}G_{12}\beta_{12,3}\\\nonumber
    &&\spacer\qquad+3\mu_{1234,5}^2F_{1234}-\mu_{234,5}^2F_{234}\beta_{1,234}-\mu_{34,5}^2F_{34}G_{12}\beta_{12,34}-\mu_{4,5}^2G_{123}\beta_{123,4}\big)\\\nonumber
    &&\spacer\,+\,c_{5,0}\big(3F_{345}+3F_{23}F_{45}\big)\,+\,c_{6,0}\big(\mu_{1,2}^2F_{345}+2\mu_{1,23}^2F_{23}F_{45}+2\mu_{1,234}^2F_{234}+\mu_{12,34}^2F_{12}F_{34}\big)\\\nonumber
    &&\spacer\,+\,c_{7,0}\big(3\mu_{1,2}\mu_{1,345}\mu_{2,345}F_{345}+3\mu_{1,23}\mu_{1,45}\mu_{23,45}F_{23}F_{45}\big)\\\nonumber
    &&\spacer\,+\,c_{8,0}\big(2\mu_{12,3}\mu_{12,45}\mu_{3,45}F_{12}F_{45}-2\mu_{2,3}\mu_{2,45}\mu_{3,45}F_{45}\beta_{1,2}+2\mu_{123,4}\mu_{123,5}\mu_{4,5}F_{123}-\mu_{23,4}\mu_{23,5}\mu_{4,5}F_{23}\beta_{1,23}\\\nonumber
    &&\spacer\qquad-\mu_{3,4}\mu_{3,5}\mu_{4,5}G_{12}\beta_{12,3}\big)\\\nonumber
    &&\spacer\,+\,c_{9,0}\big(\mu_{12,3}^2F_{12}F_{45}-\mu_{2,3}^2F_{45}\beta_{1,2}+\mu_{12,34}^2F_{12}F_{34}-\mu_{2,34}^2F_{34}\beta_{1,2}+2\mu_{123,4}^2F_{123}-\mu_{23,4}^2F_{23}\beta_{1,23}-\mu_{3,4}^2G_{12}\beta_{12,3}\big)\\\nonumber
    &&\spacer\,+\,c_{10,0}\big(-\frac{1}{2}\mu_{3,45}^2F_{45}G_{12}\beta_{12,3}+\frac{1}{2}\mu_{3,45}^2F_{45}\beta_{1,23}\beta_{2,3}+\mu_{123,45}^2F_{123}F_{45}-\mu_{23,45}^2F_{23}F_{45}\beta_{1,23}-2\mu_{234,5}^2F_{234}\beta_{1,234}\\\nonumber
    &&\spacer\qquad+3\mu_{1234,5}^2F_{1234}+\frac{1}{2}\mu_{34,5}^2F_{34}\beta_{1,234}\beta_{2,34}+\frac{1}{2}\mu_{4,5}^2G_{12}\beta_{12,34}\beta_{3,4}+\frac{1}{2}\mu_{4,5}^2G_{23}\beta_{1,234}\beta_{23,4}\\\nonumber
    &&\spacer\qquad-\frac{3}{2}\mu_{34,5}^2F_{34}G_{12}\beta_{12,34}-\mu_{4,5}^2G_{123}\beta_{123,4}\big)\\\nonumber
    &&\spacer\\\nonumber
    &&\spacer\,+\,c_{11,0}\big(4\mu_{12,345}^2F_{12}F_{345}-4\mu_{2,345}^2F_{345}\beta_{1,2}-2\mu_{12,45}^2F_{12}F_{45}\beta_{3,45}-2\mu_{12,5}^2F_{12}G_{34}\beta_{34,5}+2\mu_{2,45}^2F_{45}\beta_{1,2}\beta_{3,45}\\\nonumber
    &&\spacer\qquad+2\mu_{2,5}^2G_{34}\beta_{1,2}\beta_{34,5}\big)\\\nonumber
    &&\spacer\,+\,c_{12,0}\big(\mu_{12,34}\mu_{12,5}\mu_{34,5}F_{12}F_{34}-\mu_{12,4}\mu_{12,5}\mu_{4,5}F_{12}\beta_{3,4}-\mu_{2,34}\mu_{2,5}\mu_{34,5}F_{34}\beta_{1,2}+\mu_{2,4}\mu_{2,5}\mu_{4,5}\beta_{1,2}\beta_{3,4}\big)\\\nonumber
    &&\spacer\,+\,c_{13,0}\big(\mu_{12,34}^2F_{12}F_{34}-\mu_{12,4}^2F_{12}\beta_{3,4}-\mu_{2,34}^2F_{34}\beta_{1,2}+\mu_{2,4}^2\beta_{1,2}\beta_{3,4}\big)\\\nonumber
    &&\spacer\,+\,c_{14,0}\big(-\frac{1}{2}\mu_{3,4}\mu_{3,5}\mu_{4,5}G_{12}\beta_{12,3}+\frac{1}{2}\mu_{3,4}\mu_{3,5}\mu_{4,5}\beta_{1,23}\beta_{2,3}+\mu_{123,4}\mu_{123,5}\mu_{4,5}F_{123}-\mu_{23,4}\mu_{23,5}\mu_{4,5}F_{23}\beta_{1,23}\big)\\\nonumber
    &&\spacer\,+\,c_{15,0}\big(-\frac{1}{2}\mu_{3,4}^2G_{12}\beta_{12,3}+\frac{1}{2}\mu_{3,4}^2\beta_{1,23}\beta_{2,3}+\mu_{123,4}^2F_{123}-\mu_{23,4}^2F_{23}\beta_{1,23}\big)\\\nonumber
    &&\spacer\,+\,c_{16,0}\big(-\frac{1}{2}\mu_{3,45}^2F_{45}G_{12}\beta_{12,3}+\frac{1}{2}\mu_{3,45}^2F_{45}\beta_{1,23}\beta_{2,3}+\frac{1}{2}\mu_{3,5}^2G_{12}\beta_{12,3}\beta_{4,5}\\\nonumber
    &&\spacer\qquad-\frac{1}{2}\mu_{3,5}^2\beta_{1,23}\beta_{2,3}\beta_{4,5}+\mu_{123,45}^2F_{123}F_{45}-\mu_{123,5}^2F_{123}\beta_{4,5}-\mu_{23,45}^2F_{23}F_{45}\beta_{1,23}+\mu_{23,5}^2F_{23}\beta_{1,23}\beta_{4,5}\big)\\\nonumber
    &&\spacer\\\nonumber
    &&\spacer\,+\,c_{17,0}\big(\frac{1}{12}\mu_{4,5}^2G_{12}\beta_{12,34}\beta_{3,4}-\frac{1}{12}\mu_{4,5}^2\beta_{1,234}\beta_{2,34}\beta_{3,4}+\frac{1}{2}\mu_{1234,5}^2F_{1234}-\frac{1}{2}\mu_{234,5}^2F_{234}\beta_{1,234}\\\nonumber
    &&\spacer\qquad-\frac{1}{4}\mu_{34,5}^2F_{34}G_{12}\beta_{12,34}+\frac{1}{4}\mu_{34,5}^2F_{34}\beta_{1,234}\beta_{2,34}-\frac{1}{6}\mu_{4,5}^2G_{123}\beta_{123,4}+\frac{1}{6}\mu_{4,5}^2G_{23}\beta_{1,234}\beta_{23,4}\big)\\\nonumber
    &&\spacer\opNoP{\,+\,4c_{18,0}F_{45}}\opNoP{\,+\,c_{19,0}\big(\mu_{1,2}\mu_{1,3}\mu_{2,3}F_{45}+3\mu_{1,2}\mu_{1,34}\mu_{2,34}F_{34}\big)}\opNoP{\,+\,c_{20,0}\big(2\mu_{1,2}^2F_{45}+2\mu_{1,23}^2F_{23}\big)}\\\nonumber
    &&\spacer\opNoP{\,+\,4c_{21,0}\mu_{1,2}^2\mu_{3,45}^2F_{45}}\opNoP{\,+\,c_{22,0}\big(\mu_{12,3}^2F_{12}-\mu_{2,3}^2\beta_{1,2}\big)}\opNoP{\,+\,c_{23,0}\big(\mu_{12,3}^2\mu_{4,5}^2F_{12}-\mu_{2,3}^2\mu_{4,5}^2\beta_{1,2}\big)}\\\nonumber
    &&\spacer\opNoP{\,+\,c_{24,0}\big(\mu_{12,3}\mu_{12,4}\mu_{3,4}F_{12}-\mu_{2,3}\mu_{2,4}\mu_{3,4}\beta_{1,2}\big)}\opNoP{\,+\,c_{25,0}}\opNoP{\,+\,c_{26,0}\mu_{1,2}\mu_{1,3}\mu_{2,3}}\opNoP{\,+\,c_{27,0}\mu_{1,2}^2}\\\nonumber
    &&\spacer\opNoP{\,+\,c_{28,0}\mu_{4,5}^2\mu_{1,2}\mu_{1,3}\mu_{2,3}}\opNoP{\,+\,c_{29,0}\mu_{1,2}^2\mu_{3,4}^2}.
\eeq
}\noindent

Similarly, the NLG kernels up to third order in fields are given by
{\footnotesize
\beq
    &&K_1^{\rm NLG}[c_{1,1}](\vk) =  -c_{1,1}k^2\\\nonumber
    &&K_2^{\rm NLG}[c_{1,1},\ldots, c_{5,1}](\vq_1,\vq_2) = -c_{1,1}q_{12}^2F_{12}-c_{2,1}q_{12}^2\mu_{1,2}^2-c_{3,1}q_{12}^2-c_{4,1}q_{12}^2\mu_{12,2}^2-c_{5,1}q_2^2\\\nonumber
    &&K_3^{\rm NLG}[c_{1,1},\ldots, c_{17,1},\tilde c_{1,1},\tilde c_{2,1}](\vq_1,\vq_2,\vq_3) = -c_{1,1}q_{123}^2F_{123}-2c_{2,1}q_{123}^2\mu_{1,23}^2F_{23}-2c_{3,1}q_{123}^2F_{23}\\\nonumber
    &&\spacer\,-\,c_{4,1}q_{123}^2\big(\mu_{123,1}^2F_{23}+\mu_{123,12}^2F_{12}\big)\,-\,c_{5,1}\big(q_{23}^2F_{23}+q_3^2F_{12}\big)\\\nonumber
    &&\spacer\,-\,c_{6,1}q_{123}^2\big(\mu_{12,3}^2F_{12}-\mu_{2,3}^2\beta_{1,2}\big)\\\nonumber
    &&\spacer\,-\,c_{7,1}q_{123}^2\big(\mu_{1,23}\mu_{123,1}\mu_{123,23}F_{23}-\mu_{1,3}\mu_{123,1}\mu_{123,3}\beta_{2,3}\big)\\\nonumber
    &&\spacer\opNoP{\,-\,c_{8,1}q_{123}^2\,-\,c_{9,1}q_{123}^2\mu_{123,1}^2\,-\,c_{10,1}q_{123}^2\mu_{1,2}^2\,-\,c_{11,1}q_{123}^2\mu_{123,1}^2\mu_{2,3}^2}\\\nonumber
    &&\spacer\opNoP{\,-\,c_{12,1}q_{123}^2\mu_{1,2}\mu_{1,3}\mu_{2,3}\,-\,c_{13,1}q_{123}^2\mu_{123,1}\mu_{1,2}\mu_{123,2}}\\\nonumber
    &&\spacer\opNoP{\,-\,c_{14,1}q_3^2\,-\,c_{15,1}q_3^2\mu_{2,3}^2\,-\,c_{16,1}q_1q_3\mu_{1,3}\mu_{2,3}^2}\\\nonumber
    &&\spacer\opNoP{\,+\,c_{17,1}\Big(2q_{123}^2\big(\mu_{1,23}\mu_{123,1}\mu_{123,23}F_{23}-\mu_{1,3}\mu_{123,1}\mu_{123,3}\beta_{2,3}\big)-\big(q_3^2+q_{123}^2\big)\big(\mu_{12,3}^2F_{12}-\mu_{2,3}^2\beta_{1,2}\big)\Big)}\\\nonumber
    &&\spacer\opNL{\,+\,\tfrac{1}{2}\tilde c_{1,1}q_{123}^2\beta_{1,23}\Big(F_{23}\big(1-\mu_{123,23}^2\big)-\beta_{2,3}\big(\mu_{23,3}^2-\mu_{123,3}^2\big)\Big)}\\\nonumber
    &&\spacer\opNLNoP{\,-\,\tilde c_{2,1}q_{123}^2\Big(\big(1-\mu_{123,1}^2\big)\big(F_{23}-\beta_{2,3}\mu_{23,3}^2\big)-\tfrac{7}{5}\big(\mu_{1,23}\mu_{123,1}\mu_{123,23}F_{23}-\mu_{1,3}\mu_{123,1}\mu_{123,3}\beta_{2,3}\big)\Big)}.
\eeq
}\noindent
Finally, the single NNLG kernel is
{\footnotesize
\beq
    K_1^{\rm NNLG}[c_{1,2}](\vk) &=& c_{1,2}k^4.
\eeq
}\noindent

\subsection{Redshift-Space}\label{app: kernels-rsd}
\noindent The LG redshift-space kernels can be computed from the mapping \eqref{eq: rsd-mapping} and take the following form up to fifth order:
{\footnotesize
\beq
    &&Z^{\rm LG}_1(\vk) = K^{\rm LG}_1+f\mu^2\\\nonumber
    &&Z^{\rm LG}_2(\vq_1,\vq_2) = K^{\rm LG}_{12}+fk\mu\Big(K^{\rm LG}_1\frac{\nu_2}{q_2}+\frac{\nu_{12}}{q_{12}}G_{12}\Big)+\frac{(fk\mu)^2}{2}\frac{\nu_1}{q_1}\frac{\nu_2}{q_2}\\\nonumber
    &&Z^{\rm LG}_3(\vq_1,\vq_2,\vq_3) = K^{\rm LG}_{123}+fk\mu\Big(K^{\rm LG}_{12}\frac{\nu_3}{q_3}+K^{\rm LG}_1\frac{\nu_{23}}{q_{23}}G_{23}+\frac{\nu_{123}}{q_{123}}G_{123}\Big)+\frac{(fk\mu)^2}{2}\Big(K^{\rm LG}_1\frac{\nu_2}{q_2}\frac{\nu_3}{q_3}+2\frac{\nu_{12}}{q_{12}}G_{12}\frac{\nu_3}{q_3}\Big)+\frac{(fk\mu)^3}{6}\frac{\nu_1}{q_1}\frac{\nu_2}{q_2}\frac{\nu_3}{q_3}\\\nonumber
    &&Z^{\rm LG}_4(\vq_1,\vq_2,\vq_3,\vq_4) = K^{\rm LG}_{1234}+fk\mu\Big(K^{\rm LG}_{123}\frac{\nu_4}{q_4}+K^{\rm LG}_{12}\frac{\nu_{34}}{q_{34}}G_{34}+K^{\rm LG}_1\frac{\nu_{234}}{q_{234}}G_{234}+\frac{\nu_{1234}}{q_{1234}}G_{1234}\Big)\\\nonumber
    &&\qquad\qquad+\frac{(fk\mu)^2}{2}\Big(K^{\rm LG}_{12}\frac{\nu_3}{q_3}\frac{\nu_4}{q_4}+2K^{\rm LG}_1\frac{\nu_{23}}{q_{23}}G_{23}\frac{\nu_4}{q_4}+2\frac{\nu_{123}}{q_{123}}G_{123}\frac{\nu_4}{q_4}+\frac{\nu_{12}}{q_{12}}G_{12}\frac{\nu_{34}}{q_{34}}G_{34}\Big)\\\nonumber
    &&\qquad\qquad+\frac{(fk\mu)^3}{6}\Big(K^{\rm LG}_1\frac{\nu_2}{q_2}\frac{\nu_3}{q_3}\frac{\nu_4}{q_4}+3\frac{\nu_{12}}{q_{12}}G_{12}\frac{\nu_3}{q_3}\frac{\nu_4}{q_4}\Big)+\frac{(fk\mu)^4}{24}\frac{\nu_1}{q_1}\frac{\nu_2}{q_2}\frac{\nu_3}{q_3}\frac{\nu_4}{q_4}\\\nonumber
    &&Z^{\rm LG}_5(\vq_1,\vq_2,\vq_3,\vq_4,\vq_5) = K^{\rm LG}_{12345}+fk\mu\Big(K^{\rm LG}_{1234}\frac{\nu_5}{q_5}+K^{\rm LG}_{123}\frac{\nu_{45}}{q_{45}}G_{45}+K^{\rm LG}_{12}\frac{\nu_{345}}{q_{345}}G_{345}+K^{\rm LG}_1\frac{\nu_{2345}}{q_{2345}}G_{2345}+\frac{\nu_{12345}}{q_{12345}}G_{12345}\Big)\\\nonumber
    &&\qquad\qquad+\frac{(fk\mu)^2}{2}\Big(K^{\rm LG}_{123}\frac{\nu_4}{q_4}\frac{\nu_5}{q_5}+2K^{\rm LG}_{12}\frac{\nu_{34}}{q_{34}}G_{34}\frac{\nu_5}{q_5}+2K^{\rm LG}_1\frac{\nu_{234}}{q_{234}}G_{234}\frac{\nu_5}{q_5}+K^{\rm LG}_1\frac{\nu_{23}}{q_{23}}G_{23}\frac{\nu_{45}}{q_{45}}G_{45}+2\frac{\nu_{1234}}{q_{1234}}G_{1234}\frac{\nu_5}{q_5}+2\frac{\nu_{123}}{q_{123}}G_{123}\frac{\nu_{45}}{q_{45}}G_{45}\Big)\\\nonumber
    &&\qquad\qquad+\frac{(fk\mu)^3}{6}\Big(K^{\rm LG}_{12}\frac{\nu_3}{q_3}\frac{\nu_4}{q_4}\frac{\nu_5}{q_5}+3K^{\rm LG}_1\frac{\nu_{23}}{q_{23}}G_{23}\frac{\nu_4}{q_4}\frac{\nu_5}{q_5}+3\frac{\nu_{123}}{q_{123}}G_{123}\frac{\nu_4}{q_4}\frac{\nu_5}{q_5}+3\frac{\nu_{12}}{q_{12}}G_{12}\frac{\nu_{34}}{q_{34}}G_{34}\frac{\nu_5}{q_5}\Big)\\\nonumber
    &&\qquad\qquad+\frac{(fk\mu)^4}{24}\Big(K^{\rm LG}_1\frac{\nu_2}{q_2}\frac{\nu_3}{q_3}\frac{\nu_4}{q_4}\frac{\nu_5}{q_5}+4\frac{\nu_{12}}{q_{12}}G_{12}\frac{\nu_3}{q_3}\frac{\nu_4}{q_4}\frac{\nu_5}{q_5}\Big)+\frac{(fk\mu)^5}{120}\frac{\nu_1}{q_1}\frac{\nu_2}{q_2}\frac{\nu_3}{q_3}\frac{\nu_4}{q_4}\frac{\nu_5}{q_5},
\eeq
}\noindent
writing $\nu_i\equiv\hq_i\cdot\hn$, $\mu\equiv\hk\cdot\hn$ and $K_{i_1\cdots i_n} \equiv K_n(\vq_{i_1},\ldots,\vq_{i_n})$. 

Up to third-order in fields, the NLG kernels are
{\footnotesize
\beq
    &&Z^{\rm NLG}_1(\vk) = K^{\rm NLG}_1+fk^2\mu^2(c^{(1)}_{1,1}-\tfrac12 fc^{(2)}_{2,1})-\tfrac12f^2k^2\mu^4(c^{(2)}_{3,1}+fc^{(2)}_{1,1})\\\nonumber
    &&Z^{\rm NLG}_2(\vq_1,\vq_2) = K^{\rm NLG}_{12}+fk\mu\,K^{\rm NLG}_1\frac{\nu_2}{q_2}+fk\mu\Big[c^{(1)}_{1,1}\big[k\mu F_{12}+fk\mu\,q_1\nu_1\frac{\nu_2}{q_2}\big]+c^{(1)}_{2,1}k\mu \mu_{1,2}^2+c^{(1)}_{3,1}k\mu\\\nonumber
    &&\qquad\qquad +c^{(1)}_{4,1}k\big(\mu F_{12}-\nu_2\mu_{12,2}\beta_{1,2}\big)\opNL{+\tilde c^{(1)}_{1,1}k\mu \big(F_{12}-\mu_{12,2}^2\beta_{1,2}\big)}\Big]\\\nonumber
    &&\qquad\qquad -\frac{(fk\mu)^2}{2}\Big[c^{(2)}_{1,1}\big[fk\mu\,\frac{\nu_{12}}{q_{12}}G_{12}+\frac{(fk\mu)^2}{2}\,\frac{\nu_1}{q_1}\frac{\nu_2}{q_2}\big]+c^{(2)}_{2,1}\big[F_{12}+fk\mu\,\frac{\nu_2}{q_2}\big]\\\nonumber
    &&\qquad\qquad +c^{(2)}_{3,1}\big[\mu^2F_{12}+fk\mu\,\nu_1^2\frac{\nu_2}{q_2}\big]+c^{(2)}_{4,1}\big(\mu^2F_{12}-\nu_2^2\beta_{1,2}\big)+c^{(2)}_{5,1}\nu_2^2+c^{(2)}_{6,1}\nu_1\nu_2\mu_{1,2}\opStar{+c^{(2)}_{7,1}\mu_{1,2}^2}\opStar{+c^{(2)}_{8,1}}\Big]\\\nonumber
    &&Z^{\rm NLG}_3(\vq_1,\vq_2,\vq_3) = K^{\rm NLG}_{123}+fk\mu\Big(K^{\rm NLG}_{12}\frac{\nu_3}{q_3}+K^{\rm NLG}_1\frac{\nu_{23}}{q_{23}}G_{23}\Big)+\frac{(fk\mu)^2}{2}\,K^{\rm NLG}_1\frac{\nu_2}{q_2}\frac{\nu_3}{q_3}+fk\mu\Big[\\\nonumber
    &&\qquad\qquad c^{(1)}_{1,1}\big[k\mu F_{123}+fk\mu\,q_1\nu_1\frac{\nu_{23}}{q_{23}}G_{23}+fk\mu\,q_{12}\nu_{12}F_{12}\frac{\nu_3}{q_3}+\frac{(fk\mu)^2}{2}\,q_1\nu_1\frac{\nu_2}{q_2}\frac{\nu_3}{q_3}\big]\\\nonumber
    &&\qquad\qquad +c^{(1)}_{2,1}\big[k\mu \big(\mu_{1,23}^2F_{23}+\mu_{12,3}^2F_{12}\big)+fk\mu\,q_{12}\nu_{12}\mu_{1,2}^2\frac{\nu_3}{q_3}\big]+c^{(1)}_{3,1}\big[k\mu \big(F_{12}+F_{23}\big)+fk\mu\,q_{12}\nu_{12}\frac{\nu_3}{q_3}\big]\\\nonumber
    &&\qquad\qquad +c^{(1)}_{4,1}\big[k\big(2\mu F_{123}-\nu_3\mu_{123,3}\beta_{12,3}G_{12}-\nu_{23}\mu_{123,23}\beta_{1,23}F_{23}\big)+fk\mu\,q_{12}\frac{\nu_3}{q_3}\big(\nu_{12}F_{12}-\nu_2\mu_{12,2}\beta_{1,2}\big)\big]\\\nonumber
    &&\qquad\qquad +c^{(1)}_{5,1}k\mu \big(\mu_{12,3}^2F_{12}-\mu_{2,3}^2\beta_{1,2}\big)\\\nonumber
    &&\qquad\qquad +c^{(1)}_{6,1}k\big(\mu F_{123}+\tfrac{1}{2}\nu_3\mu_{123,3}\beta_{1,23}\beta_{2,3}-\nu_{23}\mu_{123,23}\beta_{1,23}F_{23}-\tfrac{1}{2}\nu_3\mu_{123,3}\beta_{12,3}G_{12}\big)\opNoP{+c^{(1)}_{7,1}k\mu}\opNoP{+c^{(1)}_{8,1}k\mu \mu_{1,2}^2}\\\nonumber
    &&\qquad\qquad \opNoP{+c^{(1)}_{9,1}k\mu \mu_{1,2}\mu_{1,3}\mu_{2,3}}\opNoP{+c^{(1)}_{10,1}q_{12}\nu_{12}\mu_{1,2}^2}\opNoP{+c^{(1)}_{11,1}q_{12}\nu_3\mu_{12,3}\mu_{1,2}^2}\opNoP{+c^{(1)}_{12,1}q_{23}\big(\nu_{23}F_{23}-\nu_3\mu_{23,3}\beta_{2,3}\big)}\\\nonumber
    &&\qquad\qquad \opNoP{+c^{(1)}_{13,1}\big[k\big(\tfrac{10}{13}\nu_{23}\mu_{123,23}\beta_{1,23}F_{23}+\tfrac{35}{39}\mu \mu_{2,3}^2\beta_{1,2}+\tfrac{5}{13}\nu_3\mu_{123,3}\beta_{12,3}G_{12}-\tfrac{10}{13}\mu F_{123}-\tfrac{35}{39}\mu \mu_{12,3}^2F_{12}}\\\nonumber
    &&\qquad\qquad \opNoP{\qquad-\tfrac{5}{13}\nu_3\mu_{123,3}\beta_{1,23}\beta_{2,3}\big)+q_3\nu_3\big(\mu_{12,3}^2F_{12}-\mu_{2,3}^2\beta_{1,2}\big)\big]}\opNoP{+c^{(1)}_{14,1}q_{23}\nu_1\big(\mu_{1,23}F_{23}-\mu_{1,3}\mu_{23,3}\beta_{2,3}\big)}\\\nonumber
    &&\qquad\qquad \opNL{+\tilde c^{(1)}_{1,1}\big[k\mu \big(2F_{123}-\mu_{123,23}^2\beta_{1,23}F_{23}-\mu_{123,3}^2\beta_{12,3}G_{12}\big)+fk\mu\,q_{12}\nu_{12}\frac{\nu_3}{q_3}\big(F_{12}-\mu_{12,2}^2\beta_{1,2}\big)\big]}\\\nonumber
    &&\qquad\qquad \opNL{+\tilde c^{(1)}_{2,1}q_{23}\mu \big(\mu_{1,123}\mu_{1,23}\mu_{23,3}^2\beta_{2,3}-\mu_{1,123}\mu_{1,3}\mu_{23,3}\beta_{2,3}\big)}\\\nonumber
    &&\qquad\qquad \opNLNoP{+\tilde c^{(1)}_{3,1}k\big(\nu_{23}\mu_{123,23}\beta_{1,23}F_{23}+\tfrac{1}{2}\nu_3\mu_{123,3}\beta_{12,3}G_{12}+\tfrac{1}{2}\mu \mu_{123,3}^2\beta_{1,23}\beta_{2,3}-\mu \mu_{123,23}^2\beta_{1,23}F_{23}-\tfrac{1}{2}\mu \mu_{123,3}^2\beta_{12,3}G_{12}}\\\nonumber
    &&\qquad\qquad \opNLNoP{\qquad-\tfrac{1}{2}\nu_3\mu_{123,3}\beta_{1,23}\beta_{2,3}\big)}\\\nonumber
    &&\qquad\qquad \opNLNoP{+\tilde c^{(1)}_{4,1}\big[k\mu \big(\mu_{12,3}^2F_{12}+\mu_{123,23}^2F_{23}-\mu_{123,3}^2\beta_{2,3}-\mu_{2,3}^2\beta_{1,2}\big)+q_3\nu_3\big(2\mu_{2,3}^2\beta_{1,2}-2\mu_{12,3}^2F_{12}\big)\big]}\\\nonumber
    &&\qquad\qquad \opNLNoP{+\tilde c^{(1)}_{5,1}\big[k\mu \big(\mu_{123,23}\mu_{1,123}\mu_{1,23}F_{23}-\mu_{123,3}\mu_{1,123}\mu_{1,3}\beta_{2,3}\big)+q_3\nu_3\big(\mu_{2,3}^2\beta_{1,2}-\mu_{12,3}^2F_{12}\big)\big]}\\\nonumber
    &&\qquad\qquad \opNLNoP{+\tilde c^{(1)}_{6,1}k\mu \mu_{1,123}^2\mu_{2,3}^2}\\\nonumber
    &&\qquad\qquad \opNLNoP{+\tilde c^{(1)}_{7,1}\big[k\mu \big(F_{23}+\mu_{1,123}^2\mu_{23,3}^2\beta_{2,3}-\mu_{1,123}^2F_{23}-\mu_{23,3}^2\beta_{2,3}\big)+q_3\nu_3\big(\tfrac{7}{5}\mu_{2,3}^2\beta_{1,2}-\tfrac{7}{5}\mu_{12,3}^2F_{12}\big)\big]}\\\nonumber
    &&\qquad\qquad \opNL{+c_{1,0}\tilde c^{(1)}_{1,1}q_{23}\nu_{23}\big(F_{23}-\mu_{23,3}^2\beta_{2,3}\big)}\Big]\\\nonumber
    &&\qquad\qquad -\frac{(fk\mu)^2}{2}\Big[c^{(2)}_{1,1}\big[fk\mu\,\frac{\nu_{123}}{q_{123}}G_{123}+(fk\mu)^2\,\frac{\nu_{12}}{q_{12}}G_{12}\frac{\nu_3}{q_3}+\frac{(fk\mu)^3}{6}\,\frac{\nu_1}{q_1}\frac{\nu_2}{q_2}\frac{\nu_3}{q_3}\big]\\\nonumber
    &&\qquad\qquad +c^{(2)}_{2,1}\big[F_{123}+fk\mu\,F_{12}\frac{\nu_3}{q_3}+fk\mu\,\frac{\nu_{23}}{q_{23}}G_{23}+\frac{(fk\mu)^2}{2}\,\frac{\nu_2}{q_2}\frac{\nu_3}{q_3}\big]\\\nonumber
    &&\qquad\qquad +c^{(2)}_{3,1}\big[\mu^2F_{123}+fk\mu\,\nu_1^2\frac{\nu_{23}}{q_{23}}G_{23}+fk\mu\,\nu_{12}^2F_{12}\frac{\nu_3}{q_3}+\frac{(fk\mu)^2}{2}\,\nu_1^2\frac{\nu_2}{q_2}\frac{\nu_3}{q_3}\big]\\\nonumber
    &&\qquad\qquad +c^{(2)}_{4,1}\big[2\mu^2F_{123}-\nu_3^2\beta_{12,3}G_{12}-\nu_{23}^2\beta_{1,23}F_{23}+fk\mu\,\frac{\nu_3}{q_3}\big(\nu_{12}^2F_{12}-\nu_2^2\beta_{1,2}\big)\big]\\\nonumber
    &&\qquad\qquad +c^{(2)}_{5,1}\big[\nu_3^2F_{12}+\nu_{23}^2F_{23}+fk\mu\,\nu_2^2\frac{\nu_3}{q_3}\big]+c^{(2)}_{6,1}\big[\nu_1\nu_{23}\mu_{1,23}F_{23}+\nu_{12}\nu_3\mu_{12,3}F_{12}+fk\mu\,\nu_1\nu_2\mu_{1,2}\frac{\nu_3}{q_3}\big]\\\nonumber
    &&\qquad\qquad \opStar{+c^{(2)}_{7,1}\big[\mu_{1,23}^2F_{23}+\mu_{12,3}^2F_{12}+fk\mu\,\mu_{1,2}^2\frac{\nu_3}{q_3}\big]}\opStar{+c^{(2)}_{8,1}\big[F_{12}+F_{23}+fk\mu\,\frac{\nu_3}{q_3}\big]}\\\nonumber
    &&\qquad\qquad \opStar{+c^{(2)}_{9,1}\nu_1\big(\nu_{23}\mu_{1,23}F_{23}-\nu_3\mu_{1,3}\beta_{2,3}\big)}\opStar{+c^{(2)}_{10,1}\big(\nu_{23}^2F_{23}-\nu_3^2\beta_{2,3}\big)}\opStar{+c^{(2)}_{11,1}\big(\mu_{12,3}^2F_{12}-\mu_{2,3}^2\beta_{1,2}\big)}\\\nonumber
    &&\qquad\qquad \opNoP{+c^{(2)}_{12,1}\big(\mu^2F_{123}+\tfrac{101}{30}\nu_1\nu_{23}\mu_{1,23}F_{23}+\tfrac{7}{3}\nu_3^2\beta_{2,3}+\tfrac{7}{6}\mu_{2,3}^2\beta_{1,2}+\tfrac{1}{2}\nu_3^2\beta_{1,23}\beta_{2,3}-\tfrac{7}{3}\nu_{23}^2F_{23}-\tfrac{7}{6}\mu_{12,3}^2F_{12}}\\\nonumber
    &&\qquad\qquad \opNoP{\qquad-\nu_{23}^2\beta_{1,23}F_{23}-\tfrac{101}{30}\nu_1\nu_3\mu_{1,3}\beta_{2,3}-\tfrac{1}{2}\nu_3^2\beta_{12,3}G_{12}\big)}\opNoP{+c^{(2)}_{13,1}\nu_2\nu_3\mu_{2,3}}\opNoP{+c^{(2)}_{14,1}}\opNoP{+c^{(2)}_{15,1}\mu_{1,2}^2}\\\nonumber
    &&\qquad\qquad \opNoP{+c^{(2)}_{16,1}\mu_{1,2}\mu_{1,3}\mu_{2,3}}\opNoP{+c^{(2)}_{17,1}\nu_3^2\mu_{1,2}^2}\opNoP{+c^{(2)}_{18,1}\nu_3^2}\Big],
\eeq
}\noindent
while the single NNLG kernel is
{\footnotesize
\beq
    &&Z^{\rm NNLG}_1(\vk) = K^{\rm NNLG}_1-fk^4\mu^2(c^{(1)}_{1,2}-\opStar{\tfrac12 fc^{(2)}_{2,2}})+f^2k^4\mu^4(\tfrac12 c^{(2)}_{1,2}-\opStar{\tfrac16 fc^{(3)}_{2,2}}+\opStar{\tfrac1{24} f^2c^{(4)}_{2,2}})-f^3k^4\mu^6(\tfrac16 c^{(3)}_{1,2}-\opStar{\tfrac1{24} fc^{(4)}_{3,2}}-\opStar{\tfrac1{24} f^2c^{(4)}_{1,2}}).
\eeq
}\noindent
As before, terms in \opNoP{gray} and \opNLNoP{light green} do not contribute to the two-loop power spectrum; terms in \opStar{blue} are degenerate with other power spectrum contributions at fixed $f(z)$, and can be omitted unless higher-order statistics are added.

\subsection{Correlators}\label{app: kernels-correlators}
\noindent We close by listing the deterministic contributions (including counterterms) for all statistics fixed by the fifth-order EFT theory developed in this work. First, the two-loop redshift-space two-loop power spectrum reads (from \eqref{eq: Pdef}):
\newcommand{\Pir}[1]{\mathcal{P}_{\rm L}^{(#1)}}
{\footnotesize
\beq\label{eq: Pdef-rsd}
    &&P^{(s),\rm det}_{gg}(\vk)=P^{(s),\rm det}_{\rm tree}(\vk)+P^{(s),\rm det}_{\rm 1-loop}(\vk)+P^{(s),\rm det}_{\rm 2-loop}(\vk)\\\nonumber
    &&P^{(s),\rm det}_{\rm tree}(\vk) = |Z_{1}^{\rm LG}(\vk)|^2\Pir{2}(\vk)\\\nonumber
    &&P^{(s),\rm det}_{\rm 1-loop}(\vk) = \int_{\vp}\left[6Z_{1}^{\rm LG}(\vk)[Z_3^{\rm LG}](\vk,\vp,-\vp)\Pir{1}(\vk)\Pir{1}(\vp) + 2|Z_2^{\rm LG}(\vp,\vk-\vp)|^2\Pir{1}(\vp)\Pir{1}(\vk-\vp)\right]\\\nonumber
    &&\qquad\qquad\qquad\,+\,2Z_{1}^{\rm NLG}(\vk)Z_{1}^{\rm LG}(\vk)\Pir{1}(\vk)\\\nonumber
    &&P^{(s),\rm det}_{\rm 2-loop}(\vk) = \int_{\vp\,\vq}\left[30Z_{1}^{\rm LG}(\vk)[[Z_5^{\rm LG}]](\vk,\vp,-\vp,\vq,-\vq)\Pir{0}(\vk)\Pir{0}(\vp)\Pir{0}(\vq)\right.\\\nonumber
    &&\qquad\qquad\qquad\qquad\,+\,24Z_2^{\rm LG}(\vp,\vk-\vp)[Z_4^{\rm LG}](-\vp,\vp-\vk,\vq,-\vq)\Pir{0}(\vk-\vp)\Pir{0}(\vp)\Pir{0}(\vq)\\\nonumber
    &&\qquad\qquad\qquad\qquad\,+\,9[Z_3^{\rm LG}](\vk,\vp,-\vp)[Z_3^{\rm LG}](\vk,\vq,-\vq)\Pir{0}(\vk)\Pir{0}(\vp)\Pir{0}(\vq)\\\nonumber
    &&\qquad\qquad\qquad\qquad\,+\,\left.6\left|Z_3^{\rm LG}(\vp,\vq,\vk-\vp-\vq)\right|^2\Pir{0}(\vk-\vp-\vq)\Pir{0}(\vp)\Pir{0}(\vq)\right]\\\nonumber
    &&\qquad\qquad\qquad\,+\,\int_{\vp}\left[6\left(Z_1^{\rm NLG}(\vk)[Z_3^{\rm LG}](\vk,\vp,-\vp)+Z_1^{\rm LG}(\vk)[Z_3^{\rm NLG}](\vk,\vp,-\vp)\right)\Pir{0}(\vk)\Pir{0}(\vp)\right.\\\nonumber
    &&\qquad\qquad\qquad\qquad\,+\,\left.4Z_{2}^{\rm LG}(\vp,\vk-\vp)Z_2^{\rm NLG}(\vp,\vk-\vp)\Pir{0}(\vp)\Pir{0}(\vk-\vp)\right]\\\nonumber
    &&\qquad\qquad\qquad\,+\,\left(2Z_1^{\rm NNLG}(\vk)Z_1^{\rm LG}(\vk)+Z_1^{\rm NLG}(\vk)Z_1^{\rm NLG}(\vk)\right)\Pir{0}(\vk),
\eeq
}\noindent
where $[Z_n]$ and $[[Z_n]]$ are renormalized kernels (\S\ref{sec: renorm-rsd}), and $\Pir{i}$ are damped linear power spectra (\S\ref{sec: ir-resum}). In practice, this must be supplemented with the stochastic contributions given in \eqref{eq: stoch-Pk}, with the tadpole $k\to 0$ contributions subtracted. The one-loop model used in previous works is recovered by omitting $P^{(s),\rm det}_{2-\rm loop}$ and replacing $\Pir{n}$ with $\Pir{n-1}$. The one-loop bispectrum can be computed similarly:
{\footnotesize
\beq\label{eq: Bdef-rsd}
    &&B_{ggg}^{(s),\rm det}(\vk_1,\vk_2,\vk_3) = B_{\rm tree}^{(s),\rm det}(\vk_1,\vk_2,\vk_3) + B_{\rm 1-loop}^{(s),\rm det}(\vk_1,\vk_2,\vk_3)\\\nonumber
    &&B_{\rm tree}^{(s),\rm det}(\vk_1,\vk_2,\vk_3) = 2Z^{\rm LG}_2(\vk_1,\vk_2)Z^{\rm LG}_1(\vk_1)Z^{\rm LG}_1(\vk_2)\Pir{1}(\vk_1)\Pir{1}(\vk_2)+\text{2 perms.}\\\nonumber
    &&B_{\rm 1-loop}^{(s),\rm det}(\vk_1,\vk_2,\vk_3) = 8\int_{\vp}Z_2^{\rm LG}(\vp,\vk_1-\vp)Z_2^{\rm LG}(-\vp,\vk_2+\vp)Z_2^{\rm LG}(\vp-\vk_1,-\vk_2-\vp)\Pir{0}(\vp)\Pir{0}(\vk_1-\vp)\Pir{0}(\vk_2+\vp)\\\nonumber
    &&\quad\qquad\qquad\qquad\qquad\,+\,6\int_{\vp}Z_1^{\rm LG}(\vk_1)Z_2^{\rm LG}(\vp,\vk_2-\vp)Z_3^{\rm LG}(-\vk_1,-\vp,\vp-\vk_2)\Pir{0}(\vk_1)\Pir{0}(\vp)\Pir{0}(\vk_2-\vp)+\text{5 perms.}\\\nonumber
    &&\quad\qquad\qquad\qquad\qquad\,+\,6Z_2^{\rm LG}(\vk_1,\vk_2)Z_1^{\rm LG}(\vk_2)\Pir{0}(\vk_1)\Pir{0}(\vk_2)\int_{\vp}[Z_3^{\rm LG}](\vk_1,\vp,-\vp)\Pir{0}(\vp)+\text{5 perms.}\\\nonumber
    &&\quad\qquad\qquad\qquad\qquad\,+\,12Z_1^{\rm LG}(\vk_1)Z_1^{\rm LG}(\vk_2)\Pir{0}(\vk_1)\Pir{0}(\vk_2)\int_{\vp}[Z_4^{\rm LG}](-\vk_1,-\vk_2,\vp,-\vp)\Pir{0}(\vp)+\text{2 perms.}\\\nonumber
    &&\quad\qquad\qquad\qquad\qquad\,+\,2\left(Z_2^{\rm NLG}(\vk_1,\vk_2)Z_1^{\rm LG}(\vk_1)Z_1^{\rm LG}(\vk_2)+Z_2^{\rm LG}(\vk_1,\vk_2)\left[Z_1^{\rm NLG}(\vk_1)Z_1^{\rm LG}(\vk_2)+Z_1^{\rm LG}(\vk_1)Z_1^{\rm NLG}(\vk_2)\right]\right)\\\nonumber
    &&\quad\qquad\qquad\qquad\qquad\qquad\qquad\times\,\Pir{0}(\vk_1)\Pir{0}(\vk_2)+\text{2 perms.},
\eeq
}\noindent
matching \citep{Philcox:2022frc,DAmico:2022ukl}, except with a fully anisotropic treatment of IR resummation. The one-loop trispectrum is given by
{\footnotesize
\beq\label{eq: Tdef-rsd}
    &&T_{gggg}^{(s),\rm det}(\vk_1,\vk_2,\vk_3,\vk_4) = T_{\rm tree}^{(s),\rm det}(\vk_1,\vk_2,\vk_3,\vk_4) + T_{\rm 1-loop}^{(s),\rm det}(\vk_1,\vk_2,\vk_3,\vk_4)\\\nonumber
    &&T_{\rm tree}^{(s),\rm det} = 6Z_1^{\rm LG}(\vk_1)Z_1^{\rm LG}(\vk_2)Z_1^{\rm LG}(\vk_3)Z_3^{\rm LG}(\vk_1,\vk_2,\vk_3)\Pir{1}(\vk_1)\Pir{1}(\vk_2)\Pir{1}(\vk_3)+\text{3 perms.}\\\nonumber
    &&\qquad\qquad\quad\,+\,4Z_1^{\rm LG}(\vk_1)Z_2^{\rm LG}(-\vk_1,\vk_{12})Z_2^{\rm LG}(\vk_{12},\vk_4)Z_1^{\rm LG}(\vk_4)\Pir{1}(\vk_1)\Pir{1}(\vk_{12})\Pir{1}(\vk_4)+\text{11 perms.}\\\nonumber
    &&T_{\rm 1-loop}^{(s),\rm det} = 60\int_{\vp}Z_1^{\rm LG}(\vk_1)Z_1^{\rm LG}(\vk_2)Z_1^{\rm LG}(\vk_3)[Z_5^{\rm LG}](\vk_1,\vk_2,\vk_3,\vp,-\vp)\Pir{0}(\vk_1)\Pir{0}(\vk_2)\Pir{0}(\vk_3)\Pir{0}(\vp)+\text{3 perms.}\\\nonumber
    &&\qquad\qquad\quad\,+\,24\int_{\vp}Z_1^{\rm LG}(\vk_1)Z_1^{\rm LG}(\vk_2)Z_2^{\rm LG}(-\vk_1,\vk_{13})[Z_4^{\rm LG}](\vk_2,\vk_{13},\vp,-\vp)\Pir{0}(\vk_1)\Pir{0}(\vk_2)\Pir{0}(\vk_{13})\Pir{0}(\vp)+\text{23 perms.}\\\nonumber
    &&\qquad\qquad\quad\,+\,24\int_{\vp}Z_1^{\rm LG}(\vk_1)Z_1^{\rm LG}(\vk_2)Z_2^{\rm LG}(\vk_3-\vp,\vp)Z_4^{\rm LG}(\vk_1,\vk_2,\vk_3-\vp,\vp)\Pir{0}(\vk_1)\Pir{0}(\vk_2)\Pir{0}(\vk_3-\vp)\Pir{0}(\vp)+\text{11 perms.}\\\nonumber
    &&\qquad\qquad\quad\,+\,18\int_{\vp}Z_1^{\rm LG}(\vk_1)Z_1^{\rm LG}(\vk_2)[Z_3^{\rm LG}](\vk_3,\vp,-\vp)Z_3^{\rm LG}(\vk_1,\vk_2,\vk_3)\Pir{0}(\vk_1)\Pir{0}(\vk_2)\Pir{0}(\vk_3)\Pir{0}(\vp)+\text{11 perms.}\\\nonumber
    &&\qquad\qquad\quad\,+\,18\int_{\vp}Z_1^{\rm LG}(\vk_1)Z_1^{\rm LG}(\vk_2)Z_3^{\rm LG}(-\vk_1,\vk_{13}-\vp,\vp)Z_3^{\rm LG}(\vk_2,\vk_{13}-\vp,\vp)\Pir{0}(\vk_1)\Pir{0}(\vk_2)\Pir{0}(\vk_{13}-\vp)\Pir{0}(\vp)+\text{11 perms.}\\\nonumber
    &&\qquad\qquad\quad\,+\,12\int_{\vp}Z_1^{\rm LG}(\vk_1)Z_2^{\rm LG}(-\vk_1,\vk_{12})[Z_3^{\rm LG}](\vk_3,\vp,-\vp)Z_2^{\rm LG}(\vk_{12},\vk_3)\Pir{0}(\vk_1)\Pir{0}(\vk_{12})\Pir{0}(\vk_3)\Pir{0}(\vp)+\text{23 perms.}\\\nonumber
    &&\qquad\qquad\quad\,+\,12\int_{\vp}Z_1^{\rm LG}(\vk_1)Z_2^{\rm LG}(-\vk_1,\vk_{12})Z_2^{\rm LG}(\vk_3-\vp,\vp)Z_3^{\rm LG}(\vk_{12},\vk_3-\vp,\vp)\Pir{0}(\vk_1)\Pir{0}(\vk_{12})\Pir{0}(\vk_3-\vp)\Pir{0}(\vp)+\text{23 perms.}\\\nonumber
    &&\qquad\qquad\quad\,+\,24\int_{\vp}Z_1^{\rm LG}(\vk_1)Z_2^{\rm LG}(\vp-\vk_3,\vk_{23}-\vp)Z_2^{\rm LG}(\vk_3-\vp,\vp)Z_3^{\rm LG}(\vk_1,\vk_{23}-\vp,\vp)\\\nonumber
    &&\qquad\qquad\qquad\qquad\quad\,\times\,\Pir{0}(\vk_1)\Pir{0}(\vk_3-\vp)\Pir{0}(\vk_{23}-\vp)\Pir{0}(\vp)+\text{11 perms.}\\\nonumber
    &&\qquad\qquad\quad\,+\,16\int_{\vp}Z_2^{\rm LG}(\vk_{13}-\vp,\vp-\vk_3)Z_2^{\rm LG}(\vp-\vk_{13},-\vk_4-\vp)Z_2^{\rm LG}(\vk_3-\vp,\vp)Z_2^{\rm LG}(\vk_4+\vp,-\vp)\\\nonumber
    &&\qquad\qquad\qquad\qquad\quad\times\,\Pir{0}(\vk_{13}-\vp)\Pir{0}(\vk_3-\vp)\Pir{0}(\vk_4+\vp)\Pir{0}(\vp)+\text{2 perms.}\\\nonumber
    &&\qquad\qquad\quad\,+\,6Z_1^{\rm LG}(\vk_1)Z_1^{\rm LG}(\vk_2)Z_1^{\rm LG}(\vk_3)Z_3^{\rm NLG}(\vk_1,\vk_2,\vk_3)\Pir{0}(\vk_1)\Pir{0}(\vk_2)\Pir{0}(\vk_3)+\text{3 perms.}\\\nonumber
    &&\qquad\qquad\quad\,+\,6Z_1^{\rm NLG}(\vk_1)Z_1^{\rm LG}(\vk_2)Z_1^{\rm LG}(\vk_3)Z_3^{\rm LG}(\vk_1,\vk_2,\vk_3)\Pir{0}(\vk_1)\Pir{0}(\vk_2)\Pir{0}(\vk_3)+\text{11 perms.}\\\nonumber
    &&\qquad\qquad\quad\,+\,4Z_1^{\rm LG}(\vk_1)Z_2^{\rm NLG}(-\vk_1,\vk_{12})Z_2^{\rm LG}(\vk_{12},\vk_4)Z_1^{\rm LG}(\vk_4)\Pir{0}(\vk_1)\Pir{0}(\vk_{12})\Pir{0}(\vk_4)+\text{23 perms.}\\\nonumber
    &&\qquad\qquad\quad\,+\,4Z_1^{\rm NLG}(\vk_1)Z_2^{\rm LG}(-\vk_1,\vk_{12})Z_2^{\rm LG}(\vk_{12},\vk_4)Z_1^{\rm LG}(\vk_4)\Pir{0}(\vk_1)\Pir{0}(\vk_{12})\Pir{0}(\vk_4)+\text{23 perms.}
\eeq
}\noindent
where $\vk_{ij}\equiv\vk_i+\vk_j$ (e.g., \citep{Bertolini:2016bmt,Steele:2021lnz} in real-space). Finally, the same operators generate the tree-level quadspectrum and the pentaspectrum, which read
{\footnotesize
\beq\label{eq: Qdef-rsd}
    &&Q_{ggggg}^{(s),\rm det}(\vk_1,\ldots,\vk_5) = Q_{\rm tree}^{(s),\rm det}(\vk_1,\ldots,\vk_5)\\\nonumber
    &&Q_{\rm tree}^{(s),\rm det} = 24Z_1^{\rm LG}(\vk_1)Z_1^{\rm LG}(\vk_2)Z_1^{\rm LG}(\vk_3)Z_1^{\rm LG}(\vk_4)Z_4^{\rm LG}(\vk_1,\vk_2,\vk_3,\vk_4)\Pir{0}(\vk_1)\Pir{0}(\vk_2)\Pir{0}(\vk_3)\Pir{0}(\vk_4)+\text{4 perms.}\\\nonumber
    &&\qquad\qquad\quad\,+\,12Z_1^{\rm LG}(\vk_1)Z_1^{\rm LG}(\vk_2)Z_1^{\rm LG}(\vk_3)Z_2^{\rm LG}(-\vk_1,\vk_{14})Z_3^{\rm LG}(\vk_2,\vk_3,\vk_{14})\Pir{0}(\vk_1)\Pir{0}(\vk_2)\Pir{0}(\vk_3)\Pir{0}(\vk_{14})+\text{59 perms.}\\\nonumber
    &&\qquad\qquad\quad\,+\,8Z_1^{\rm LG}(\vk_1)Z_1^{\rm LG}(\vk_2)Z_2^{\rm LG}(-\vk_1,\vk_{13})Z_2^{\rm LG}(-\vk_2,\vk_{24})Z_2^{\rm LG}(\vk_{13},\vk_{24})\Pir{0}(\vk_1)\Pir{0}(\vk_2)\Pir{0}(\vk_{13})\Pir{0}(\vk_{24})+\text{59 perms.}
\eeq
}\noindent
and
{\footnotesize
\beq
    &&\mathcal{P}_{gggggg}^{(s),\rm det}(\vk_1,\ldots,\vk_6) = \mathcal{P}_{\rm tree}^{(s),\rm det}(\vk_1,\ldots,\vk_6)\\\nonumber
    &&\mathcal{P}_{\rm tree}^{(s),\rm det} = 120\left[\prod_{i=1}^{5}Z_1^{\rm LG}(\vk_i)\Pir{0}(\vk_i)\right]Z_5^{\rm LG}(\vk_1,\vk_2,\vk_3,\vk_4,\vk_5)+\text{5 perms.}\\\nonumber
    &&\qquad\qquad\quad\,+\,48\left[\prod_{i=1}^{4}Z_1^{\rm LG}(\vk_i)\Pir{0}(\vk_i)\right]Z_2^{\rm LG}(-\vk_1,\vk_{15})Z_4^{\rm LG}(\vk_2,\vk_3,\vk_4,\vk_{15})\Pir{0}(\vk_{15})+\text{119 perms.}\\\nonumber
    &&\qquad\qquad\quad\,+\,36\left[\prod_{i=1}^{4}Z_1^{\rm LG}(\vk_i)\Pir{0}(\vk_i)\right]Z_3^{\rm LG}(-\vk_1,-\vk_2,\vk_{125})Z_3^{\rm LG}(\vk_3,\vk_4,\vk_{125})\Pir{0}(\vk_{125})+\text{89 perms.}\\\nonumber
    &&\qquad\qquad\quad\,+\,24\left[\prod_{i=1}^{3}Z_1^{\rm LG}(\vk_i)\Pir{0}(\vk_i)\right]Z_2^{\rm LG}(-\vk_1,\vk_{14})Z_2^{\rm LG}(-\vk_2,\vk_{25})Z_3^{\rm LG}(\vk_3,\vk_{14},\vk_{25})\Pir{0}(\vk_{14})\Pir{0}(\vk_{25})+\text{719 perms.}\\\nonumber
    &&\qquad\qquad\quad\,+\,16\left[\prod_{i=1}^{2}Z_1^{\rm LG}(\vk_i)\Pir{0}(\vk_i)\right]Z_2^{\rm LG}(-\vk_1,\vk_{13})Z_2^{\rm LG}(-\vk_2,\vk_{24})Z_2^{\rm LG}(-\vk_{13},\vk_{135})Z_2^{\rm LG}(\vk_{24},\vk_{135})\\\nonumber
    &&\qquad\qquad\qquad\qquad\times\,\Pir{0}(\vk_{13})\Pir{0}(\vk_{24})\Pir{0}(\vk_{135})+\text{359 perms.}
\eeq
}\noindent
respectively, defining $\vk_{ijk}\equiv\vk_i+\vk_j+\vk_k$. As mentioned above, all correlators should be accompanied by stochastic (and mixed deterministic-stochastic) contributions.

\bibliographystyle{apsrev4-1}
\bibliography{refs}

\end{document}